\documentclass[review,12pt]{elsarticle}

\usepackage{amssymb}
\usepackage{amsmath}
\usepackage{xcolor}
\usepackage{siunitx}
\usepackage{url}
\usepackage{algpseudocode}
\usepackage{bm}
\usepackage{multicol}
\usepackage{diagbox}

\definecolor{editColor}{RGB}{0, 102, 204}

\journal{Computer Methods and Programs in Biomedicine}

\begin{document}

\begin{frontmatter}



\title{Stable EEG Source Estimation for Standardized Kalman Filter using Change Rate Tracking}


\author[label1]{Joonas Lahtinen\corref{cor1}} %% Author name
\cortext[cor1]{Corresponding author}
\affiliation[label1]{organization={Faculty of Information Technology and Communication Sciences, Tampere University},%Department and Organization
            %addressline={}, 
            city={Tampere},
            postcode={33720}, 
            %state={},
            country={Finland}}

%% Abstract
\begin{abstract}
{\em Background and Objective:} Localization of brain activity is an important guiding tool for presurgical evaluation and treatment planning, particularly in the context of various brain-related diseases such as epilepsy. Since brain activity is highly dynamic and arises from the firing of neuronal networks, advanced spatiotemporal modeling is needed to capture this time-varying behavior accurately.

{\em Methods:} A new parameter tuning and model utilizing the change rate of brain activity distribution were developed to improve the filtering-parametri\-zation-stability of the otherwise accurate estimation of the recently introduced Standardized Kalman filter. Namely, we propose a backward-differenti\-ation-based measurement model for the change rate. Simulated and real non-invasive electroencephalography data, along with realistic head models from two real subjects, were used in time-evolution tracking and localization experiments focusing on somatosensory evoked potentials. The method was compared to the original Standardized Kalman filter and Standardized Low-resolution Brain Electromagnetic Tomography (sLORETA).

{\em Results:} Results indicate that the proposed parametrization yields high localization accuracy, as the original Standardized Kalman filtering localizes 7 and 6 out of 8, and sLORETA found 8 and 6 of the literature-defined originators of short latency somatosensory evoked potentials. The proposed standardized filtering method identified 7 of 8 expected originators. The change-rate-based model exhibits greater tracking stability than filtering without it against changes in filtering parameters. In addition, the method is more stable against badly set model parameters.

{\em Conclusions:} With new model parameterization, the studied standardized methodologies provide assumably accurate and stable estimations to explore the location and dynamical properties of cortical and subcortical brain activity. The results showing correct sub-thalamic localization demonstrate the significant potential of these methods in the guidance of stereo-electroencephalography sensor placements or the placement of implant electrodes for deep-brain stimulation.
\end{abstract}


% \begin{highlights}
% \item Standardized Kalman filter (SKF) estimates surface and deep brain activity from EEG
% \item New modeling parameterization significantly improves the tracking stability of SKF
% \item Tracking the change rate of brain activity further increases the stability
% \end{highlights}

%% Keywords

\end{frontmatter}



%% Use \subsubsection, \paragraph, \subparagraph commands to 
%% start 3rd, 4th and 5th level sections.
%% Refer following link for more details.
%% https://en.wikibooks.org/wiki/LaTeX/Document_Structure#Sectioning_commands

\section{Introduction}
\label{sc:Intro}
Electroencephalography (EEG) is one of the most non-invasive measurement techniques to record electric potentials on the surface of the human head in time \cite{Knosche2022}. These potentials are caused by a population of relatively synchronously activated neurons. EEG is one of the most direct non-invasive modalities for studying brain functions and tracking the time evolution of brain activity.

For activity tracking, we need to build a model of the conductivity domain, which is the human head. To get a realistic and sufficiently accurate model, we construct the geometry from the magnetic resonance images (MRI) taken from the subject and solve the electric lead field using the finite element method (FEM), as there is no analytic solution to describe the electric properties of such a complex conductivity domain with multiple different conductivity layers in different tissue structures. A head model of this type allows the use of multiple conductivity regions in different parts of the brain and also provides an approximation of the electrical properties in complex, folded brain structures. Hence, brain activity localization is significantly more accurate with realistic head models than with analytical one-to-three-layer conductivity spheres \cite{VanrumsteBart2002realvsspherical}.

Estimating the time-varying brain activity non-invasively from EEG measurements is known to be an ill-posed problem \cite{hamalainen1993}. This is because there are many more possible locations for the source of activity and its summations within the brain than there are individual scalp measurements. Moreover, the measurements are noisy, and the solutions change significantly depending on the model's quality. This yields a linear time-varying inverse problem, and a sound solution for it could be estimated, e.g., using Kalman filtering \cite{Hamid2021,SarkkaSimo2013,Kalman1961}. For the Kalman filter, one needs a measurement model and a time-evolution model for the quantity being tracked and estimated. However, measurable brain activity is known to cause measurable signals on large neuron populations, otherwise traveling along the length of axons while producing an electric field that falls way below the measurement noise level \cite{MEGEEGprimer2017}. Modeling a time-evolution behavior of this kind is a hard task. Additional information requires incorporating measurements from other modalities into the model, making the estimation time-consuming and computationally intensive to be practical. A prominent approach to modeling brain activity is to estimate it as a spatial distribution \cite{GalkaAndreas2004KalmanEEG} rather than a point object, such as a moving dipole. To enhance the localization of the brain activity via bias reduction and increase the measurement noise robustness of the estimation, the estimation of standardized current densities instead of regular ones was proposed. The idea, originating from Pascual-Marqui \cite{PascualMarqui2002}, was then polished and verified \cite{Lahtinen2024Onbias} and applied to the Kalman filtering of distributional brain activity using EEG data \cite{Lahtinen2024SKF}. The original article on Standardized Kalman filtering demonstrated the necessity of standardization for tracking simultaneous deep and surface-level brain activity; however, due to the lack of well-established knowledge of neurodynamics, a simple random walk as the evolution model was considered a safe bet in both Kalman filtering approaches. Even with promising results, the evolution modeling parameters had a notable effect on tracking accuracy.

To address Kalman filtering instability against changes in filtering parameters, we propose a more advanced parametrization for the Standardized Kalman filter and a new variant that uses temporal rate tracking. The change rate state model has the current density part and the change rate we aim to track. Additionally, we use an ad hoc measurement model for the change rate based on backward differentiation formulas (BDF). Our claim is that the BDFs introduce additional filtering-parameter-stability to the estimation, making the evolution parameter selection less volatile. This paper serves as a technical follow-up to the Standardized Kalman filter article, using the same model and similar experiments to make the results comparable. The result, computed using simulations and real data, demonstrates an increase in the filtering-parametrization-stability with the new parameter, and change-rate tracking is found to be measurement-noise-stable even under a highly ill-advised parametrization.

\section{Methods}
The time-varying EEG measurement model is formulated as follows
\begin{equation}\label{eq:measmodel}
    {\bf y}_t=L{\bf x}_t+{\bf r}_t,
\end{equation}
where ${\bf y}_t\in\mathbb{R}^m$ is the measured data at the time step $t$, ${\bf x}_t\in\mathbb{R}^n$ is the brain activity at the inspected moment, ${\bf r}_t\in\mathbb{R}^m$ is the measurement noise, and $L\in\mathbb{R}^{m\times n}$ is so-called lead field matrix that we get as the FEM approximation. 

To solve the problem for each time-step, we employ a Bayesian filtering approach called {\em Kalma filtering} (KF) for standardized estimates ${\bf x}_t\mapsto {\bf z}_t$, where the task is to obtain Gaussian posterior distributed ${\bf z}_t\mid {\bf y}_{1:t}$, which means a random variable ${\bf z}_t$ when measurement data from the initial step to step $t$ is treated as known, i.e., fed to the algorithm. The actual estimation of the brain activity is obtained as {\em maximum a posteriori} estimate $\hat{{\bf z}}_{t\mid t}$ \cite{KaipioSomersalo}, which is, in this case, the posterior mean.

\subsection{Standardized low-resolution brain electromagnetic tomography}
Standardization is an application of the random field theory \cite{BesagJulian1974RandomFields,KemenyJohnG1976RandomFields} in which a statistical field is assigned to a spatial space, a grid with fixed source positions in this case. In the case of standardization, we assign Z-scores to each source location that indicate its likelihood of being the correct one. Even if the forward model itself introduces a depth bias in the system due to the modeled signal decay with distance, the Z-scores are highly unbiased when the sensor separation is sufficiently large \cite{Lahtinen2024Onbias}. Standardized low-resolution brain electromagnetic tomography (sLORETA) \cite{PascualMarqui2002} utilizes Gaussian process regression, also known as Minimum norm estimate \cite{HamalainenMNE} as the time-independent base-method. Let us have a model with Gaussian likelihood ${\bf y}_t\mid {\bf x}_t\sim \mathcal{N}(L{\bf x}_t,R_t)$ and a Gaussian prior written as ${\bf x}_t\sim \mathcal{N}({\bf 0},\theta I)$. As the typical estimation is of the form
\begin{equation}
    \hat{\bf x}_t=\theta L^T(\theta LL^T+R_t)^{-1}{\bf y},
\end{equation}
the standardized estimation is obtained as
\begin{equation}
    \hat{\bf z}_t=\mathrm{Diag}\left(\theta L^T(\theta LL^T+R_t)^{-1}L\right)^{-1/2}\hat{\bf x}_t,
\end{equation}
where $\mathrm{Diag}(\cdot)$ gives a diagonal matrix with the same diagonal elements as the input matrix. The standardization, i.e., the matrix multiplication with the diagonal matrix, can be interpreted as (statistical) standardization of a Gaussian random variable coming from the model evidence \cite{Lahtinen2024Onbias}.

\subsection{Standardized Kalman filtering}
With KF in general, the noise in the measurement model (\ref{eq:measmodel}) is assumed to obey a zero-mean Gaussian ${\bf r}_t\sim \mathcal{N}({\bf 0},R_t)$ with measurement noise covariance $R_t\in\mathbb{R}^{m\times m}$. An evolution model is assigned to describe the time evolution of the normal estimated variable
\begin{equation}
    {\bf x}_{t+1}=A_t{\bf x}_t+{\bf q}_{t+1},
\end{equation}
where ${\bf q}_{t+1}\sim \mathcal{N}({\bf 0},Q_{t+1})$ is also Gaussian and $A_t\in \mathbb{R}^{n\times n}$ is the {\em state-transition model}. The KF uses a Gaussian initial state ${\bf x}_0\sim\mathcal{N}({\bf m},P_0)$ to make the process fully Gaussian. Due to the said Gaussianity, the KF algorithm has this two-part form
\begin{align*}
    \hat{{\bf x}}_{t\mid t-1}&=A_t\hat{{\bf x}}_{t-1\mid t-1}\\
    P_{t\mid t-1}&=A_tP_{t-1\mid t-1}A_t^\mathrm{T}+Q_t
\end{align*}
and
\begin{align*}
    S_t&=LP_{t\mid t-1}L^\mathrm{T}+R_t\\
    K_t&=P_{t\mid t-1}L^\mathrm{T}S_t^{-1}\\
    \hat{\bf x}_{t\mid t}&=\hat{\bf x}_{t\mid t-1}+K_t\left({\bf y}_t-L\hat{\bf x}_{t\mid t-1}\right)\\
    P_{t\mid t}&=P_{t\mid t-1}-K_tS_tK_t^\mathrm{T}\\
    W_t &=\mathrm{Diag}\left(P_{t\mid t-1}^{-1/2}K_tS_tK_t^\mathrm{T}P_{t\mid t-1}^{-1/2}\right)^{-1/2}P_{t\mid t-1}^{-1/2}\\
    \hat{\bf z}_{t\mid t}&=W_t\hat{\bf x}_{t\mid t}.
\end{align*}
where the last step transforms the typical estimate to the standardized estimate \cite{Lahtinen2024SKF}. 

\subsection{Evolution models}
In the papers applying KF to the distributional estimation in a non-standardized \cite{GalkaAndreas2004KalmanEEG} and a standardized manner \cite{Lahtinen2024SKF}, the random walk model 
\begin{equation}\label{eq:RandWalk}
    {\bf x}_{t+1}={\bf x}_t+{\bf q}_{t+1},
\end{equation}
was used with similarly distributed process noise ${\bf q}_{t+1}\sim\mathcal{N}({\bf 0},q\, I)$ with fixed variance $q$. A random walk as the evolution model is a well-suited {\em a priori}  model when the underlying dynamics are not well-understood, as is the case with neural dynamics, especially when enough uncertainty has been placed into the model \cite{GBishopIntroToKF}. It is the most popular guess for state transitions. The transition model does not need to describe the true underlying evolution perfectly since the Kalman filter, as a Bayesian filter, corrects the errors in the prior state by incorporating the data into the estimation, thus forming the so-called {\em posterior estimate} for the state at each time step \cite{SarkkaSimo2013}.

Now, we introduce change rate variable ${\bf v}_t$ and the following evolution model
\begin{align}
    {\bf x}_{t+1}&={\bf x}_t+{\bf v}_t\Delta t +{\bf q}_{t+1},\\
    {\bf v}_{t+1}&={\bf v}_t+{\bf c}_{t+1},
\end{align}
where ${\bf c}_t\sim \mathcal{N}({\bf 0},2q/3\Delta t^2)$ using the following logic: Considering the subsequent time steps
\begin{align}
    {\bf x}_{t+1}&={\bf x}_t+{\bf v}_t\Delta t +{\bf q}_{t+1},\\
    {\bf x}_{t+2}&={\bf x}_{t+1}+{\bf v}_t\Delta t+{\bf c}_{t+1}\Delta t +{\bf q}_{t+2}.
\end{align}
Their difference yields
\begin{equation}
    {\bf x}_{t+2}-{\bf x}_{t+1}={\bf x}_{t+1}-{\bf x}_{t}+{\bf c}_{t+1}\Delta t +{\bf q}_{t+2}-{\bf q}_{t+1}.
\end{equation}
To make this comparable with the random walk model of equal variances $q$, we must have $\mathcal{N}({\bf 0},\Delta t^2C_{t+1}+Q_{t+2}+Q_{t+1})=\mathcal{N}(
{\bf 0},2q\, I)$, where the simplest solution is $Q_t= 2q/3\, I$ and $\Delta t^2C_{t}= 2q/3\, I$ for all $t=1,\cdots,T$.

The change rate model is adding structure to the random-walk evolution model. It makes the dynamical transition between time steps smoother, but less responsive to sudden changes. The smoothness is enforced because the change rate evolves slowly, with fluctuations that depend only on the process noise: the filter expects motion and changes to continue rather than jitter randomly, as in the random walk model. Overall, the change rate can be viewed as a latent variable that enforces continuity in the evolution of the state. To achieve the computational lightness required for high-resolution forward models, the model is designed as linear.

\subsection{Modified measurement model}
In addition to the measurement model (\ref{eq:measmodel}) posed to ${\bf x}_t$, we place another model for the change rate ${\bf v}_t$. We are using backward-differentiation formulas (BDF) \cite{Iserles_BDF} and the data itself to produce the models in the form of
\begin{equation}\label{eq:BDF}
    \sum_{k=0}^n \alpha_k{\bf y}_{t+k}=\Delta t L{\bf v}_{t+n},
\end{equation}
represented here for $n$th order. This model, although somewhat ad hoc due to the lack of change-rate measurements, will provide additional smoothness and stability to the evolution. In actuality, the forward problem that gives the EEG measurements is time-dependent and nonlinear: ${\bf y}(t)=L(t,{\bf x}(t))+{\bf r}_t$. The obtained measurements capture the accurate time-dependence of brain activity; therefore, we focus on them. Since the measurements are provided as samples with a constant sampling frequency, we approximate it at the interval $t\in\left[t_{n-2},t_n\right]$ using Lagrange interpolating polynomial:
\begin{equation}\label{eq:LaplPolyy}
    {\bf y}(t)\approx {\bf y}(t_n)+\frac{t-t_n}{\Delta t}\mathcal{D}\left\lbrace{\bf y}\right\rbrace(t_n)+\frac{(t-t_n)(t-t_{n-1})}{2\Delta t^2}\mathcal{D}^2\left\lbrace{\bf y}\right\rbrace(t_n),
\end{equation}
where $\mathcal{D}$ represents time-difference operator. Notice that the approximation is of order two, meaning that the approximation is exact for quadratic dependence on time \cite{HairerErnst1991}. Now, by derivation, we get
\begin{equation}
    L{\bf v}(t)\approx L\dot{\bf x}(t)=\dot{\bf y}(t)=\frac{1}{\Delta t}\mathcal{D}\left\lbrace{\bf y}\right\rbrace(t_n)+\frac{2t-2t_n+\Delta t}{2\Delta t^2}\mathcal{D}^2\left\lbrace{\bf y}\right\rbrace(t_n),
\end{equation}
where the left-hand side approximation, or assumption, comes from the change-rate evolution model. Computing the first time-derivative at $t_n$ yields
\begin{equation}
    L{\bf v}(t_n)\approx \frac{1}{\Delta t}\mathcal{D}\left\lbrace{\bf y}\right\rbrace(t_n)+\frac{1}{2\Delta t}\mathcal{D}^2\left\lbrace{\bf y}\right\rbrace(t_n).
\end{equation}
The equation above provides us with the second-order backward differentiation formula for the time derivative of the measurements:
\begin{equation}
    \frac{3}{2}{\bf y}_t-2{\bf y}_{t-1}+\frac{1}{2}{\bf y}_{t-2}=\Delta t L{\bf v}_{t}.
\end{equation}
Assuming ${\bf r}_t\sim \mathcal{N}({\bf 0},R)$ independent across the steps, we have for rate change noise covariance $R_v=6.5R/\Delta t^2$.

Gathering the assumptions together, we are assuming that the change in the brain activity over time can be described by an order one approximation, i.e.,
\begin{equation}
    \dot{\bf x}_t\approx {\bf v}_t,
\end{equation}
and the observations can be traced with the piecewise second-order approximation from Eq. (\ref{eq:LaplPolyy}). The intention to connect the change rate variable ${\bf v}_t$ to a smooth approximation of ${\bf y}(t)$ \cite{HairerErnst1991} provides a smooth correlation between estimated activity and the data. The approach could be expected to be stable against additive measurement noise, when a quadratic polynomial can be drawn across the points $\lbrace {\bf y}_{t-2}, {\bf y}_{t-1}, {\bf y}_{t}\rbrace$ fairly accurately for each $t=3,\cdots T$, the model provides then knowledge about the dynamical behavior of the observations. This means that the model provides more factual information about the system than solely the forward equation (\ref{eq:measmodel}). This additional information improves tracking accuracy, which in turn increases robustness against measurement noise. Additionally, the smoothness prevents sporadic, sharp changes that can only be caused by noise. Hence, the model relies more on the actual data and the quadratic approximability than the evolution model; a change away from the ideal parametrization affects it less.

\begin{figure}[tbh!]
    \centering
    \includegraphics[width=0.5\linewidth]{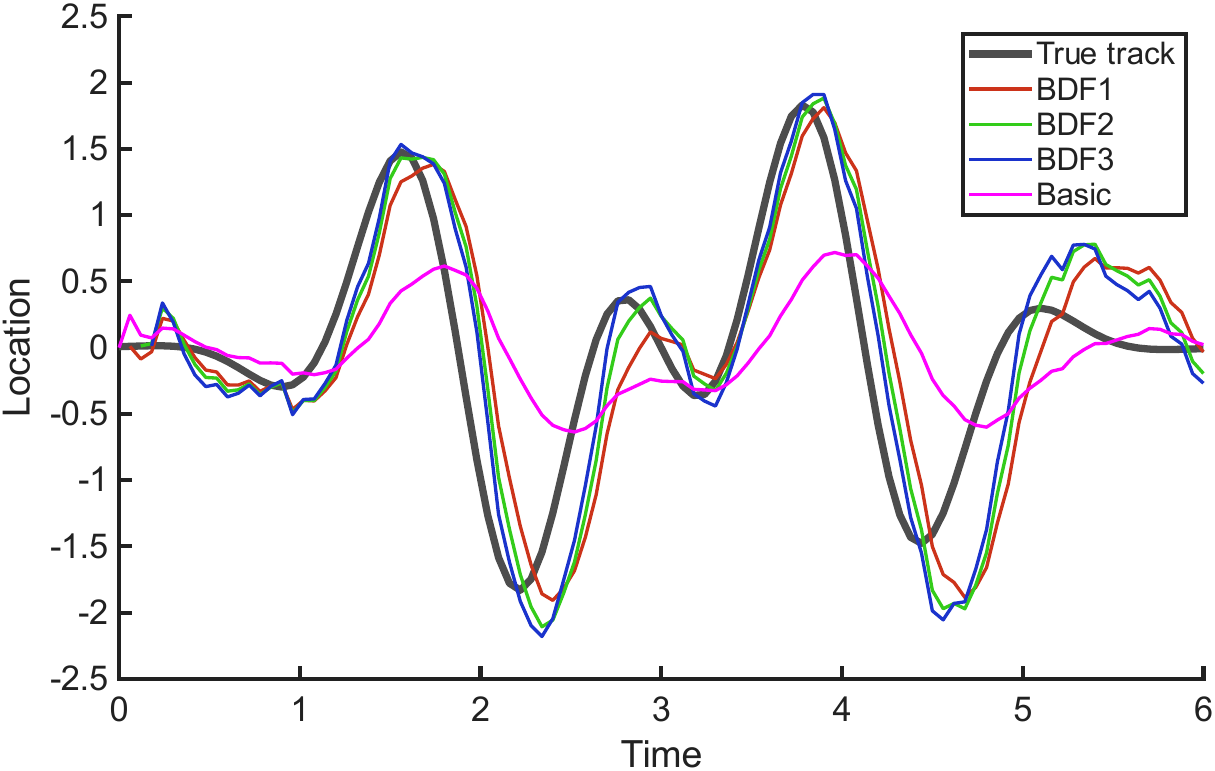}
    \caption{One-dimensional track (bold dark line) that is estimated with Kalman filters with backward differentiation formulas (BDF) of order 1, 2, and 3 (red, green, and blue curves, respectively) used in change rate part of measurement model following Equation (\ref{eq:BDF}) and the normal Kalman filter whose estimate is represented with the magenta curve. From this example, we can see how the tracking gets more accurate with a higher order of BDF.}
    \label{fig:BDFexample}
\end{figure}

From the following simple numerical example in Figure \ref{fig:BDFexample}, where we have a highly variant one-dimensional track that we try to recover using the Kalman filter with different BDF orders from 0 to 3 and without the aid of BDF, we can see that the BDF and increasing its order make the track more accurate and stable. These are welcome properties for the model, especially in this application, as it is hard, if not impossible, to determine how the evolution model -- and its parameters -- should be selected given the neurophysiological phenomena underlying neural dynamics. One of the most realistic models, the Hodgkin-Huxley model, requires high-performance computation to be utilized \cite{LandsmeerHodgkinHuxley2025}. From a clinical point of view, we focus on computationally efficient linearized modeling applicable when high computational power is not available.

Based on the example we verified with our EEG and head models, there is no need to go beyond order 2, as the stability gain is not significantly higher for order 3 than for order 2. Moreover, the known A-stability and L-stability properties of BDFs will break in order 3 and higher \cite{StewartKris1990BDFstability}. These stabilities ensure that the estimated tracks do not blow up asymptotically; instead, the solutions stay bounded.

From now on, we refer to the standardized KF using a random walk as RW-SKF, and to the rate change model with order 2 BDF change rate evolution as BDF2-CR-SKF.

\subsection{Modeling parameters}
Our goal is to use the same parameters for both Standardized Kalman filtering approaches to make the comparison as fair as possible. First of all, we assume that the initial state has no activity, so ${\bf z}_0={\bf 0}$. As initial covariance or rather variances, we use a slightly modified version of the sensitivity weighting approach proposed by Calvetti {\em et al. }\cite{Calvetti2019SensitivityWeight}. The sensitivity weighting is derived from the signal-to-noise ratio (SNR) observed in the data and the Bayesian model. For the Gaussian initial state, we have 
\begin{equation}
    \widehat{\mathrm{SNR}}=\frac{\mathbb{E}\left[\left\|{\bf y}_0\right\|_2^2\right]}{\mathbb{E}\left[\left\|{\bf r}_0\right\|_2^2\right]}=\frac{\theta_k\left\|L_k\right\|_F^2}{\mathrm{Tr}(R_0)}+1,
\end{equation}
where $\widehat{\mathrm{SNR}}$ denotes SNR over noise-contained measurements, $\theta_k$ is the source variance at $k$th mesh element, $L_k$ corresponds the sub-lead field matrix of $k$th mesh element, and $\left\|\cdot\right\|_F$ is the  Frobenius norm. Yielding source position-dependent variances:
\begin{equation}
    \theta_k=\frac{\mathrm{Tr}(R_0)(\widehat{\mathrm{SNR}}-1)}{\left\|L_k\right\|_F^2}.
\end{equation}

Here, we also use the same prior variance parameter to balance the sensitivity of sLORETA.

The process noise variance $q$ is estimated using the improved version of the approach described in \cite{Lahtinen2024SKF}. The technique assumes that the square of the activity strength changes is equal for each source between successive sampling or time steps, and that the activity follows a Wiener process. The derivation is explained in \ref{app:evolutionVar}. The purpose of the improved version is to minimize the number of hand-tunable parameters and, hence, simplify the tuning process. The variance is computed as follows:
\begin{equation}\label{eq:qparameter}
    q = \frac{10^{\rho/20}}{\left\|L\right\|_F^2\,f},
\end{equation}
where $f$ is the sampling frequency, and $\rho$ is tunable decibel value. The value of $\rho$ is suggested to be estimated by using time-independent estimations of the time series computed using simple Gaussian estimates, also known as the Minimum norm estimate \cite{HamalainenMNE}. Then, the variances of time-independent estimates are used to tune $\rho$ by matching the average variance of estimation concentration with $q$. It should be noted that the Minimum norm estimate cannot estimate the activity everywhere in the model due to its bias. With the numerical data used in part of the experiments, the $\rho$ is estimated to be 44 \unit{\decibel}. According to the definition of the variances, if one desires to use time steps larger than the data sampling frequency, $1/f$ should be replaced by the time step size.

The full Change Rate-SKF algorithm is given in \ref{app:CRSKFalgo}.

\subsection{Head models and data}
The heads were modeled as an 18-compartment volume conductor constructed using openly available T1-weighted MRI data from a healthy 49-year-old male (1st subject) \cite{piastra_maria_carla_2020_zenodo} and a healthy 32-year-old male (2nd subject). The tissue conductivities were set to 0.14 \unit{\siemens/\meter} for the white matter, 0.33 \unit{\siemens/\meter} for the grey matter and sub-cortical brain tissue, 0.0064 \unit{\siemens/\meter} for the skull, and 1.79 \unit{\siemens/\meter} for cerebrospinal fluid
(CSF) and ventricles. These values are selected following the studies of Dannhauer {\em et al.} \cite{dannhauer2011} and Shahid {\em et al.} \cite{Shahid2014subcorticalcondu}. The tissue compartments were segmented via FreeSurfer software\footnote{\url{https://surfer.nmr.mgh.harvard.edu/}} with the functions of the SPM12 package \citep{SPM12}. A head model with 1 mm tetrahedral diameter and a finite element method (FEM)- based forward solution was obtained using the Matlab-based Zeffiro Interface toolbox \citep{he2018zeffiro}. The source space contains 10,000 source locations.

The MRI datasets were measured using a MAGNETOM Prisma scanner 3.0 T (Release D13, Siemens Medical Solutions, Erlangen, Germany) with T1 and T2-weighting (T1W/T2W) fast gradient-echo pulse sequence. Somatosensory evoked potentials were measured using 80 AgCl sintered ring electrodes (EASYCAP GmbH, Herrsching, Germany) with 74 EEG channels in the standard 10–10 system. A notch filter was applied to remove interference caused by the 50 Hz power line frequency and the 60 Hz monitor frequency, which the subject watched during the video measurement, thereby reducing alpha-activity. A sampling rate of 1200 \unit{Hz} and an online low-pass filter at 300 Hz were used. A total of 1200 stimuli were recorded for montage averaging, following the guidelines for spinal and subcortical SEPs  \citep{Cruccu2008}.

\subsection{Somatosensory evoked potentials}
\begin{figure*}[t!]
\centering
    \begin{minipage}{0.2\textwidth}
    \centering
        \small{ 14 \unit{\milli\second}}
        \includegraphics[width=0.7\textwidth]{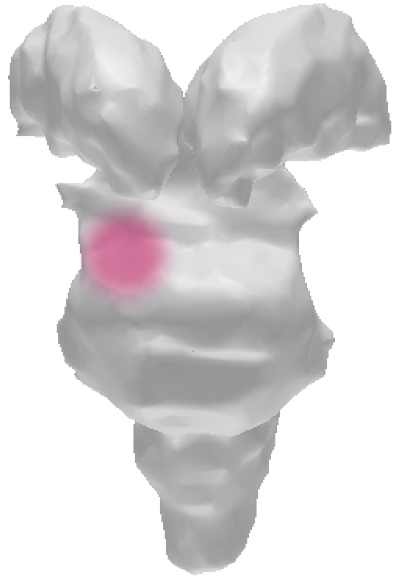}
    \end{minipage}\hspace{0.02\textwidth}\begin{minipage}{0.2\textwidth}
    \centering
    \small{ 16 \unit{\milli\second}}
        \includegraphics[width=0.7\textwidth]{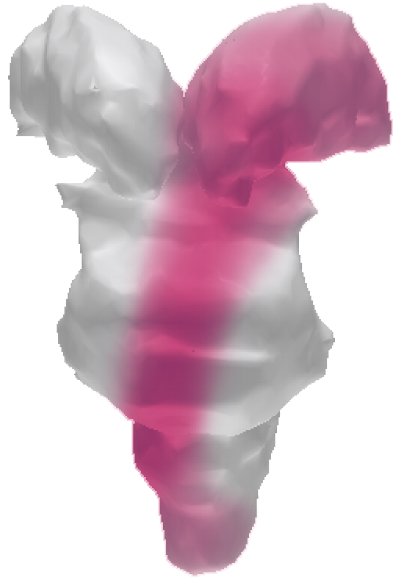}
    \end{minipage}\hspace{0.02\textwidth}\begin{minipage}{0.2\textwidth}
    \centering
    \small{ 20/22/30 \unit{\milli\second}}
        \includegraphics[width=0.7\textwidth]{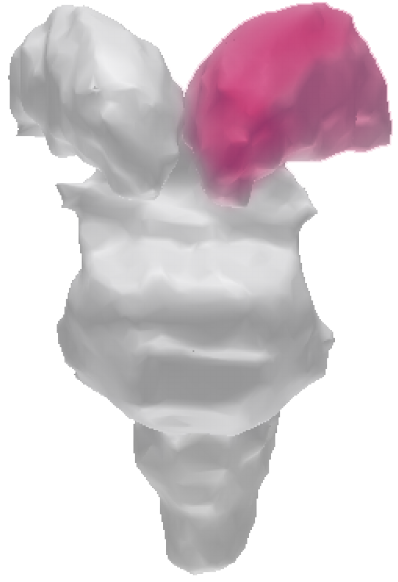}
    \end{minipage}
    
    \begin{minipage}{0.2\textwidth}
    \centering
    \small{ 20 \unit{\milli\second}}
        \includegraphics[width=\textwidth]{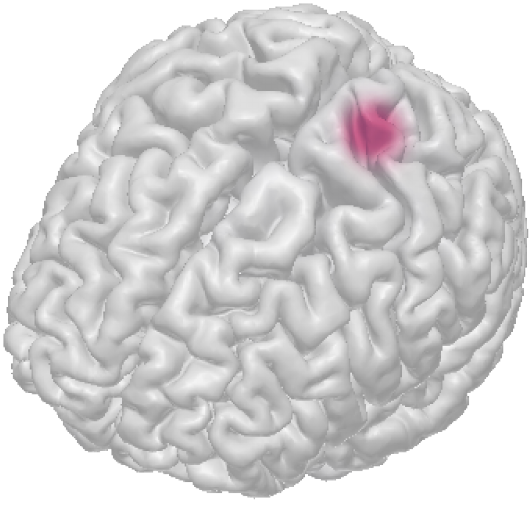}
    \end{minipage}\hspace{0.02\textwidth}\begin{minipage}{0.2\textwidth}
    \centering
    \small{ 22 \unit{\milli\second}}
        \includegraphics[width=\textwidth]{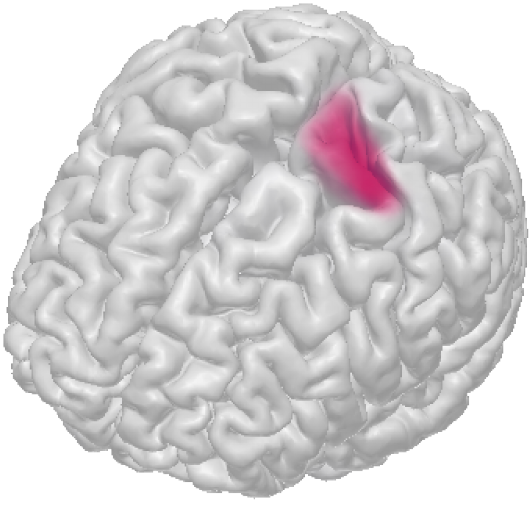}
    \end{minipage}\hspace{0.04\textwidth}\begin{minipage}{0.2\textwidth}
    \centering
    \small{ 30 \unit{\milli\second}}
        \includegraphics[width=\textwidth]{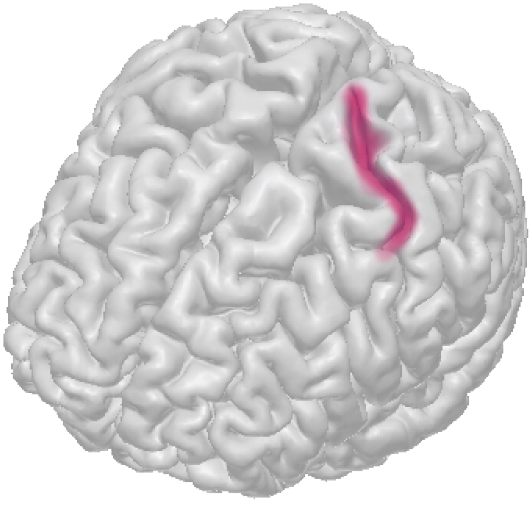}
    \end{minipage}
    \caption{Literature-based ground truth locations of somatosensory evoked potential peaks around 14, 16, 20, 22, and 30 \unit{\milli\second} after stimulus presented by coloring for deep brain structures (upper row) and cortex (bottom row).}
    \label{fig:LiteratureComponents}
\end{figure*}

In this study's experiments, we focus on the somatosensory evoked potentials (SEP), and their short-latency components around 14, 16, 20, 22, and 30 \unit{\milli\second}. The 14 \unit{\milli\second} deep component has been located to the pons (more accurately, the medial lemniscus pathway) by studying epilepsy patients having lesions at that location \cite{NoelPierre1996OoNa}. The 16 \unit{\milli\second} component is generated by a somatosensory volley that travels along the medial lemniscus pathway \cite{Valeriani2014P16}. At 20 \unit{\milli\second}, the somatosensory cortex is activated \cite{DesmedtOzaki1991P20N20,allison1991cortical,buchner1995somatotopy,FuchsManfred1998SEP}, and simultaneous thalamic activity has been found \cite{GotzTheresa2014TIPa} at the ventral posterolateral part of the thalamus. The location of the cortical 22 \unit{\milli\second} component varies across subjects, located either in Brodmann areas 1 or 4 \cite{buchner1995somatotopy}. The left thalamus is found to be activated at the same time \cite{PapadelisChristos2011BaBa}. At 30 \unit{\milli\second}, the cortical component is estimated to be located in the vicinity of the central sulcus \cite{Valeriani2000N30}, and activity in the anterior part of the left thalamus is detected \cite{cebolla2015sensorimotor}. The regions active during these components are displayed in Figure \ref{fig:LiteratureComponents}.

\subsection{Experiments}

\begin{figure*}
\centering
    \begin{minipage}{0.25\textwidth}
        \includegraphics[width=\textwidth]{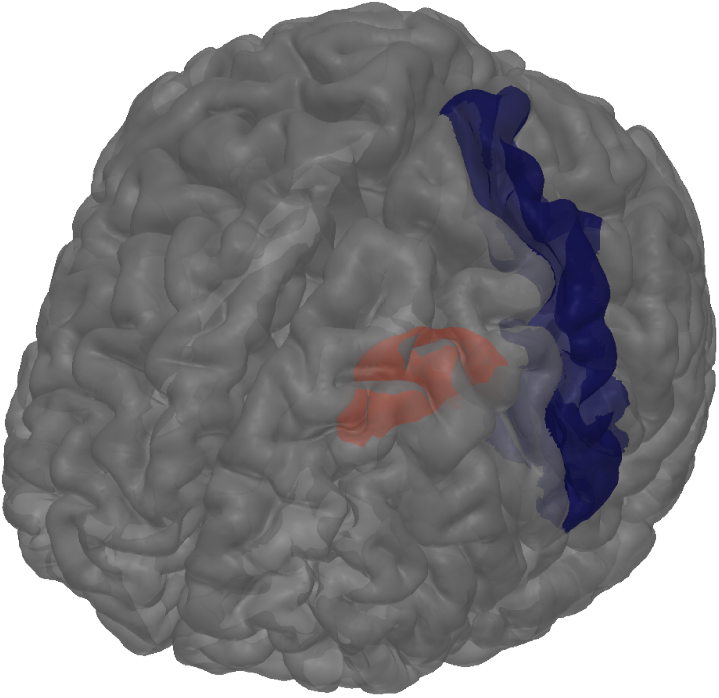}
    \end{minipage}\hspace{0.04\textwidth}\begin{minipage}{0.25\textwidth}
        \includegraphics[width=\textwidth]{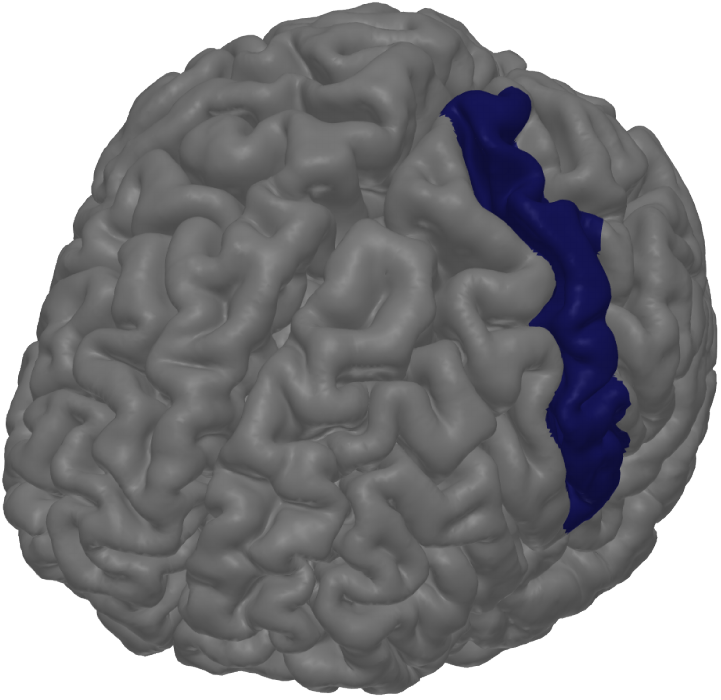}
    \end{minipage}\hspace{0.04\textwidth}\begin{minipage}{0.25\textwidth}
        \includegraphics[width=0.7\textwidth]{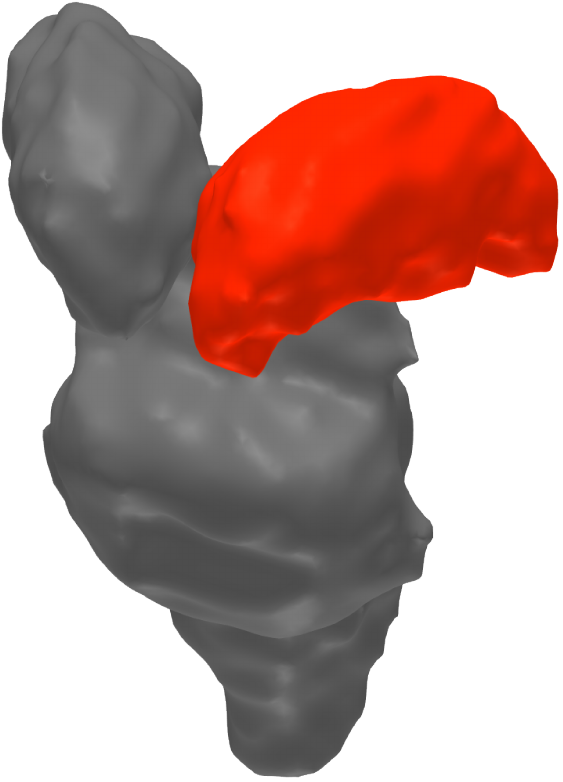}
    \end{minipage}
    \caption{Primary somatosensory cortex (dark blue) from the surface of the brain and left thalamus (red) from deep inside the brain. These are the main parts of the brain from which the short latency evoked potentials are known to emerge, namely around 20 \unit{\milli\second} after the stimulus. Due to their high reliability in activating similarly across sensations, these setups are ideal for use with a known ground truth.}
    \label{fig:BrainSetup}
\end{figure*}

\begin{figure*}
    \centering
    \includegraphics[width=0.4\linewidth]{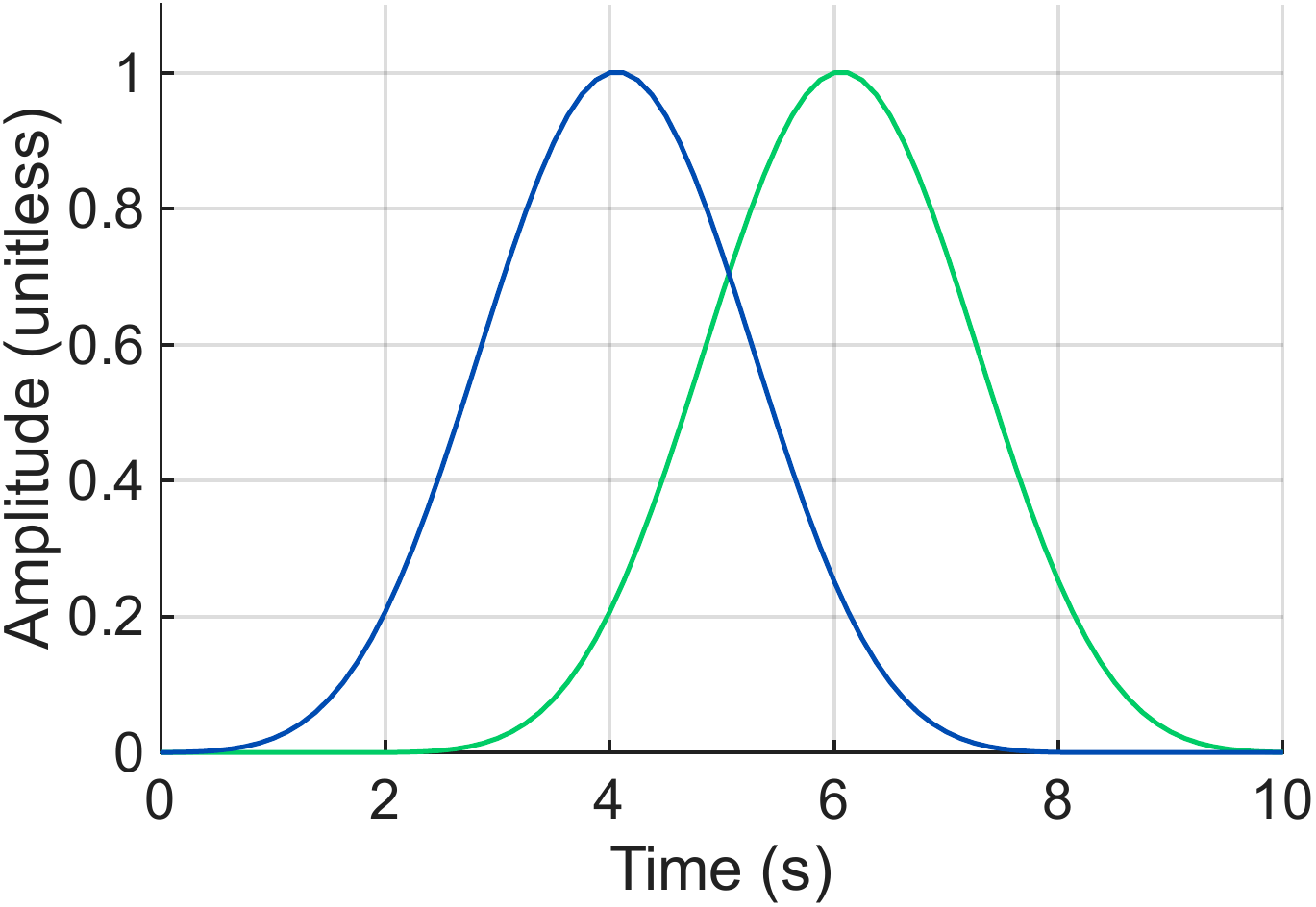}
    \caption{Activity strength tracks of deep (thalamic) activity represented by the blue curve and surface (somatosensory cortex) represented by the green curve. The pulses are modeled as Gaussian bell curves using a millisecond time scale displayed on the x-axis.}
    \label{fig:TrueTracks}
\end{figure*}

We have divided the experiments into two parts: numerical and real. In our numerical experiment, we repeat the measurement-noise-robustness experiment from the original Standardized Kalman filtering paper, simulating deep thalamic activity followed by cortical activity 2 \unit{\milli\second} apart. Then, we generate three sets of synthetic data using a fine-resolution head model with a tetrahedral mesh of 0.8 millimeters and 50,000 source positions to avoid an inversion crime. Each set contained 25 time series with realized Gaussian noise. The used noise levels are 25 \unit{\decibel} (5 \%), 15 \unit{\decibel} (18 \%), and 5 \unit{\decibel} (56 \%). The task for Kalman filters and sLORETA is to track these activity strength curves. The result curves are computed as averages over the left thalamus compartment and primary somatosensory cortex compartment identified from the MRI data using the FreeSurfer segmentation tool (Figure \ref{fig:BrainSetup}). The scaled activity strength tracks are presented in Figure \ref{fig:TrueTracks}. To test the robustness of the parametrization, we used a lower sampling rate of 1250 \unit{\hertz} and a higher sampling rate of 5000 \unit{\hertz}, with the optimal free parameter obtained at 2500 \unit{\hertz}. The lower sampling rate is half of the one used in the previous experiment, and the other one is twice that. The values are rather arbitrary and are used to illustrate that not even a wide range of sampling frequencies can cause issues with tracking when the proposed parametrization is used. The sampling rate of 1250 \unit{\hertz} causes enmeshment of deep and cortical signals with real somatosensory evoked potentials, as demonstrated by numerical experiments that match this real data; therefore, no lower sampling frequency is used, as it would make the components non-separable.

When interpreting the results, we use the following criteria: correctness of the tracking, the width of the 2.5 \% and 97.5 \% quantile interval, the height difference between the deep and surface activity peaks, and how non-aligned these tracks are from each other. The last one means that when deep activity weakens, and the surface activity rises, these tracks do not get enmeshed with each other; e.g., when the surface track rises, the deep track rises as well.

To evaluate the focality, i.e., the volume of concentrated estimation, and the localization accuracy, we use one of 25 \unit{\decibel} synthetic data sets that is estimated, and the estimations at the peak time points are visualized on the surface of the brain model. The synthetic deep and cortical components resemble those obtained at 20 \unit{\milli\second} after a median nerve stimulus. These components are most studied and, therefore, can be simulated best. This connection to real events helps us validate the results obtained from the subject's data.

Using the subjects' real data, we conduct the activity estimation experiment. Due to the low sampling frequency relative to the activity peak frequency and the number of data curves, the tracking experiment is omitted, as it would have solely anecdotal clinical value. As in the previous study, we localize five components from the data peaks at 14, 16, 20, 22, and 30 \unit{\milli\second} after stimulus. The location of the components and the literature review is provided in Figure \ref{fig:LiteratureComponents}.

\begin{figure*}[htb!]
    \centering
    \hspace{1.1cm}\begin{minipage}{0.22\textwidth}
    \centering
    {\bf 0.35 s}
        \includegraphics[width=\textwidth]{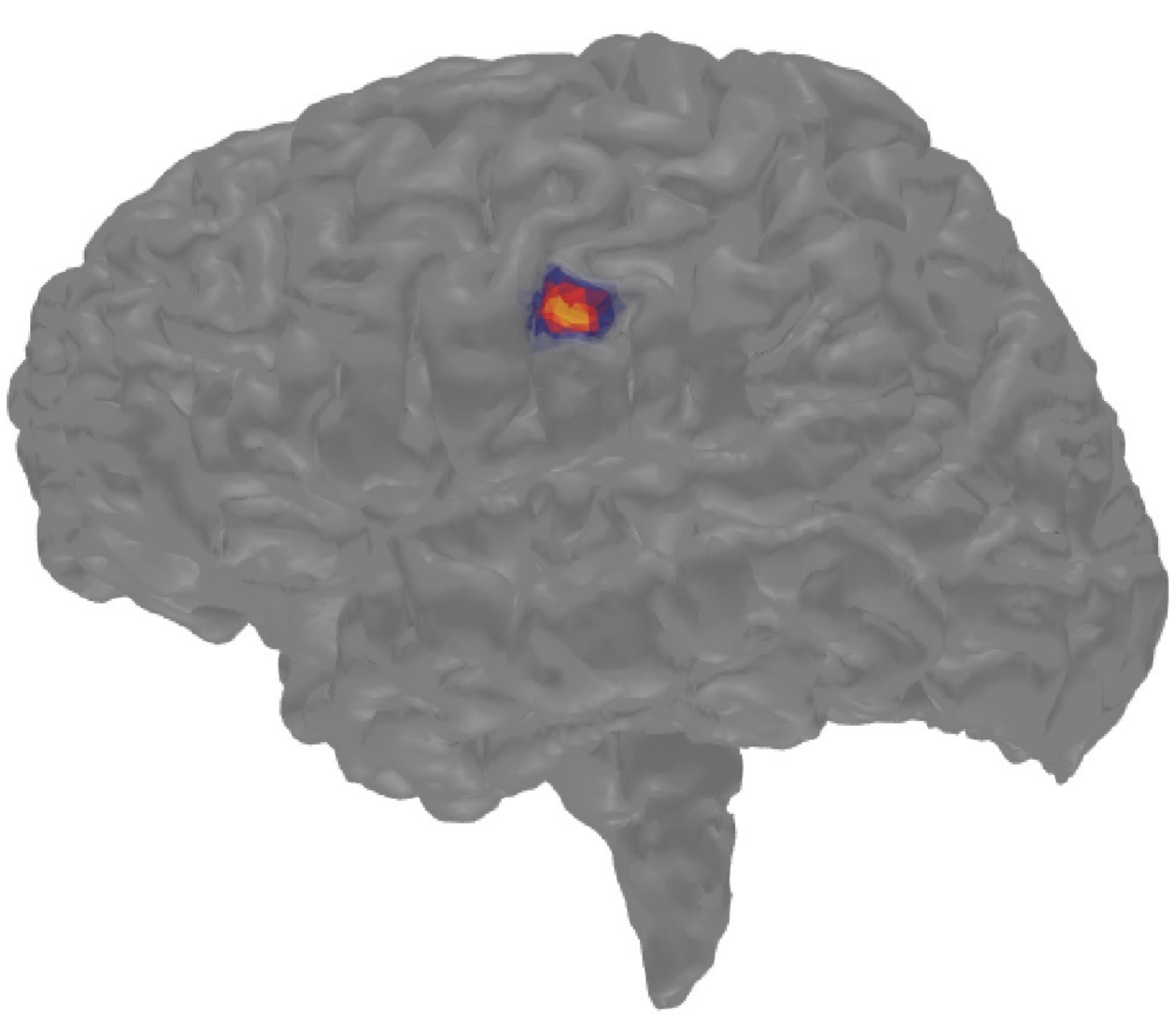}
    \end{minipage}\hspace{0.1cm}\begin{minipage}{0.22\textwidth}
    \centering
    {\bf 0.4 s} 
        \includegraphics[width=\textwidth]{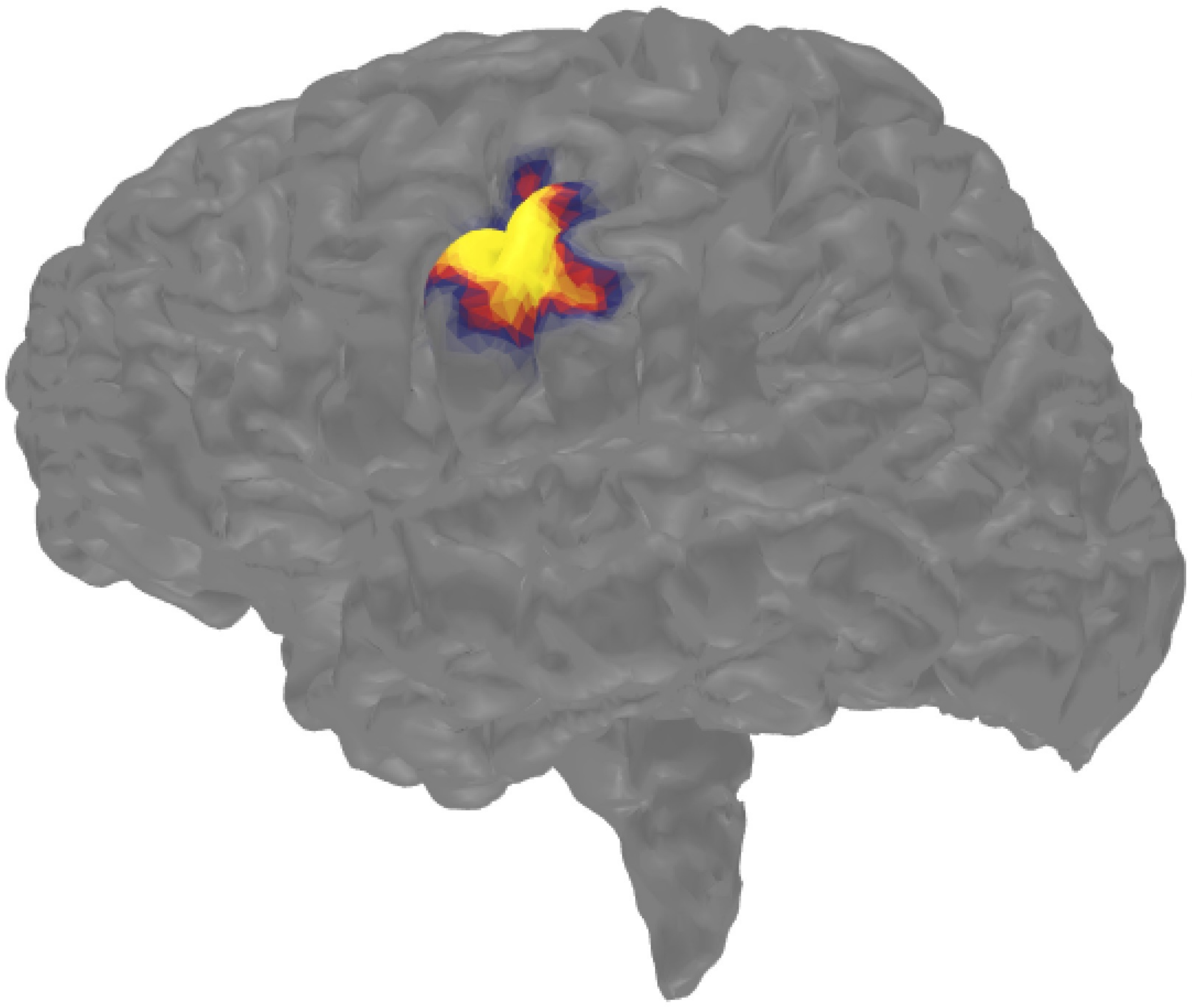}
    \end{minipage}\hspace{0.1cm}\begin{minipage}{0.22\textwidth}
    \centering
    {\bf 0.45 s} 
        \includegraphics[width=\textwidth]{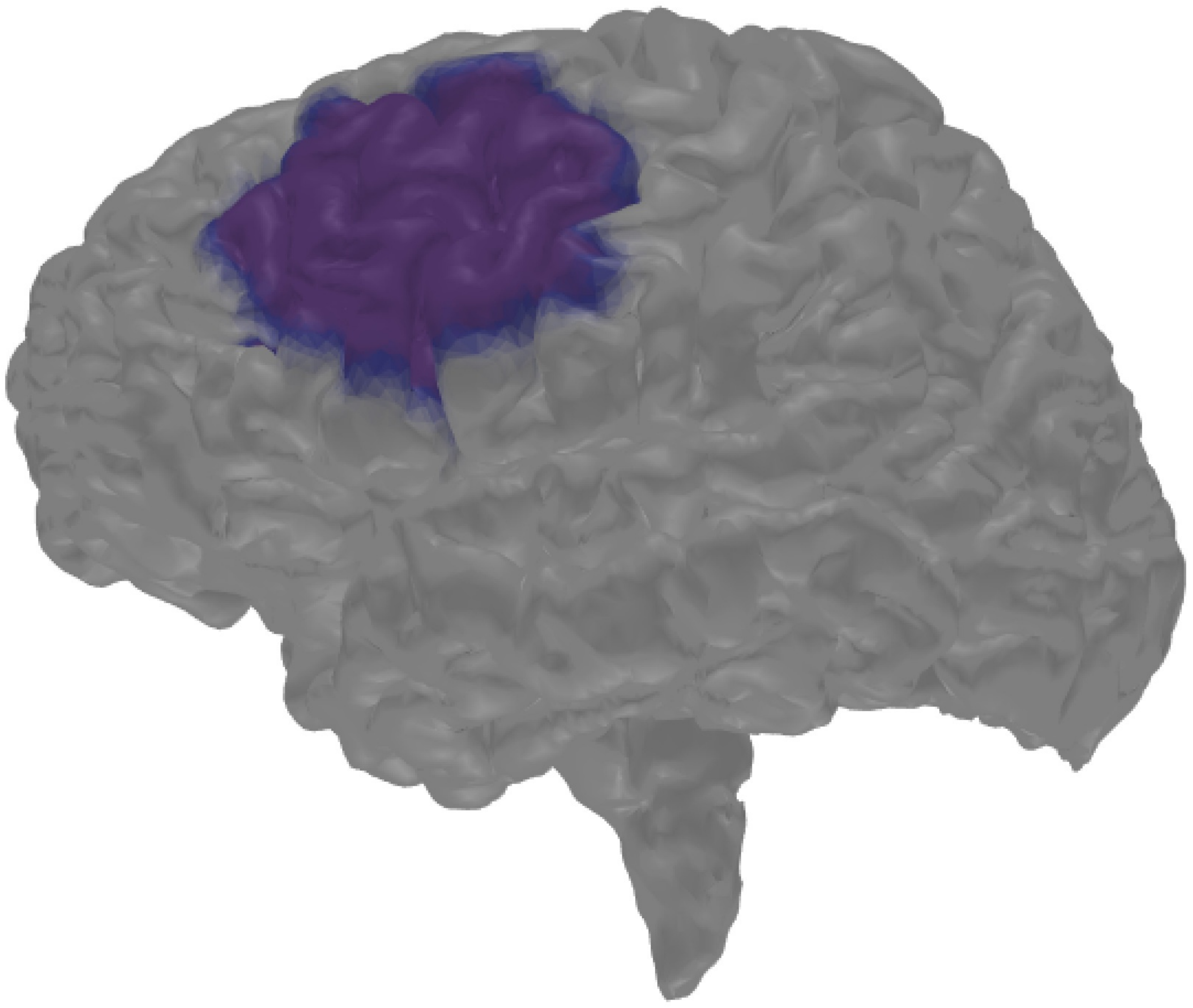}
    \end{minipage}\hspace{0.1cm}\begin{minipage}{0.22\textwidth}
    \centering
    {\bf 0.5 s} 
        \includegraphics[width=\textwidth]{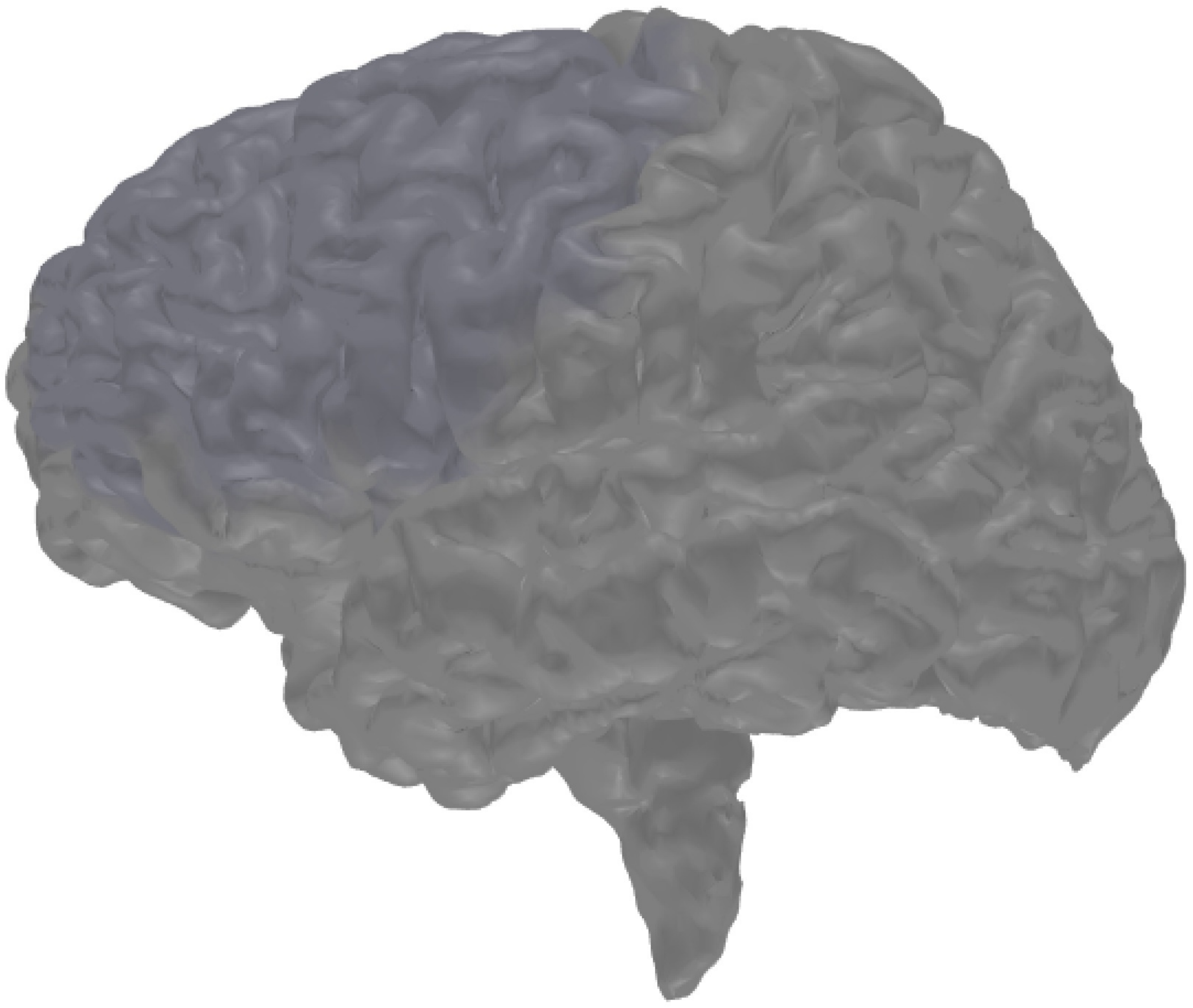}
    \end{minipage}

    \begin{minipage}{0.85\textwidth}
        \includegraphics[width=\textwidth]{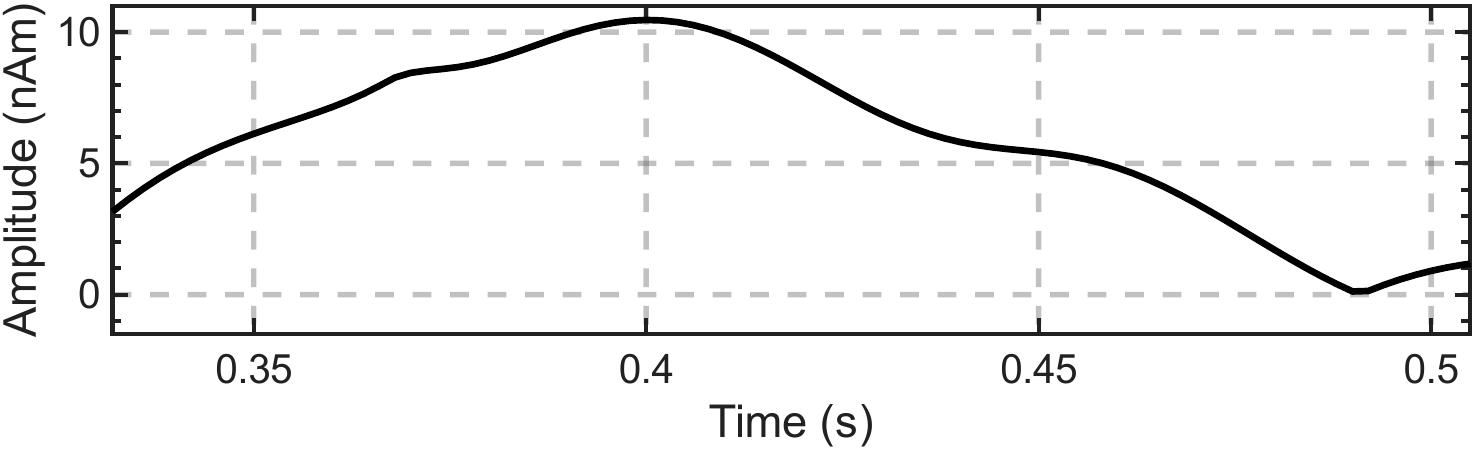}
    \end{minipage}
    \caption{The simulated epilepsy activity with batch-like activity spreading in the spatial space as shown in the top row. The total dipole moment over time is calculated using the Jansen-Rit model, producing the curve shown at the bottom. The activity starts focal and stationary, then spreads, and its mass center moves to the frontal lobe.}
    \label{fig:EpiSetup}
\end{figure*}

Lastly, we model non-stationary epileptic activity using the Jansen-Rit neural mass model designed for this purpose \cite{JansenRit1993,DavidOlivier2005JRmodel,AhmadizadehSaeed2018JRmodel,Sanchez-TodoRoser2023JRmodel}. The model describes the physiological features of the neural populations at the mesoscale by modeling the interactions of the following cortical columns: pyramidal neurons, excitatory interneurons, and inhibitory interneurons. Mathematically, the model is represented as a second-order differential equation system for excitatory and inhibitory inputs presented in \ref{app:JRNMM}. The separate neural masses are connected non-linearly through incoming membrane perturbations by a sigmoid function. The development of the simulated activity over time is shown in Figure \ref{fig:EpiSetup}.

In this experiment, we aim to demonstrate the robustness of Change-rate SKF against measurement noise in a nonlinear setting, where activity is constantly spreading across the brain, which is expected to make localization progressively harder. As a result, we draw the localization error from the activity mass centre and compute the earth mover's distance (EMD) over time.

\section{Results}

\subsection{Numerical tracking}
\begin{figure*}[htb!]
    \centering
    \begin{minipage}{0.04\textwidth}
            \rotatebox{90}{\textsl{\scriptsize 25 \unit{\decibel}}}
        \end{minipage}\begin{minipage}{0.3\textwidth}
        \begin{center}
            \scriptsize{\bf CR-SKF}
        \end{center}
        \includegraphics[width=\textwidth]{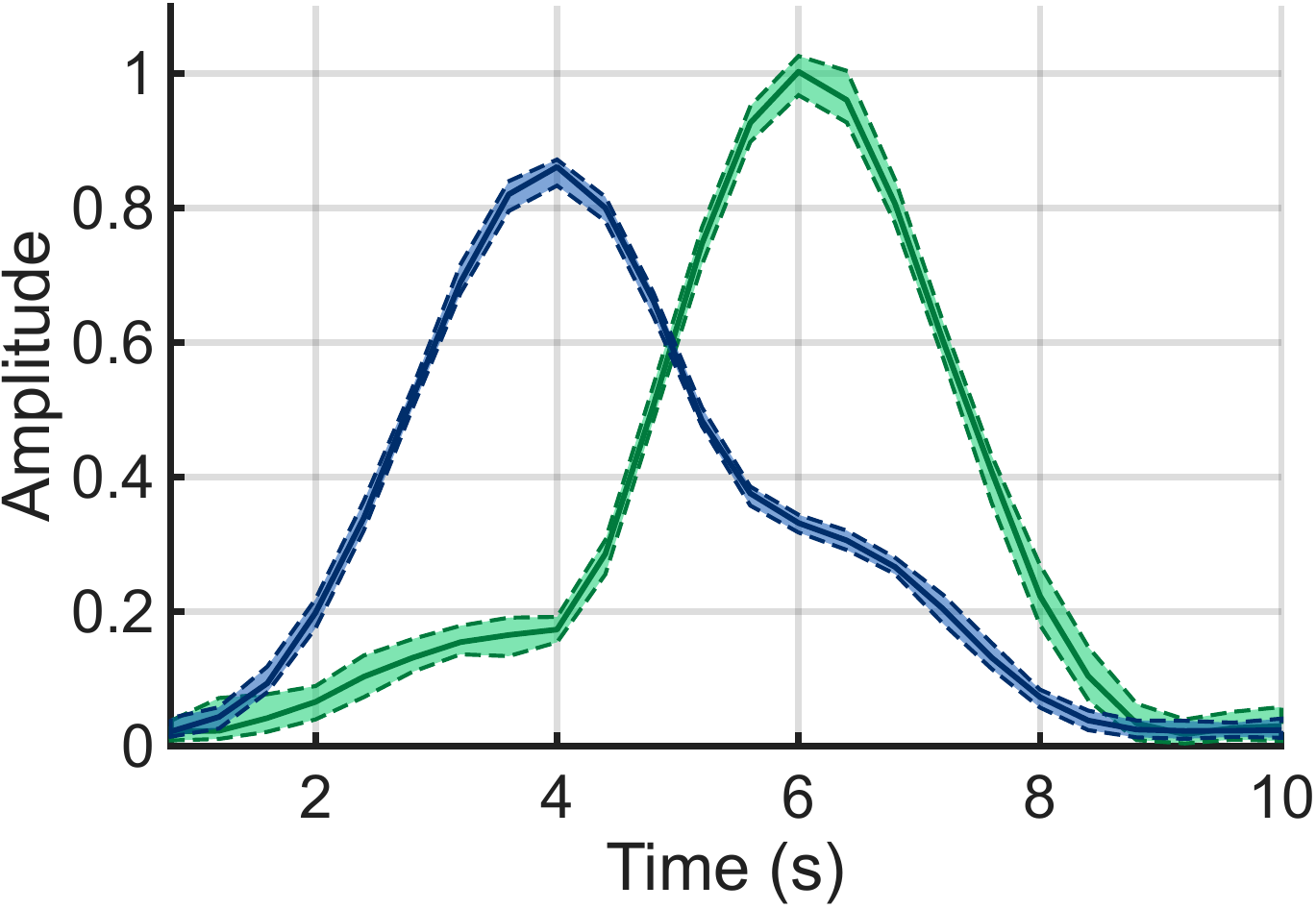}
    \end{minipage}\begin{minipage}{0.3\textwidth}
    \begin{center}
            \scriptsize{\bf RW-SKF}
        \end{center}
        \includegraphics[width=\textwidth]{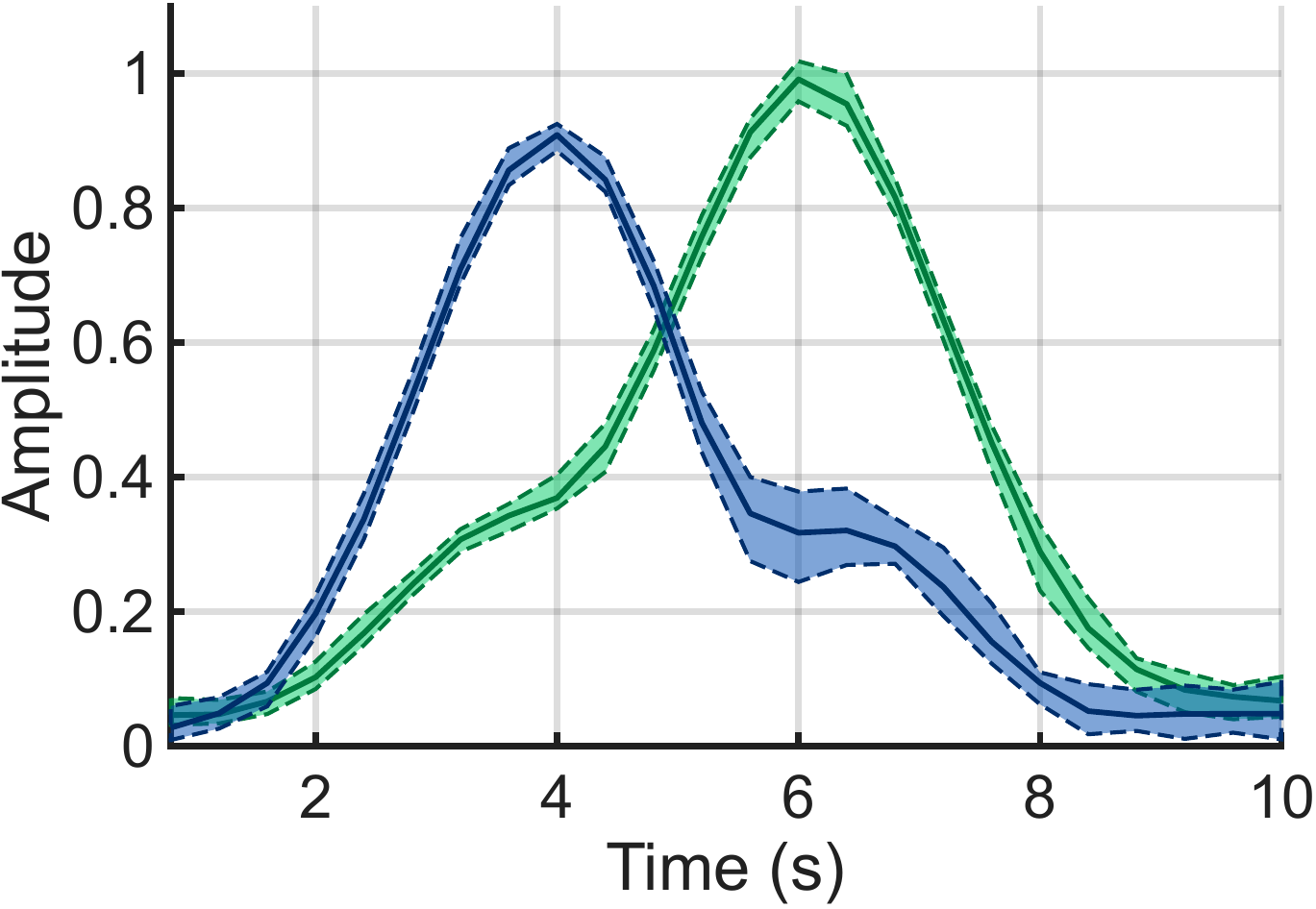}
    \end{minipage}\begin{minipage}{0.3\textwidth}
    \begin{center}
            \scriptsize{\bf sLORETA}
        \end{center}
        \includegraphics[width=\textwidth]{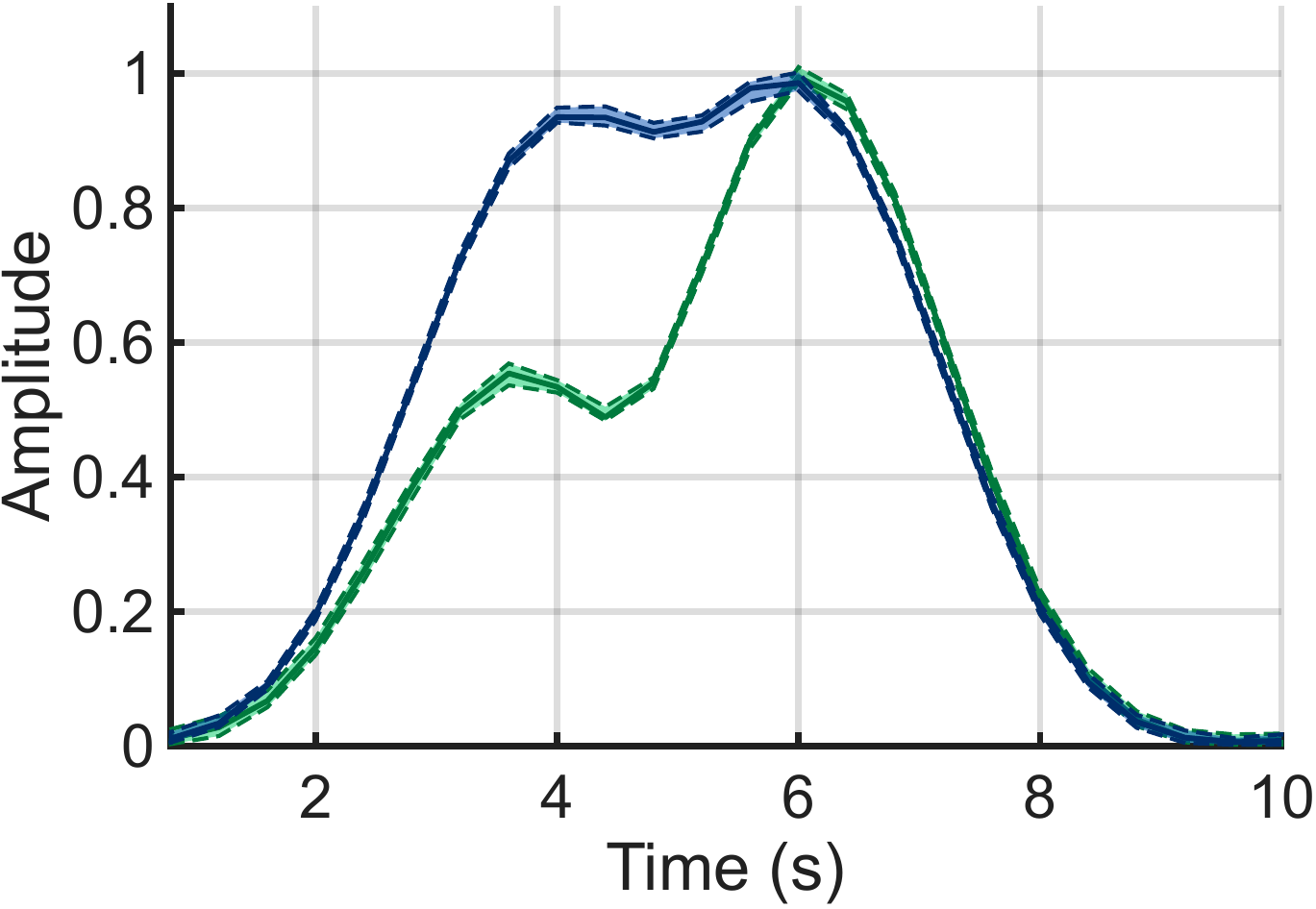}
    \end{minipage}\vspace{0.03\textwidth}
    \begin{minipage}{0.04\textwidth}
            \rotatebox{90}{\textsl{\scriptsize 15 \unit{\decibel}}}
        \end{minipage}\begin{minipage}{0.3\textwidth}
        \includegraphics[width=\textwidth]{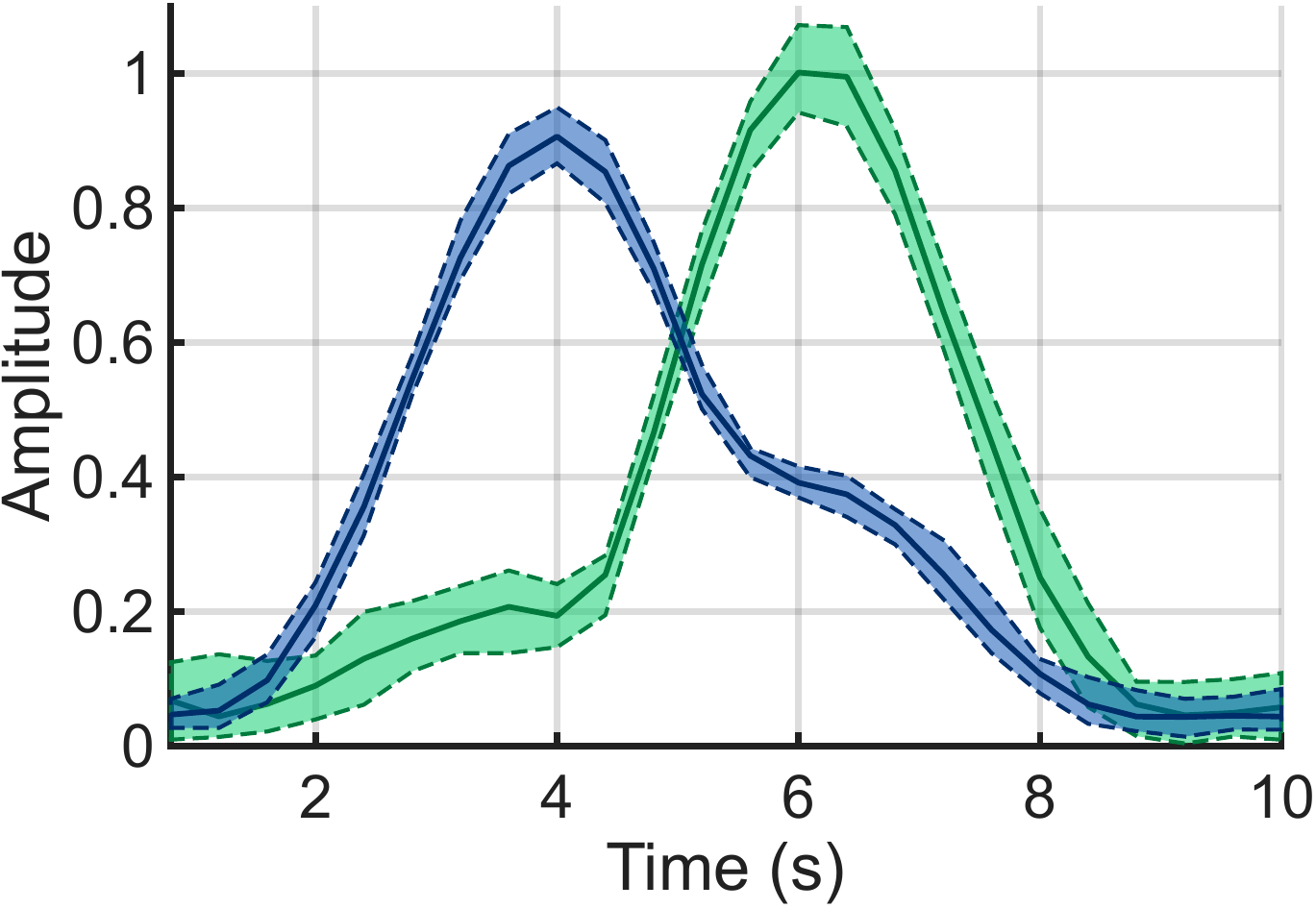}
    \end{minipage}\begin{minipage}{0.3\textwidth}
        \includegraphics[width=\textwidth]{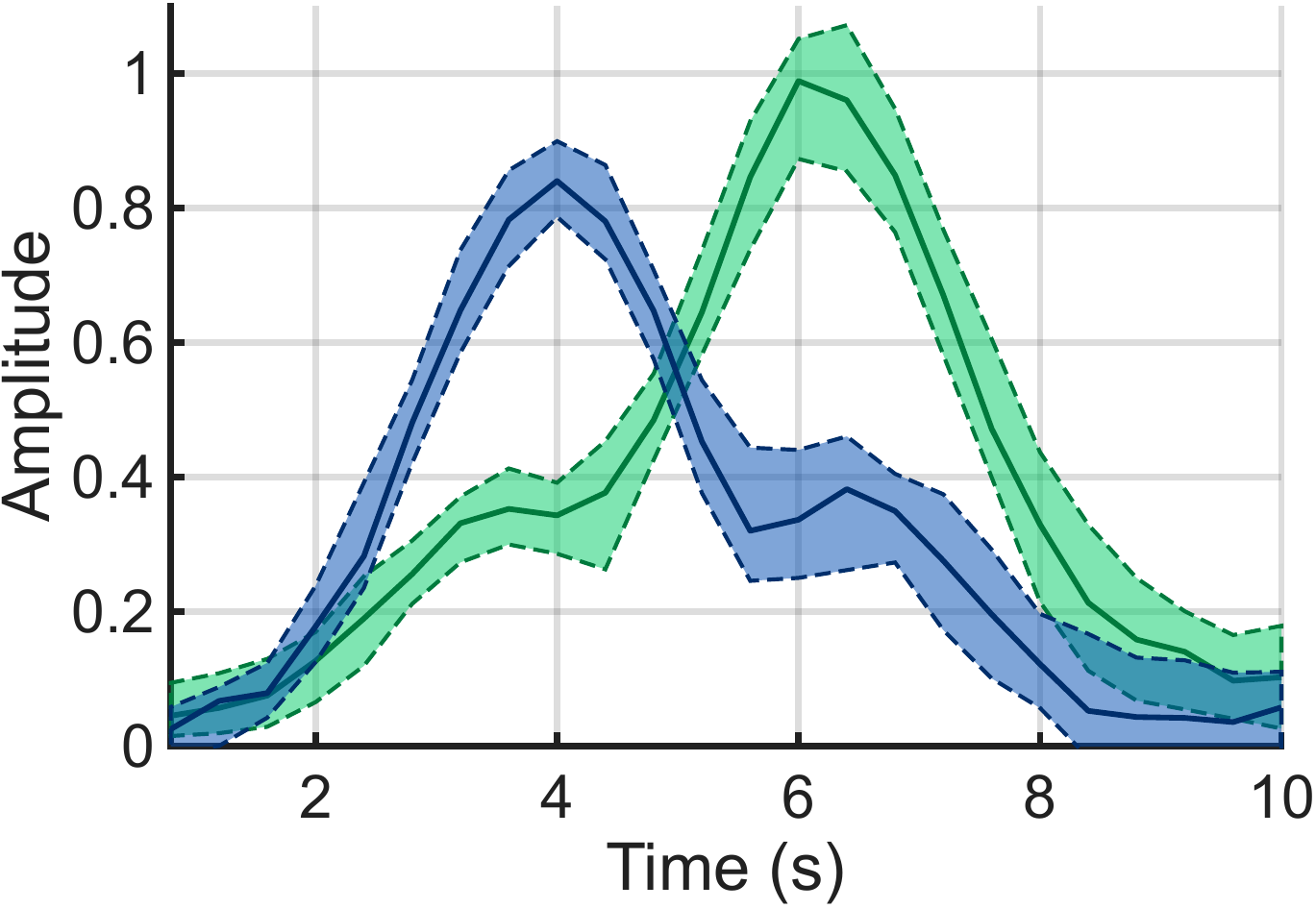}
    \end{minipage}\begin{minipage}{0.3\textwidth}
        \includegraphics[width=\textwidth]{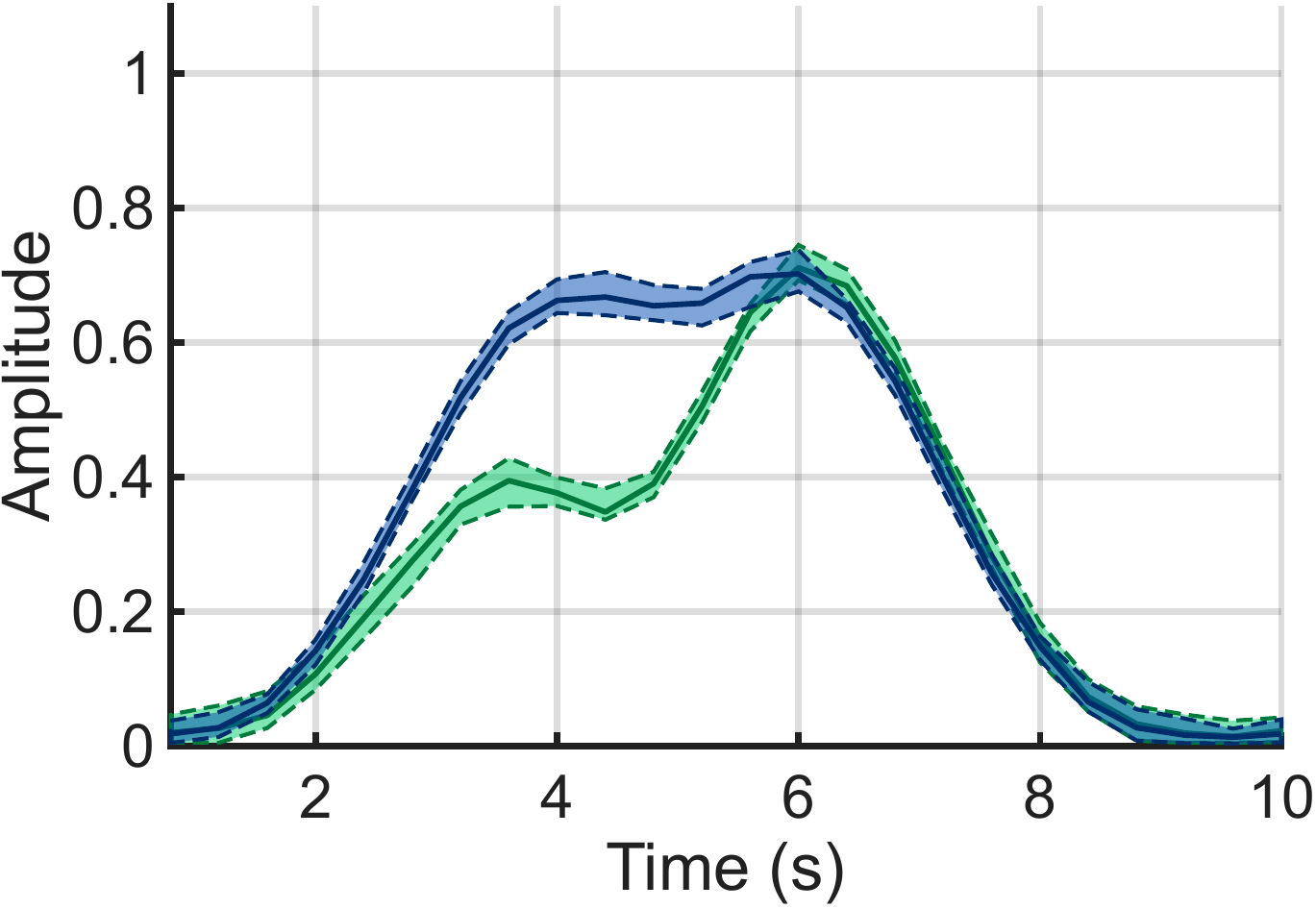}
    \end{minipage}\vspace{0.03\textwidth}
    \begin{minipage}{0.04\textwidth}
            \rotatebox{90}{\textsl{\scriptsize 5 \unit{\decibel}}}
        \end{minipage}\begin{minipage}{0.3\textwidth}
        \includegraphics[width=\textwidth]{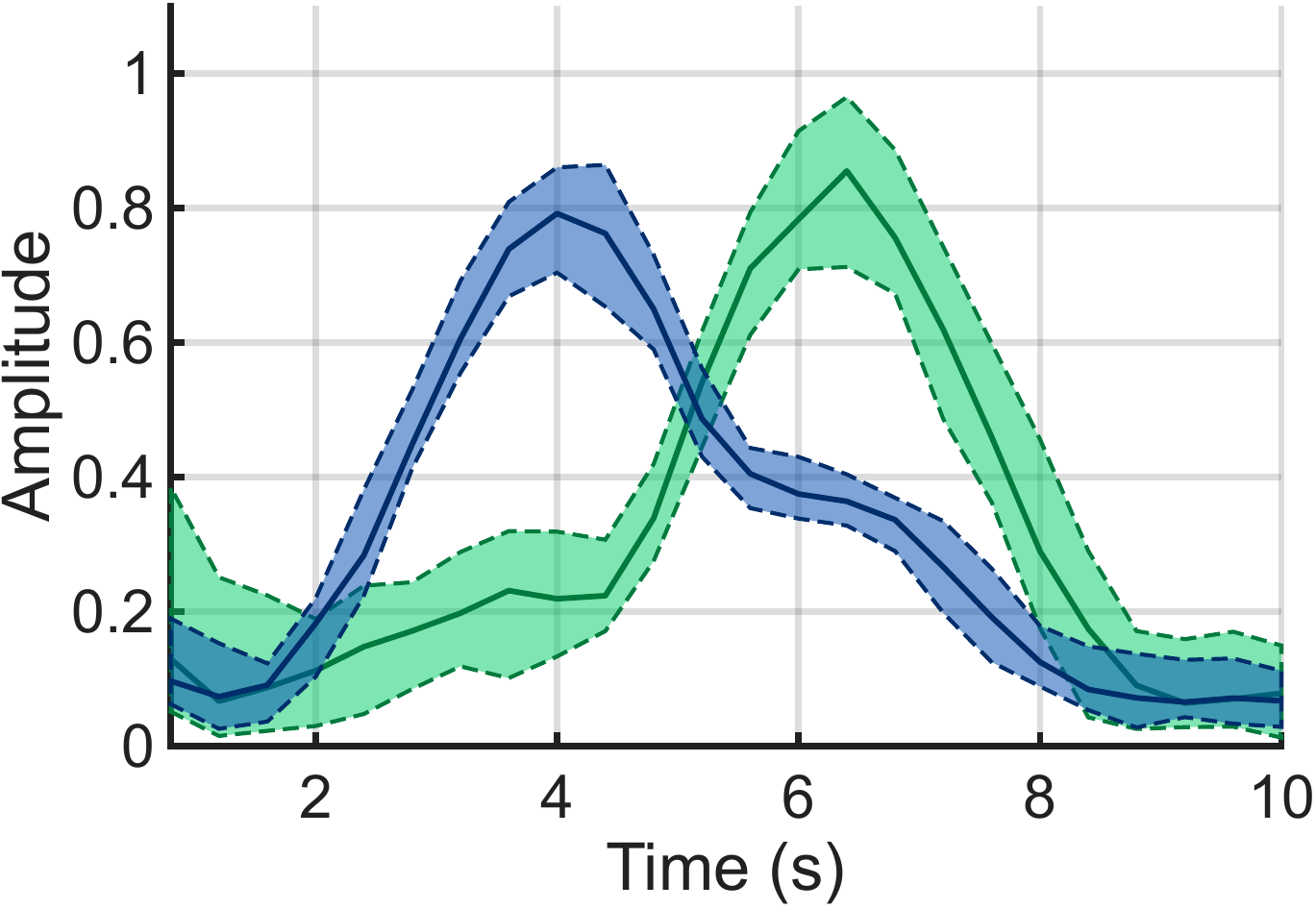}
    \end{minipage}\begin{minipage}{0.3\textwidth}
        \includegraphics[width=\textwidth]{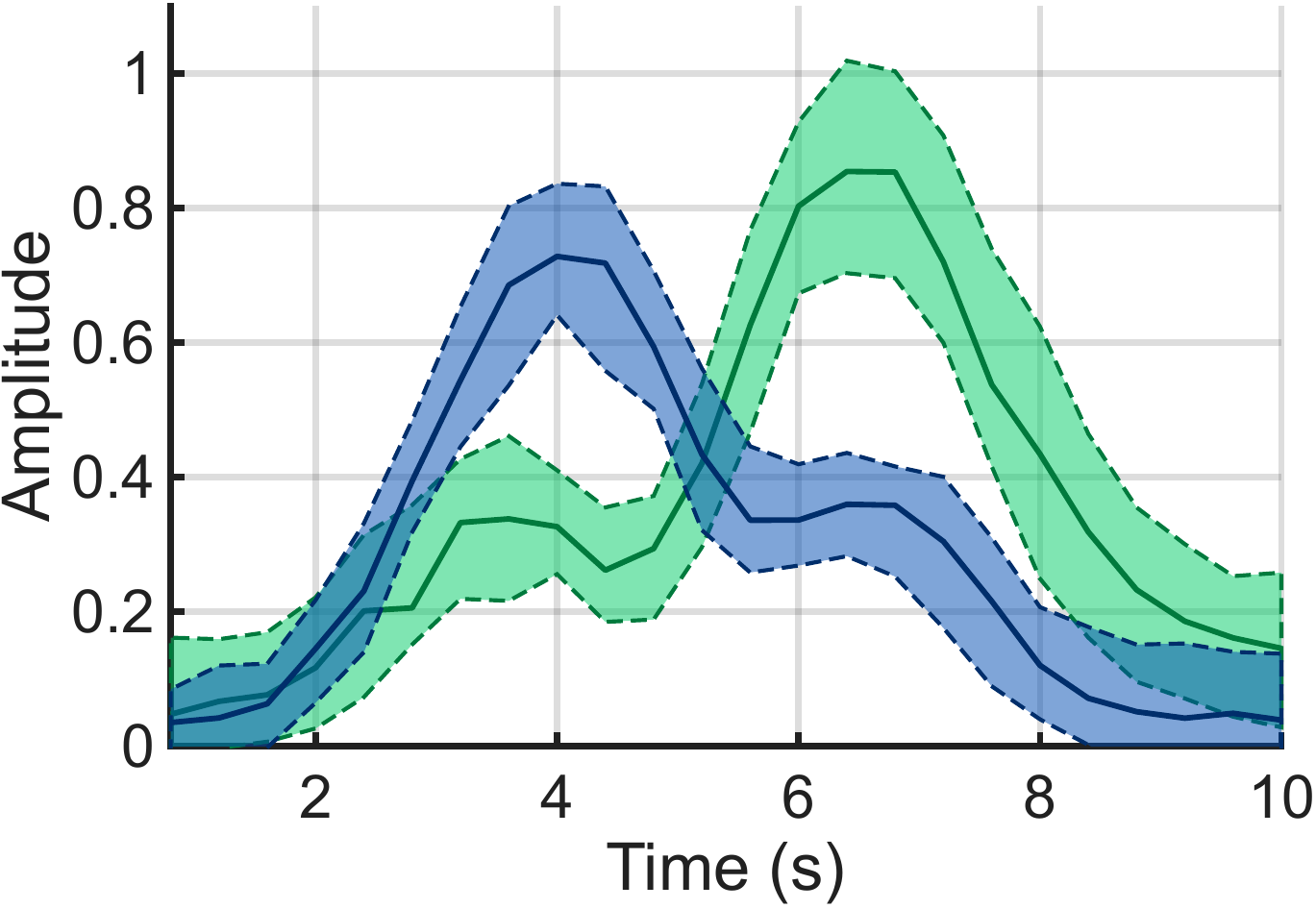}
    \end{minipage}\begin{minipage}{0.3\textwidth}
        \includegraphics[width=\textwidth]{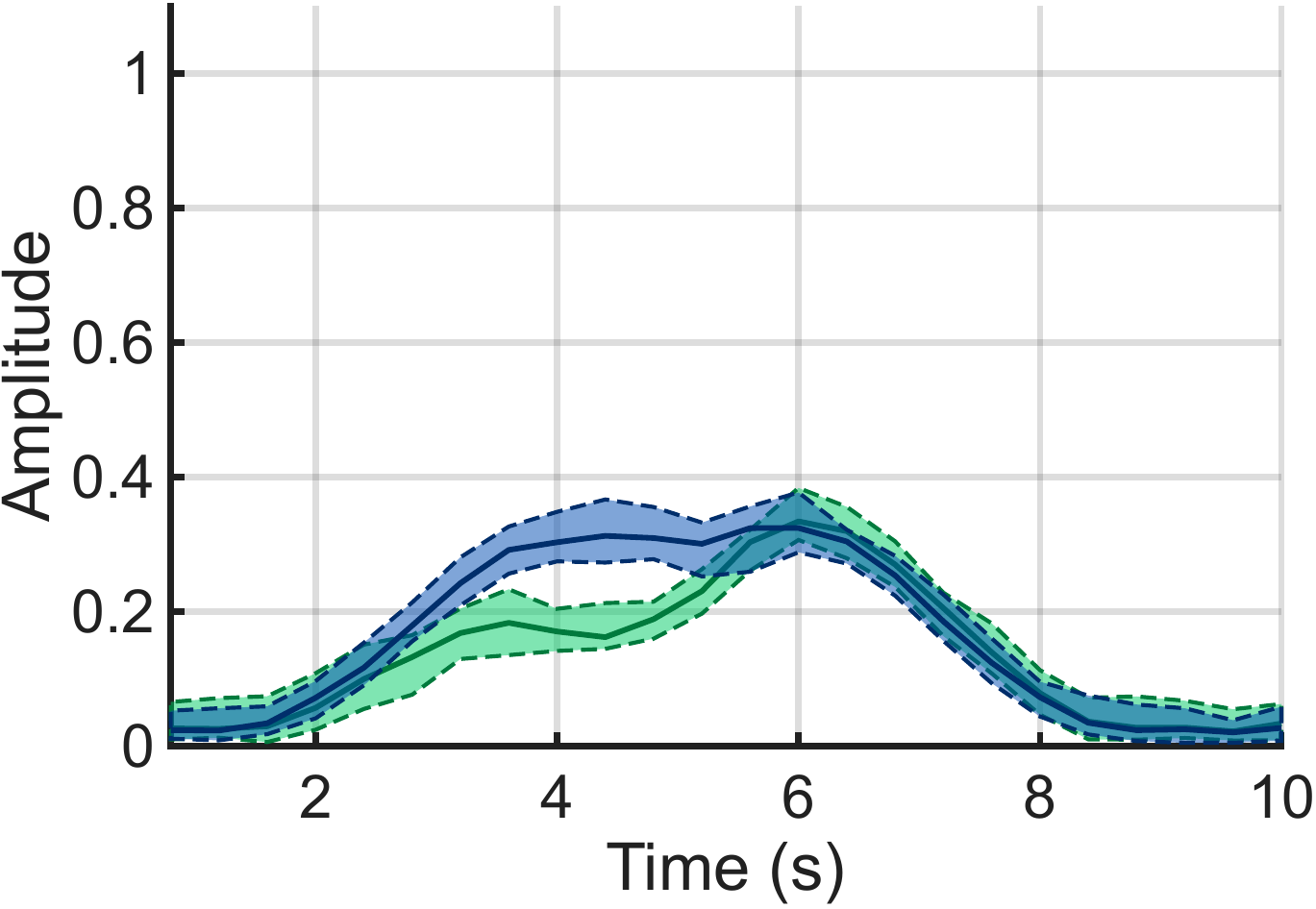}
    \end{minipage}\vspace{0.05\textwidth}
    \caption{The tracked activity strength curves presented in \ref{fig:TrueTracks} by the Change Rate Standardized Kalman filter (CR-SKF), Random walk evolution model SKF (RW-SKF), and sLORETA are displayed in the adjacent columns. Each row indicated the measurement noise level in decibels (25, 15, and 5 \unit{\decibel} from top to bottom. The blue curve indicates the deep activity at the left thalamus, and the green curve shows the surface activity at the somatosensory cortex. The solid, darker line is the mean over 25 tracks computed with simulated data of different noise realizations. The colored area limited between dashed lines represents the tracks of 2.5 \% and 97.5 \% quantile intervals.}
    \label{fig:tracks}
\end{figure*}

In the numerical tracking shown in Figure \ref{fig:tracks}, we observe highly non-enmeshed tracks for both Standardized Kalman filters (blue and green). The time-independent comparison method, sLORETA, displays enmeshed deep activity (blue track) that corresponds to deep and surface-level activity. By looking at the width of the colored areas around mean tracks that display the 2.5 \% and 97.5 \% quantile interval around the mean, the best measurement noise robustness, i.e., the smallest width of the interval, is obtained for sLORETA, next CR-SKF, and worst with RW-SKF. The peak heights are almost equal for sLORETA. CR-KF has slightly more balanced peaks than RW-SKF. For CR-SKF, the surface track dominates, while for RW-SKF, the deep track has a higher peak. Interestingly, with Standardized Kalman filtering methods, the mean height gets closer when the noise level is increased. Compared to the maximum height of the sLORETA peak at 25 \unit{\decibel}, the overall standardized activity strength diminishes along with increased noise. A similar effect can be noticed in SKFs comparing 15 \unit{\decibel} to 5 \unit{\decibel} tracking. However, the decrease is moderate.

\clearpage

\clearpage

\begin{figure*}[htb!]
    \centering
    \underline{\bf $\rho=0$ \unit{\decibel}}\vspace{0.02\textwidth}
    \begin{minipage}{0.04\textwidth}
            \rotatebox{90}{\textsl{\scriptsize{\bf CR-SKF}}}
        \end{minipage}\begin{minipage}{0.3\textwidth}
        \begin{center}
            \scriptsize{25 \unit{\decibel}}
        \end{center}
        \includegraphics[width=\textwidth]{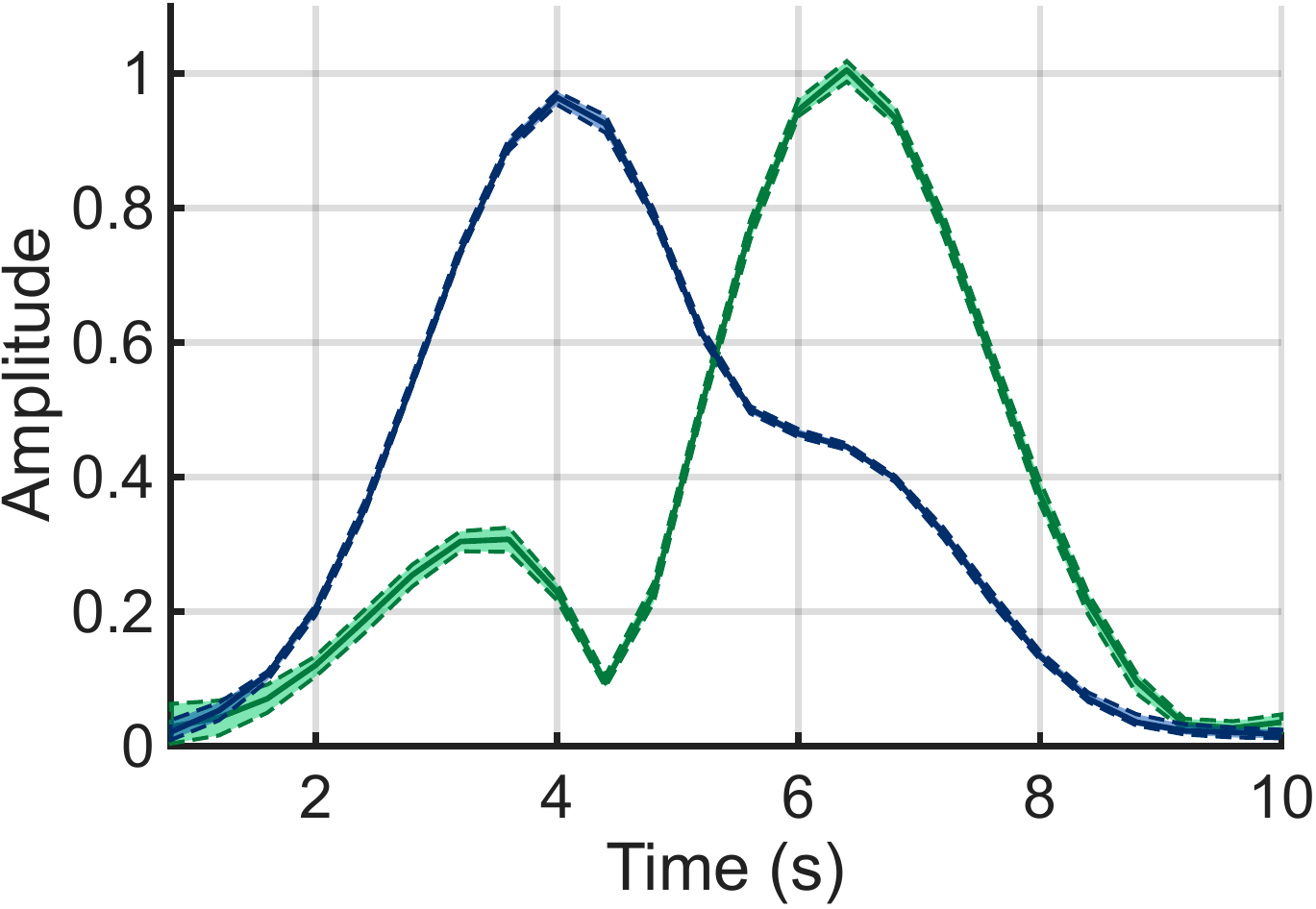}
    \end{minipage}\begin{minipage}{0.3\textwidth}
    \begin{center}
            \scriptsize{15 \unit{\decibel}}
        \end{center}
        \includegraphics[width=\textwidth]{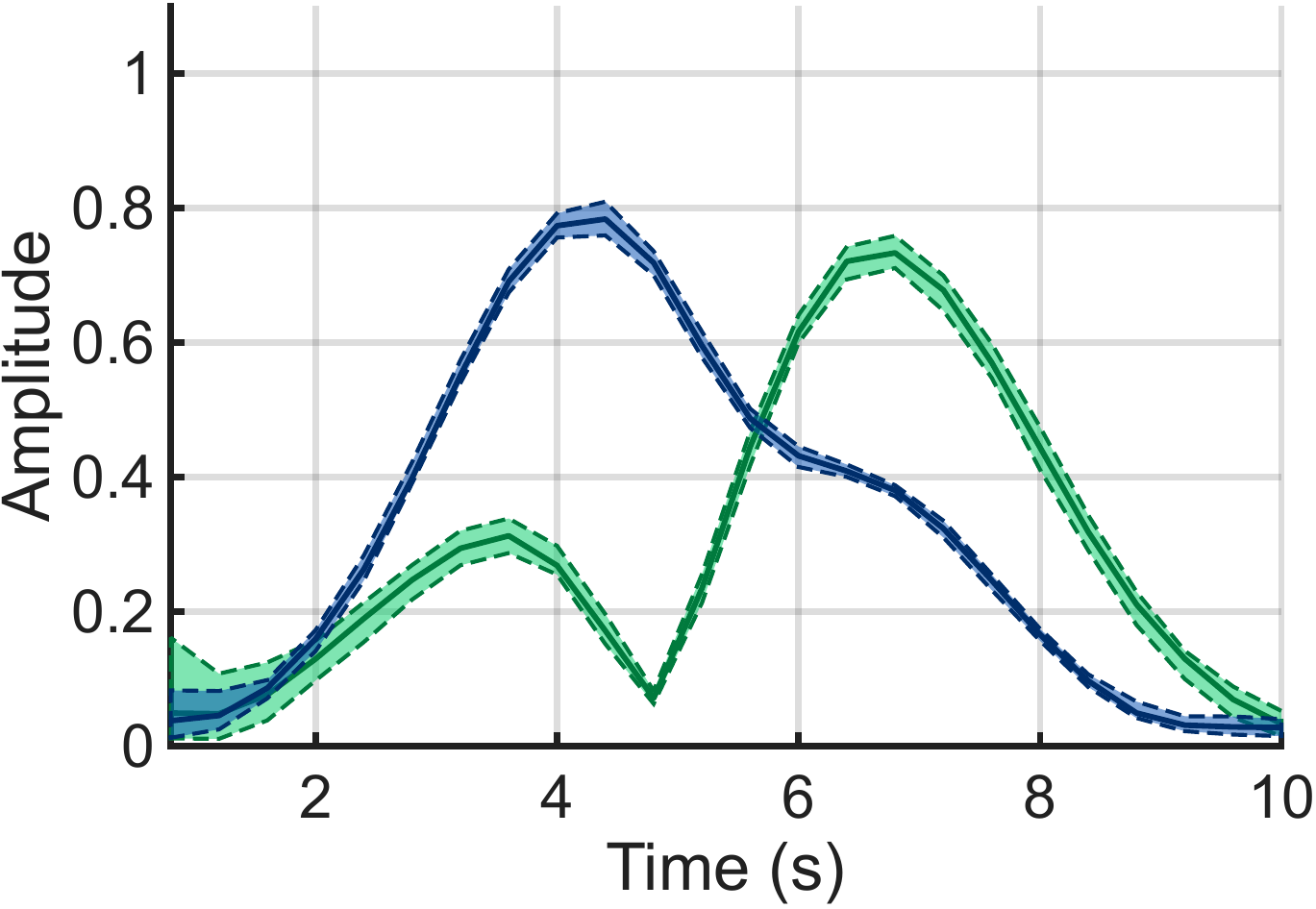}
    \end{minipage}\begin{minipage}{0.3\textwidth}
    \begin{center}
            \scriptsize{5 \unit{\decibel}}
        \end{center}
        \includegraphics[width=\textwidth]{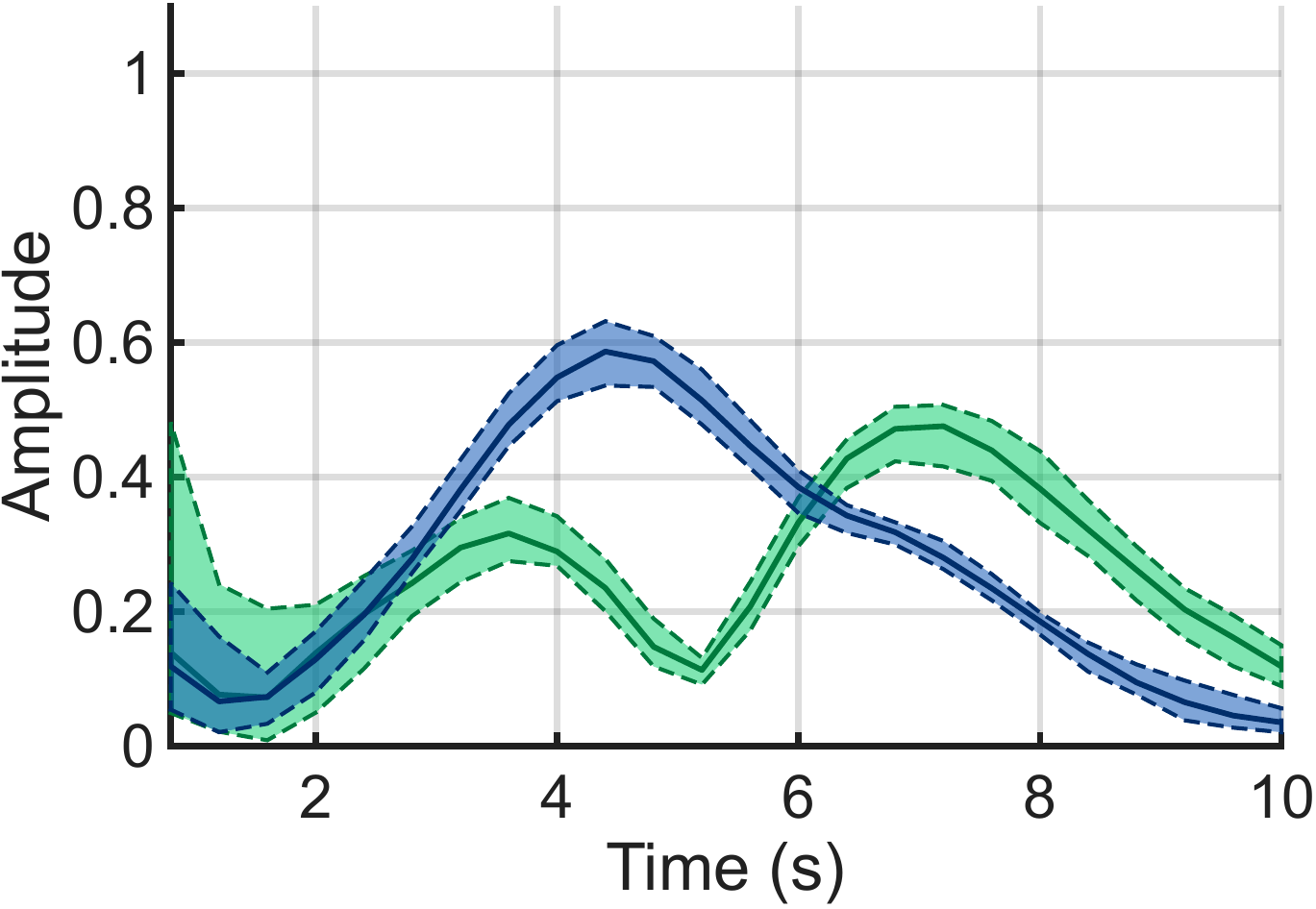}
    \end{minipage}\vspace{0.03\textwidth}
    \begin{minipage}{0.04\textwidth}
            \rotatebox{90}{\textsl{\scriptsize{\bf RW-SKF}}}
        \end{minipage}\begin{minipage}{0.3\textwidth}
        \includegraphics[width=\textwidth]{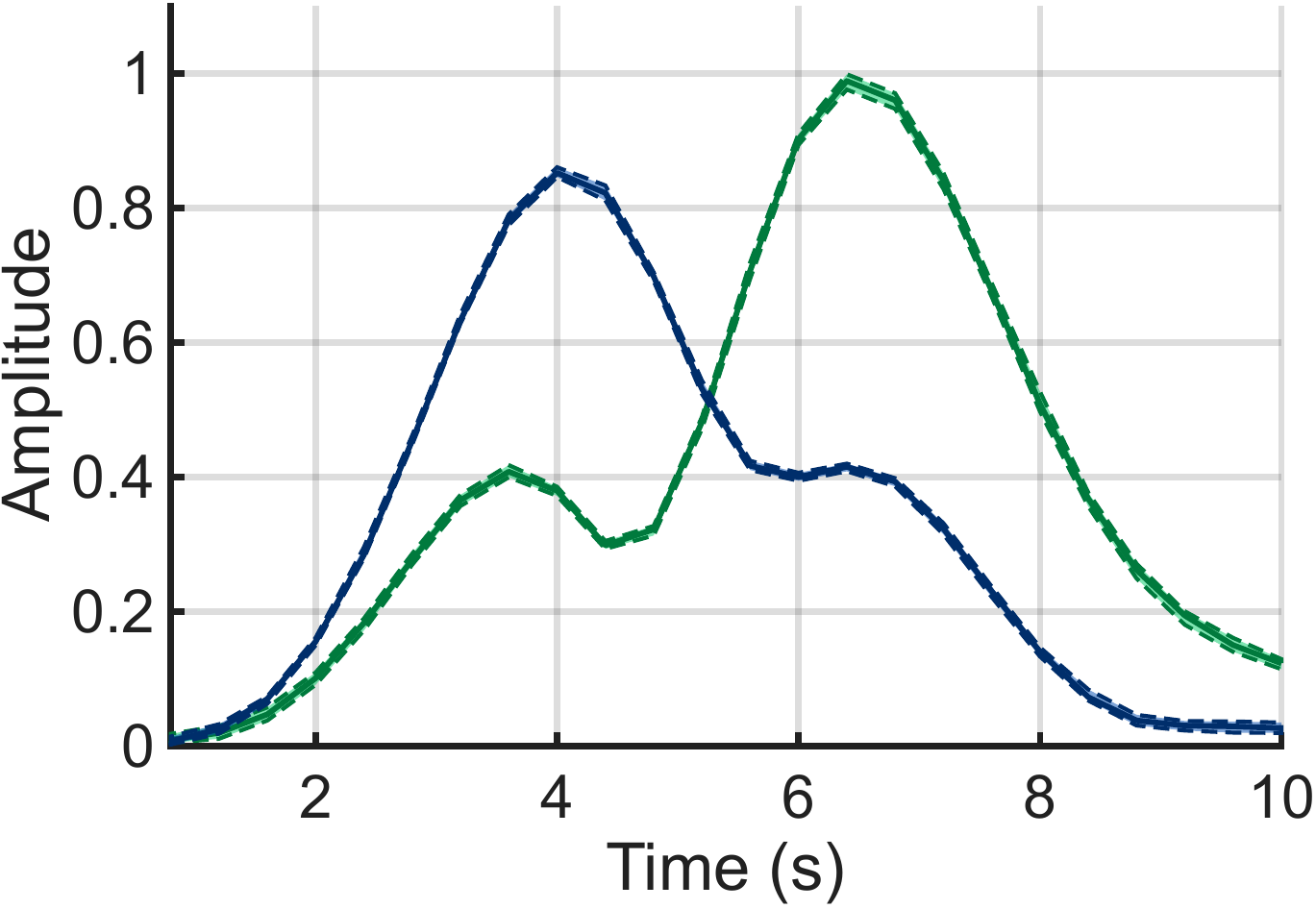}
    \end{minipage}\begin{minipage}{0.3\textwidth}
        \includegraphics[width=\textwidth]{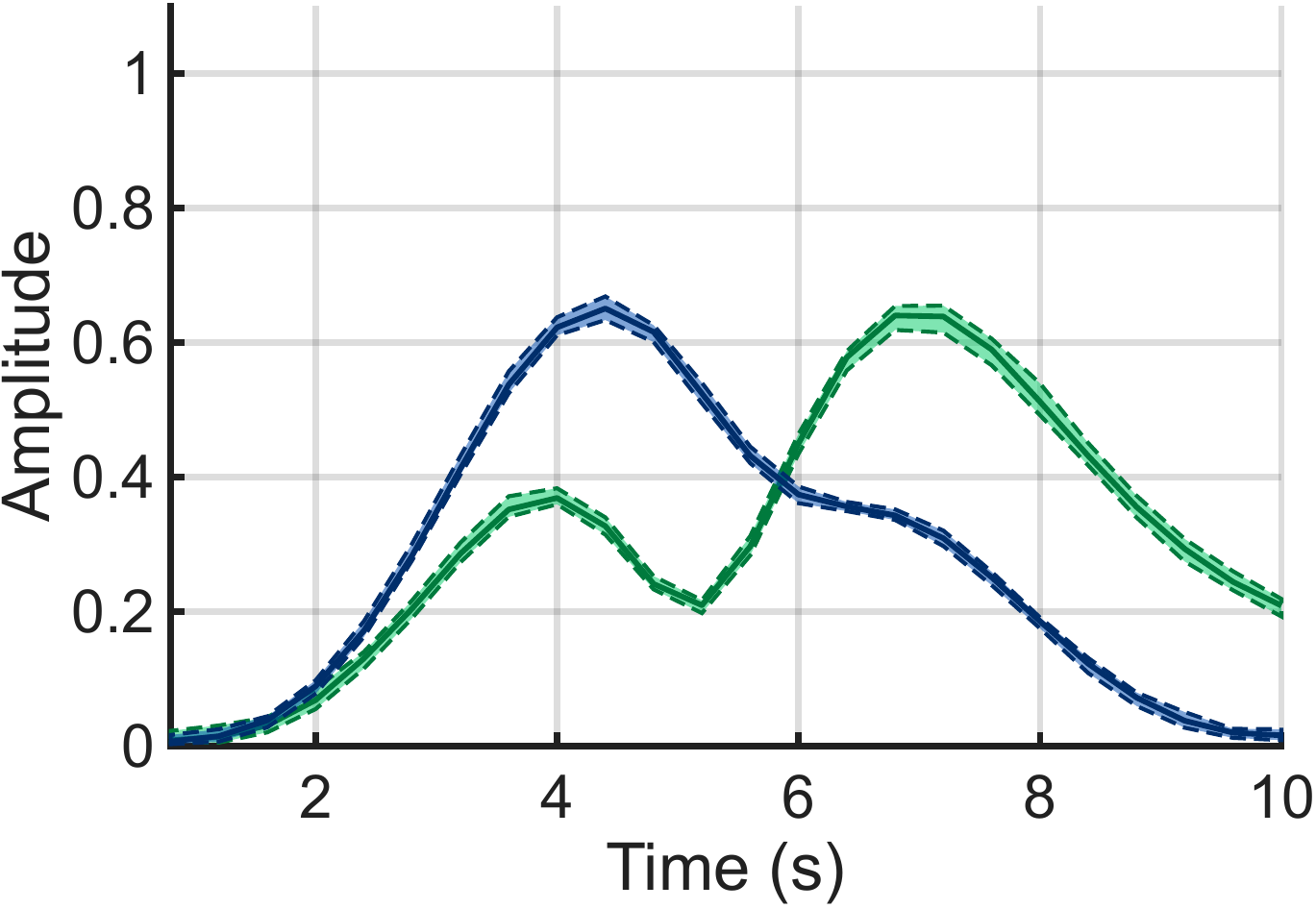}
    \end{minipage}\begin{minipage}{0.3\textwidth}
        \includegraphics[width=\textwidth]{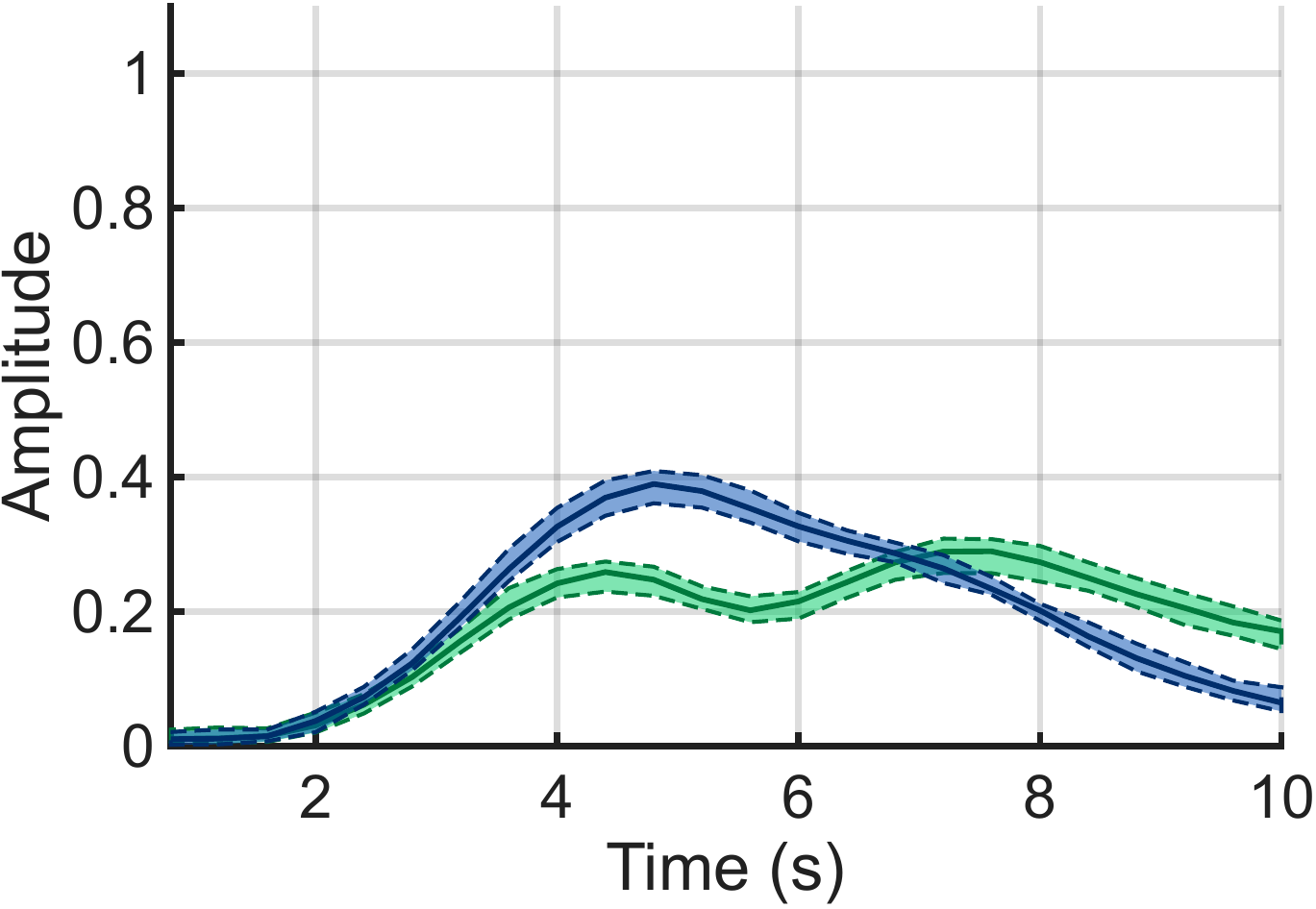}
    \end{minipage}\vspace{0.015\textwidth}
    \hrule\vspace{0.015\textwidth}
    \underline{\bf $\rho=70$ \unit{\decibel}}\vspace{0.02\textwidth}
    \begin{minipage}{0.04\textwidth}
            \rotatebox{90}{\textsl{\scriptsize{\bf CR-SKF}}}
        \end{minipage}\begin{minipage}{0.3\textwidth}
        \includegraphics[width=\textwidth]{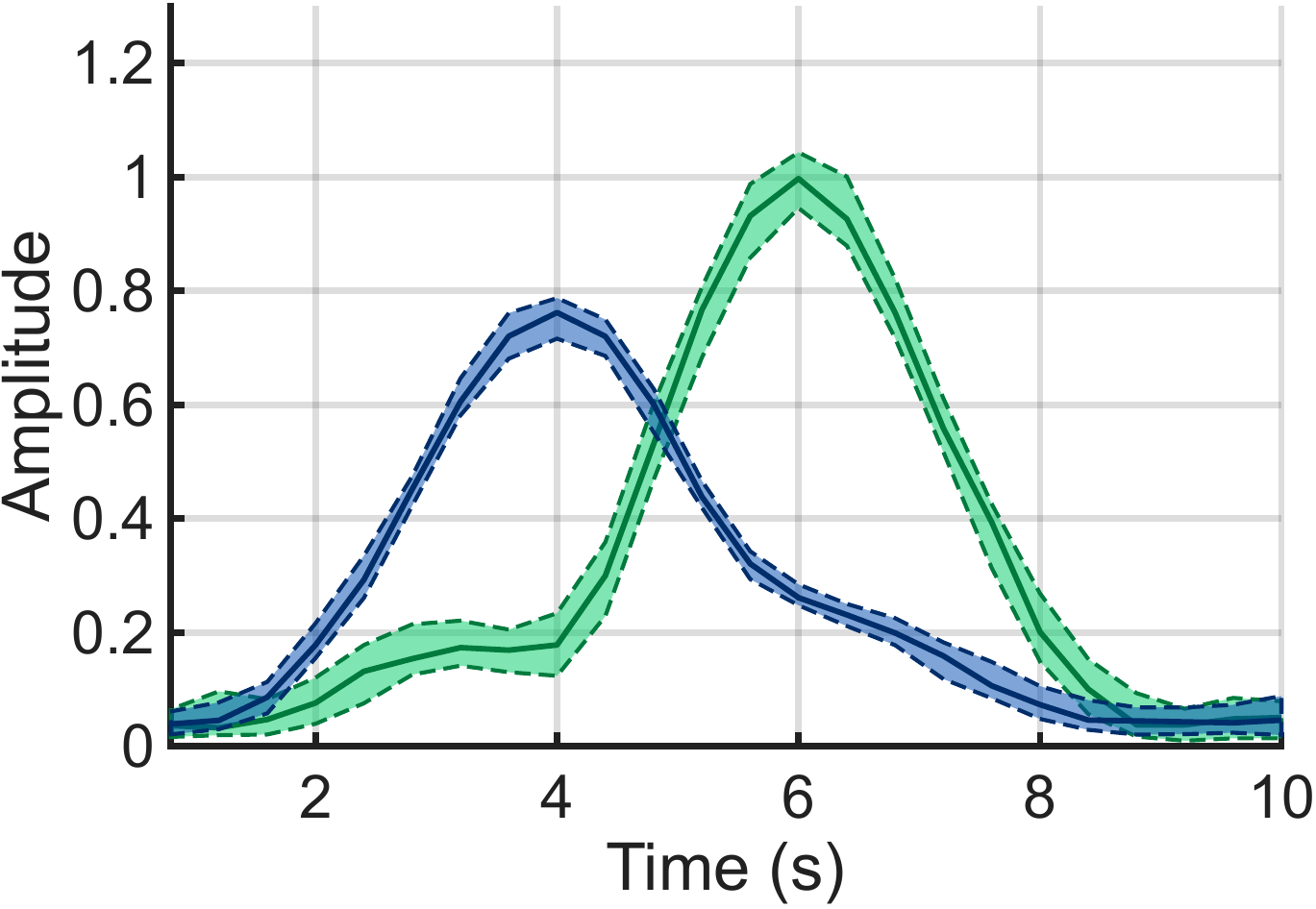}
    \end{minipage}\begin{minipage}{0.3\textwidth}
        \includegraphics[width=\textwidth]{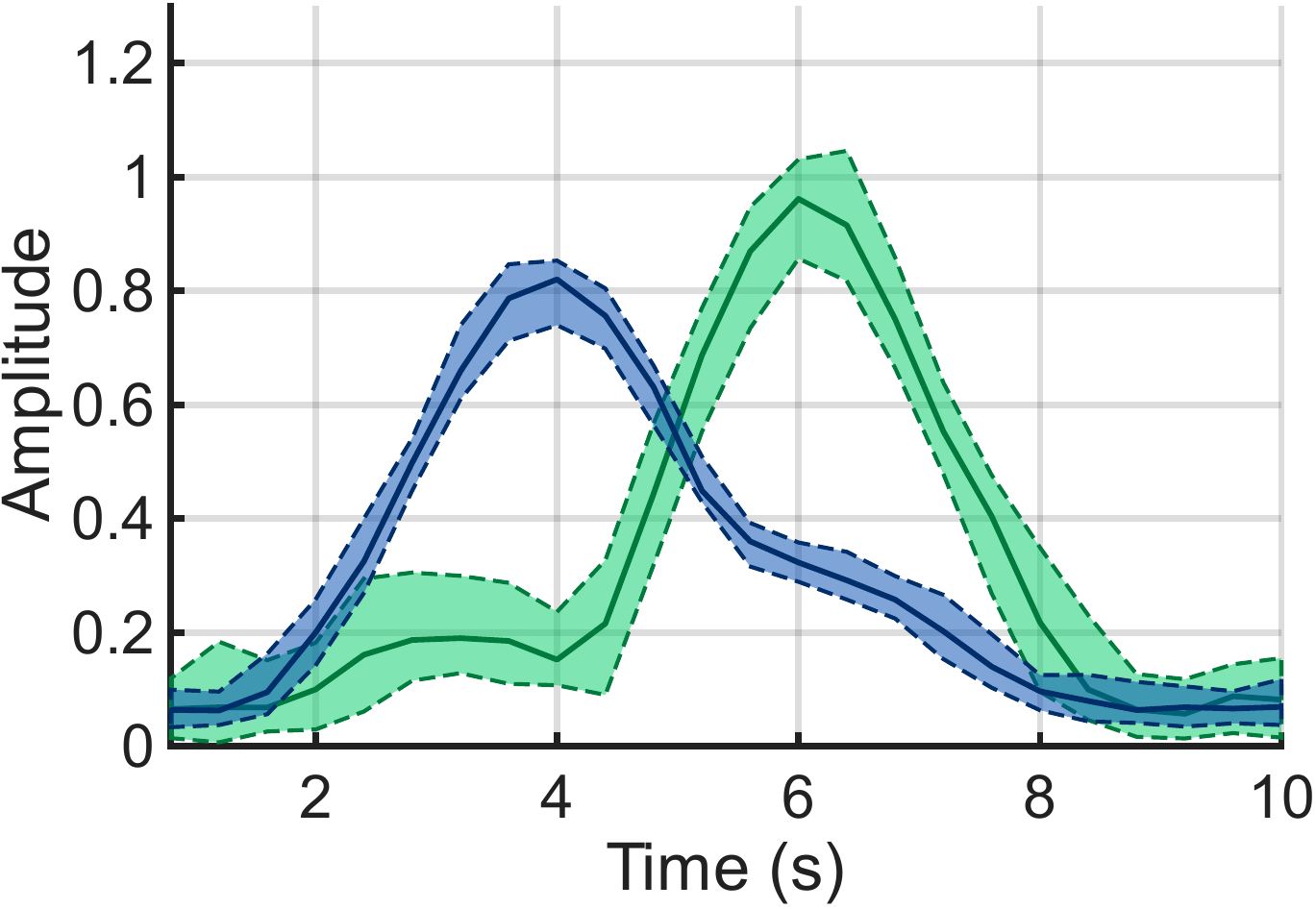}
    \end{minipage}\begin{minipage}{0.3\textwidth}
        \includegraphics[width=\textwidth]{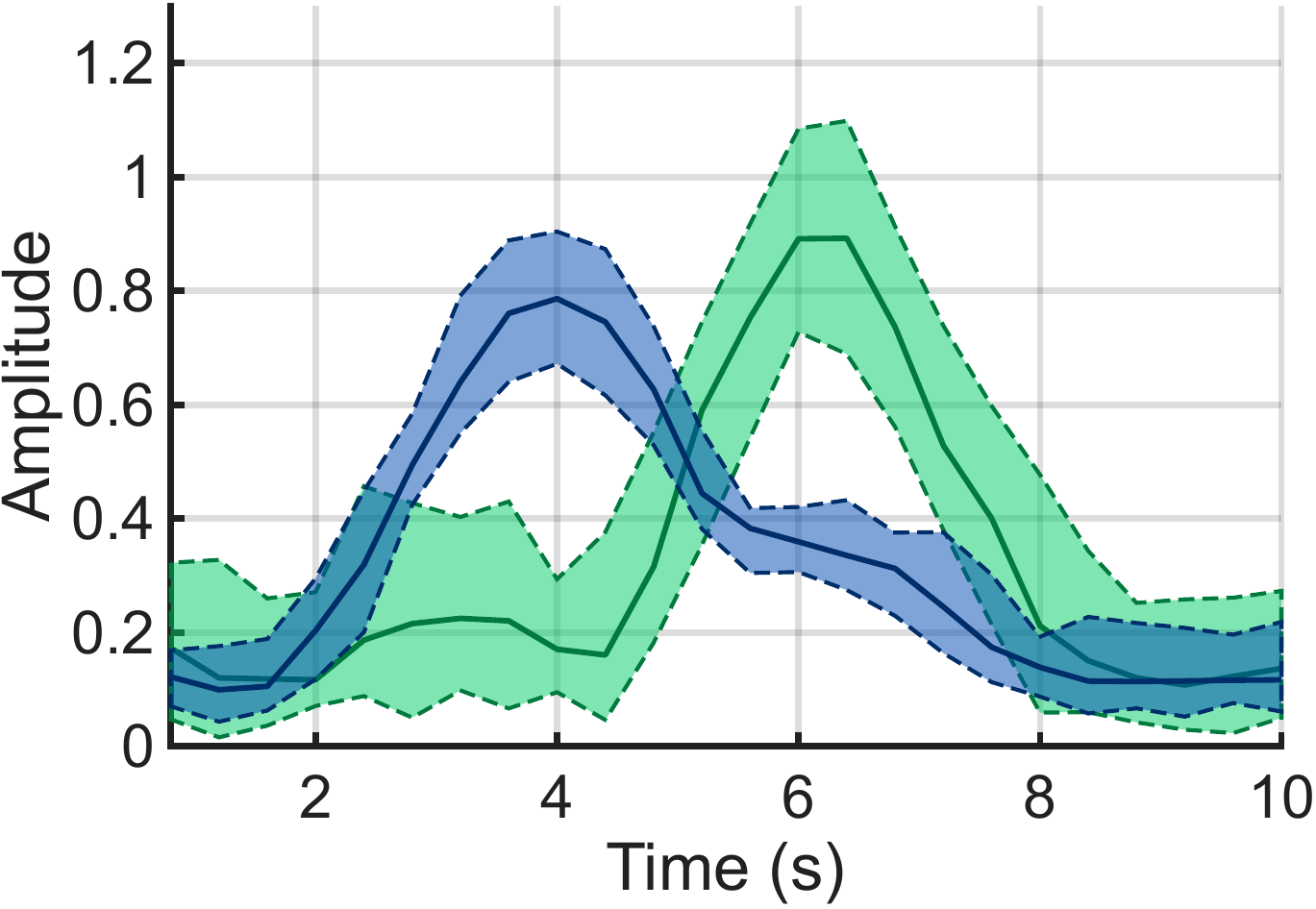}
    \end{minipage}\vspace{0.03\textwidth}
    \begin{minipage}{0.04\textwidth}
            \rotatebox{90}{\textsl{\scriptsize{\bf RW-SKF}}}
        \end{minipage}\begin{minipage}{0.3\textwidth}
        \includegraphics[width=\textwidth]{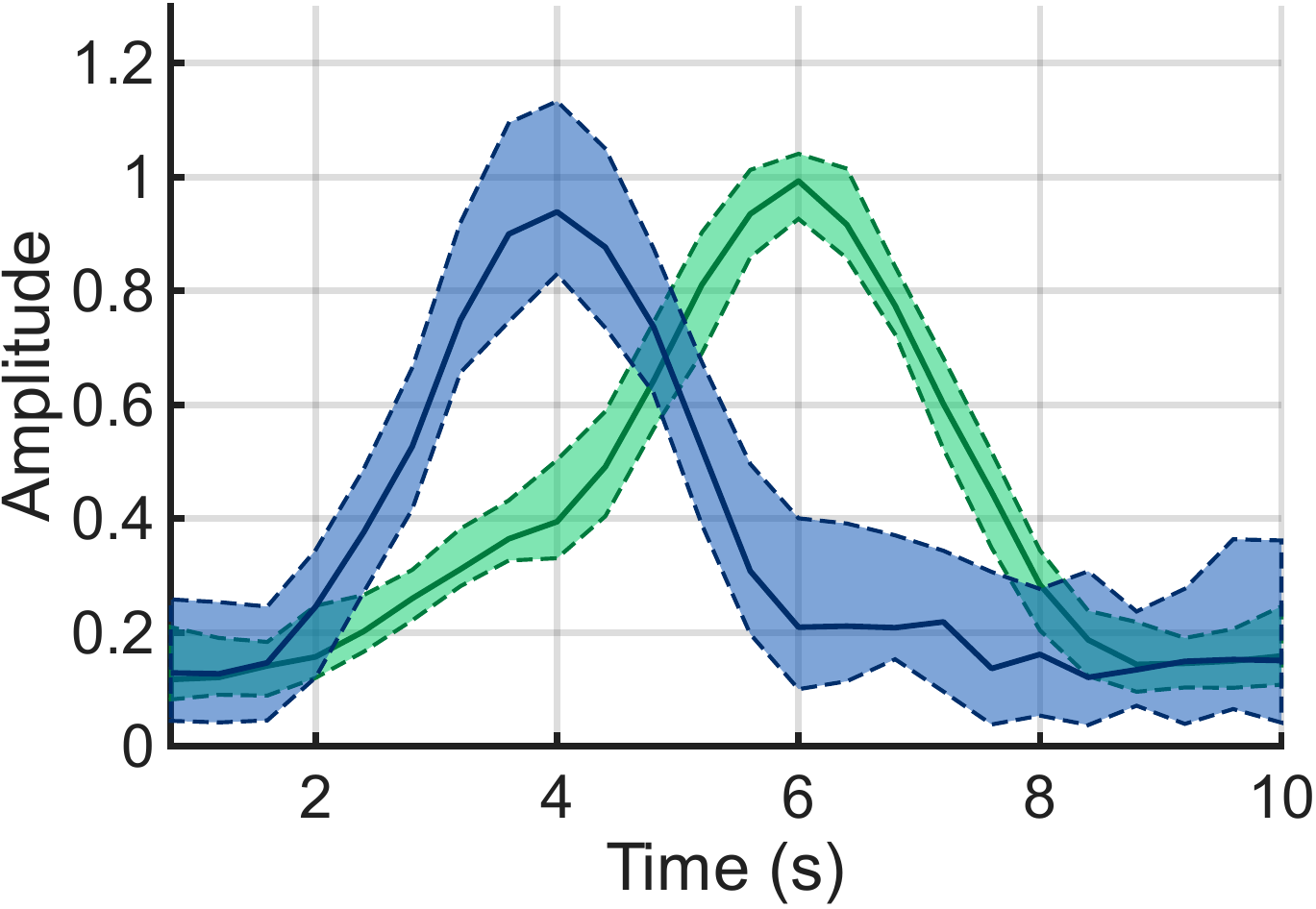}
    \end{minipage}\begin{minipage}{0.3\textwidth}
        \includegraphics[width=\textwidth]{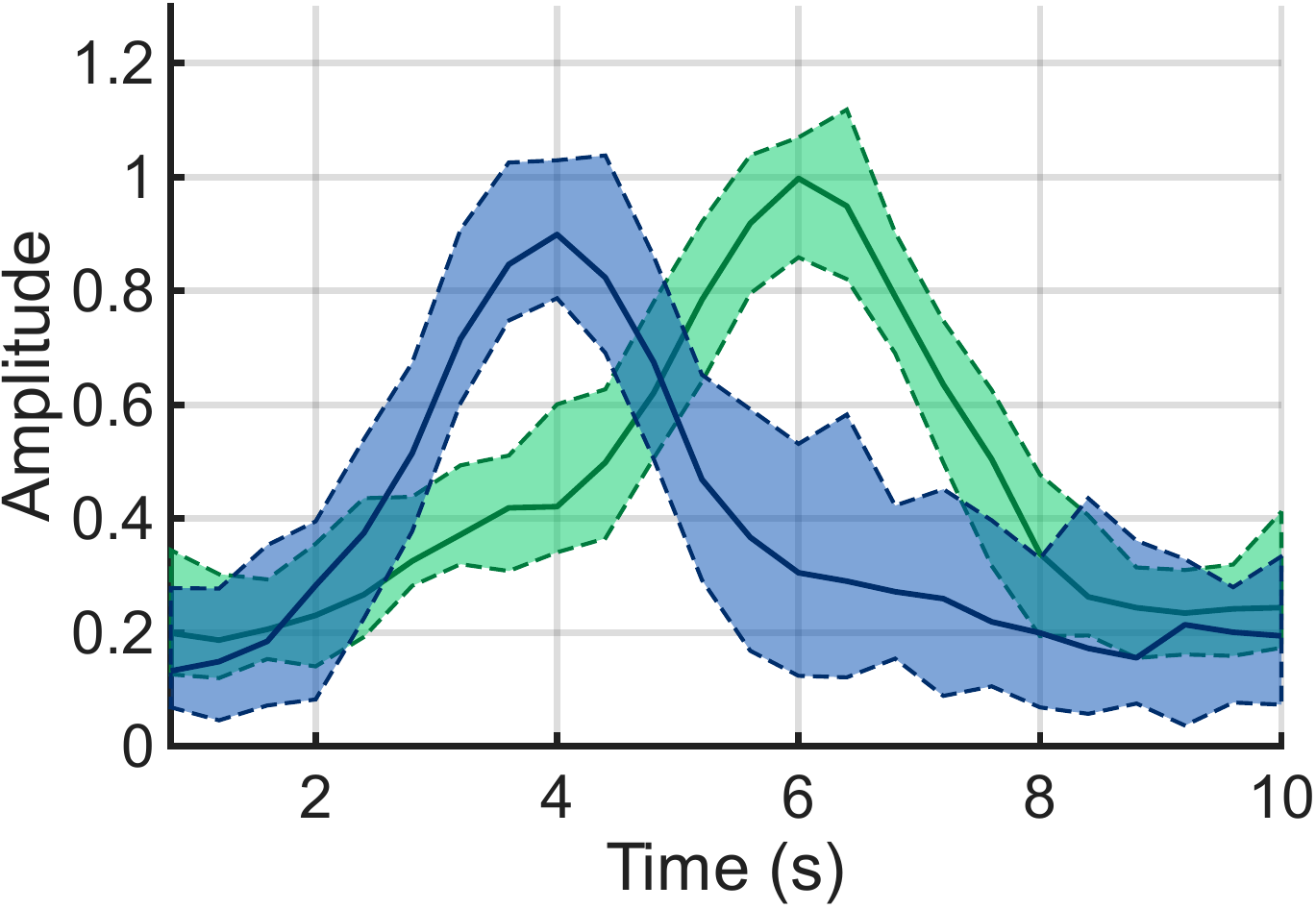}
    \end{minipage}\begin{minipage}{0.3\textwidth}
        \includegraphics[width=\textwidth]{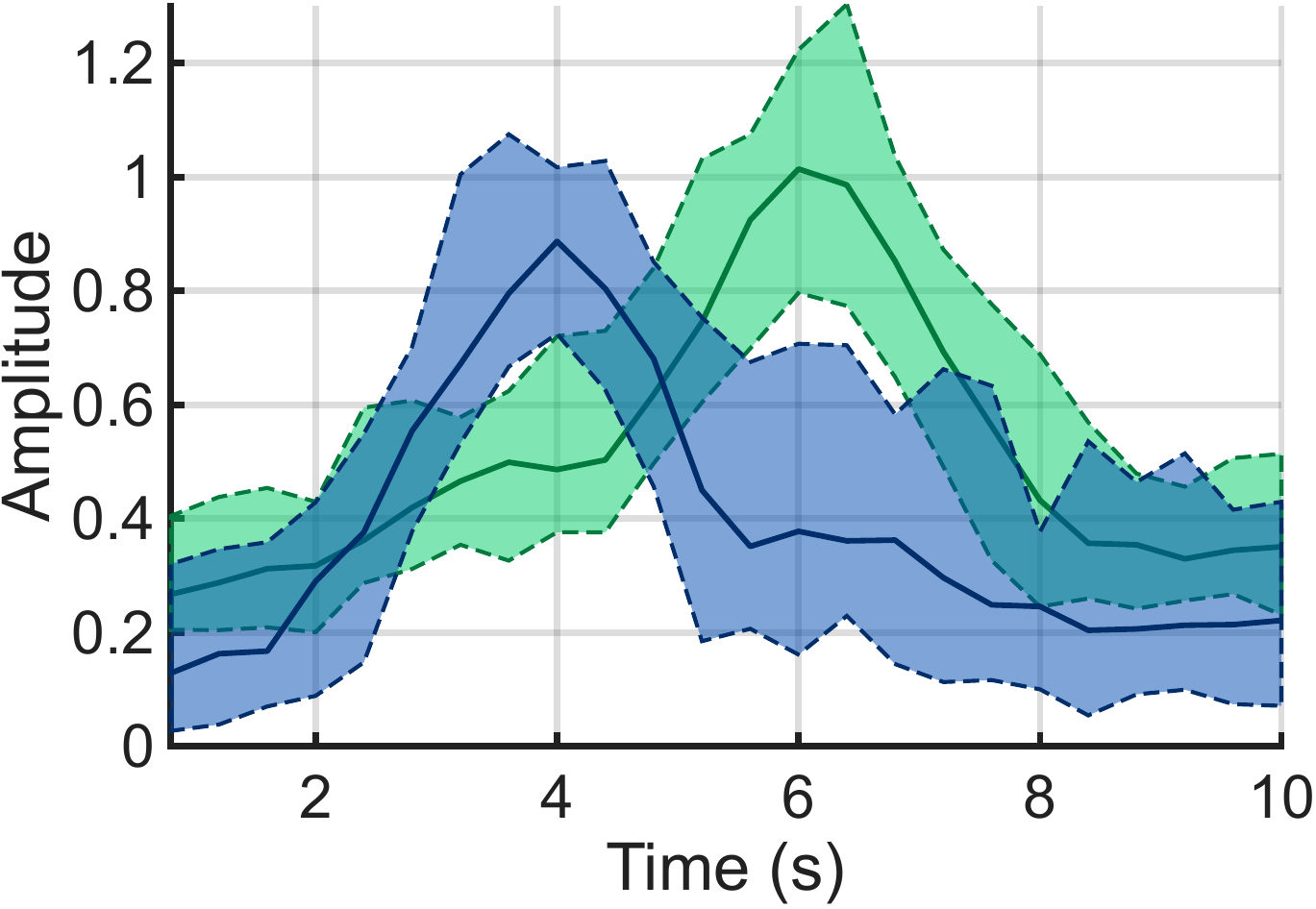}
    \end{minipage}
    \caption{Tracking results with "miscalculated" evolution prior parameters 0 and 70 \unit{\decibel} instead of the 44 \unit{\decibel}. Columns show the results for different measurement noise levels, and the rows indicate whether the track is estimated with the change rate model (CR-SKF), the random walk model (RW-SKF), and sLORETA. The blue curve indicates the deep activity, and the green curve shows the surface activity. The same 25 noise realizations are used.}
    \label{fig:mistracks}
\end{figure*}

\begin{figure}[htb!]
    \centering
    \underline{1250 \unit{\hertz}} 
    
    \vspace{0.02\textwidth}
    \begin{minipage}{0.04\textwidth}
            \rotatebox{90}{\textsl{\scriptsize{\bf CR-SKF}}}
        \end{minipage}\begin{minipage}{0.19\textwidth}
        \begin{center}
            \scriptsize{25 \unit{\decibel}}
        \end{center}
        \includegraphics[width=\textwidth]{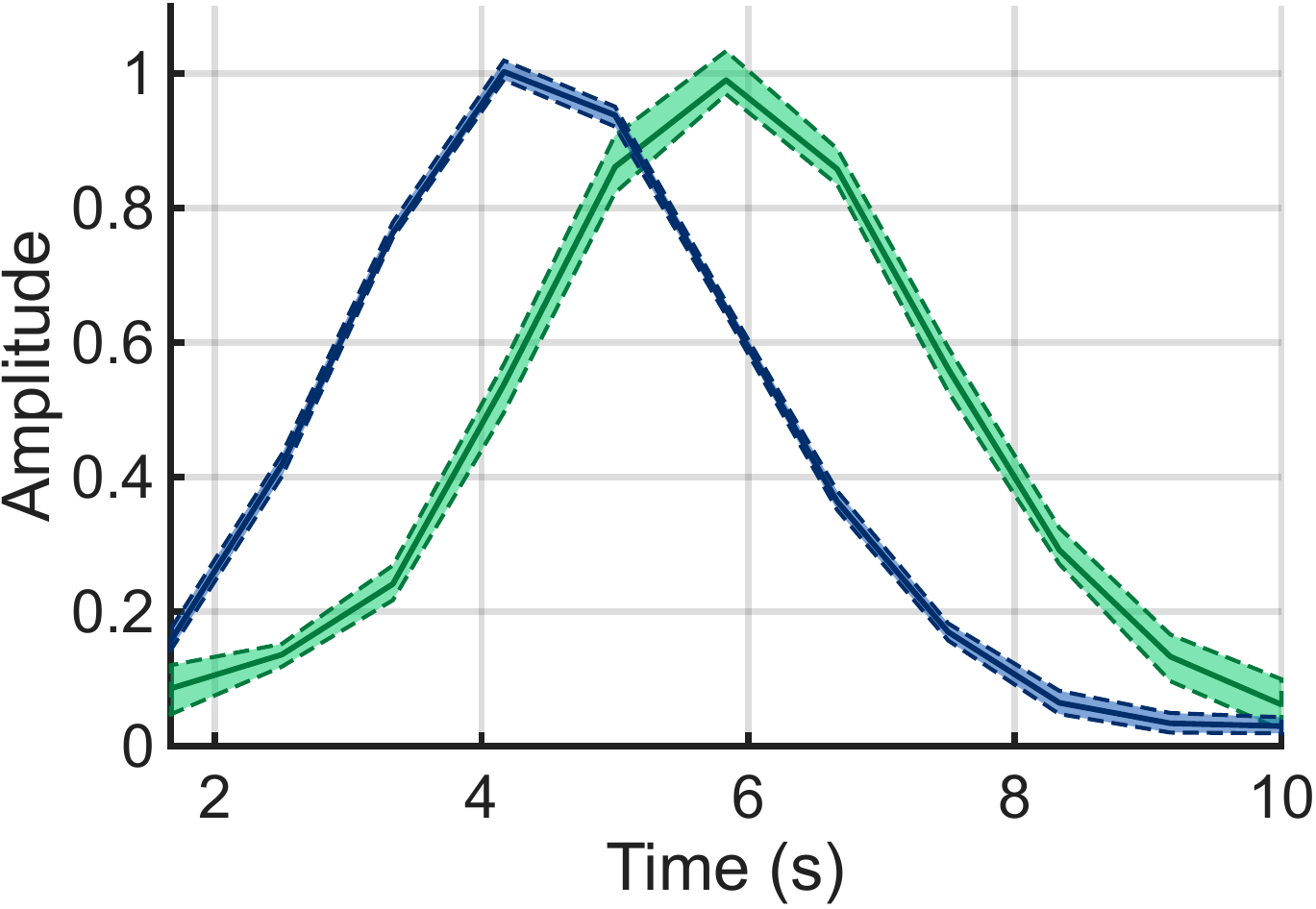}
    \end{minipage}\begin{minipage}{0.19\textwidth}
    \begin{center}
            \scriptsize{15 \unit{\decibel}}
        \end{center}
        \includegraphics[width=\textwidth]{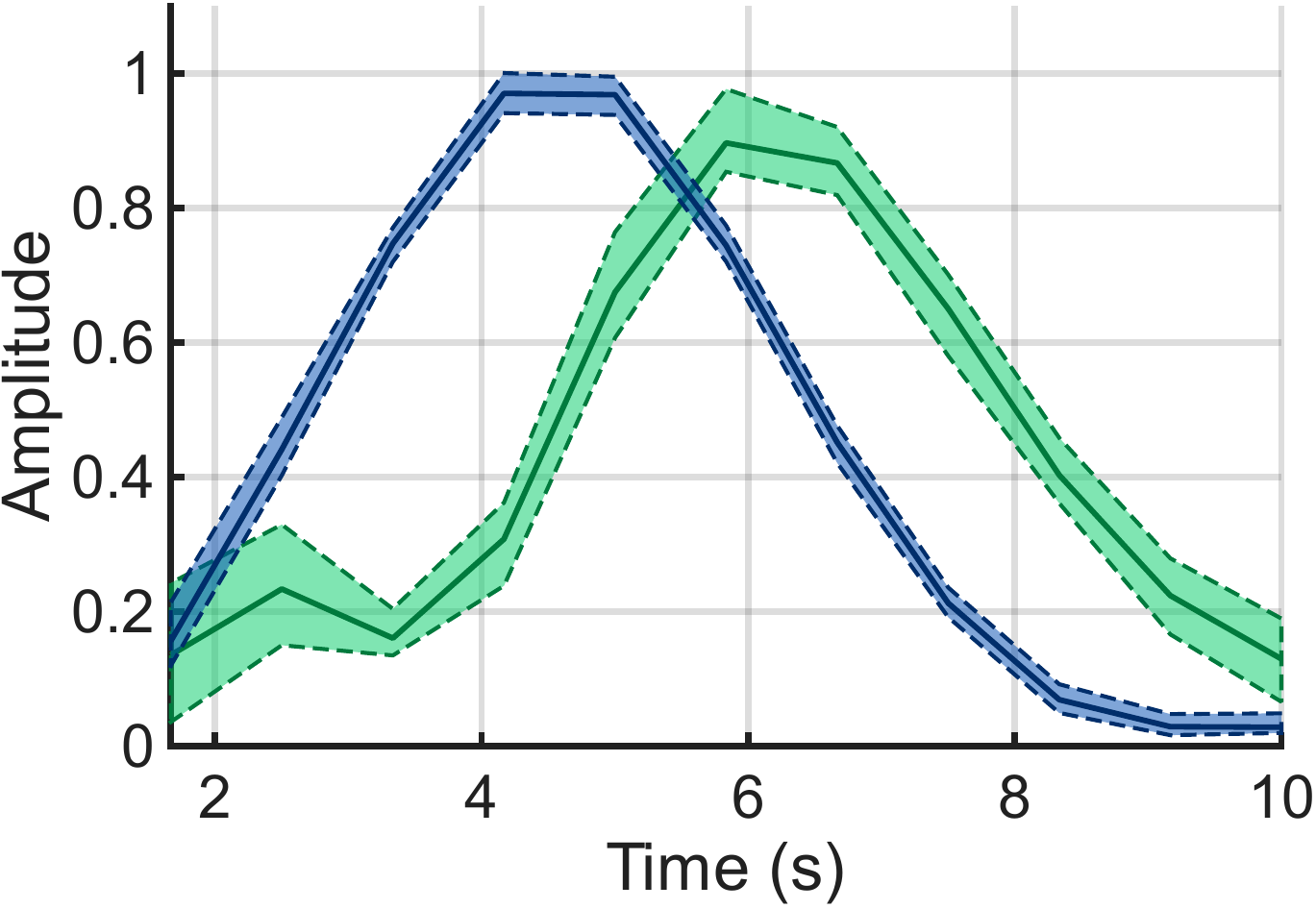}
    \end{minipage}\begin{minipage}{0.19\textwidth}
    \begin{center}
            \scriptsize{5 \unit{\decibel}}
        \end{center}
        \includegraphics[width=\textwidth]{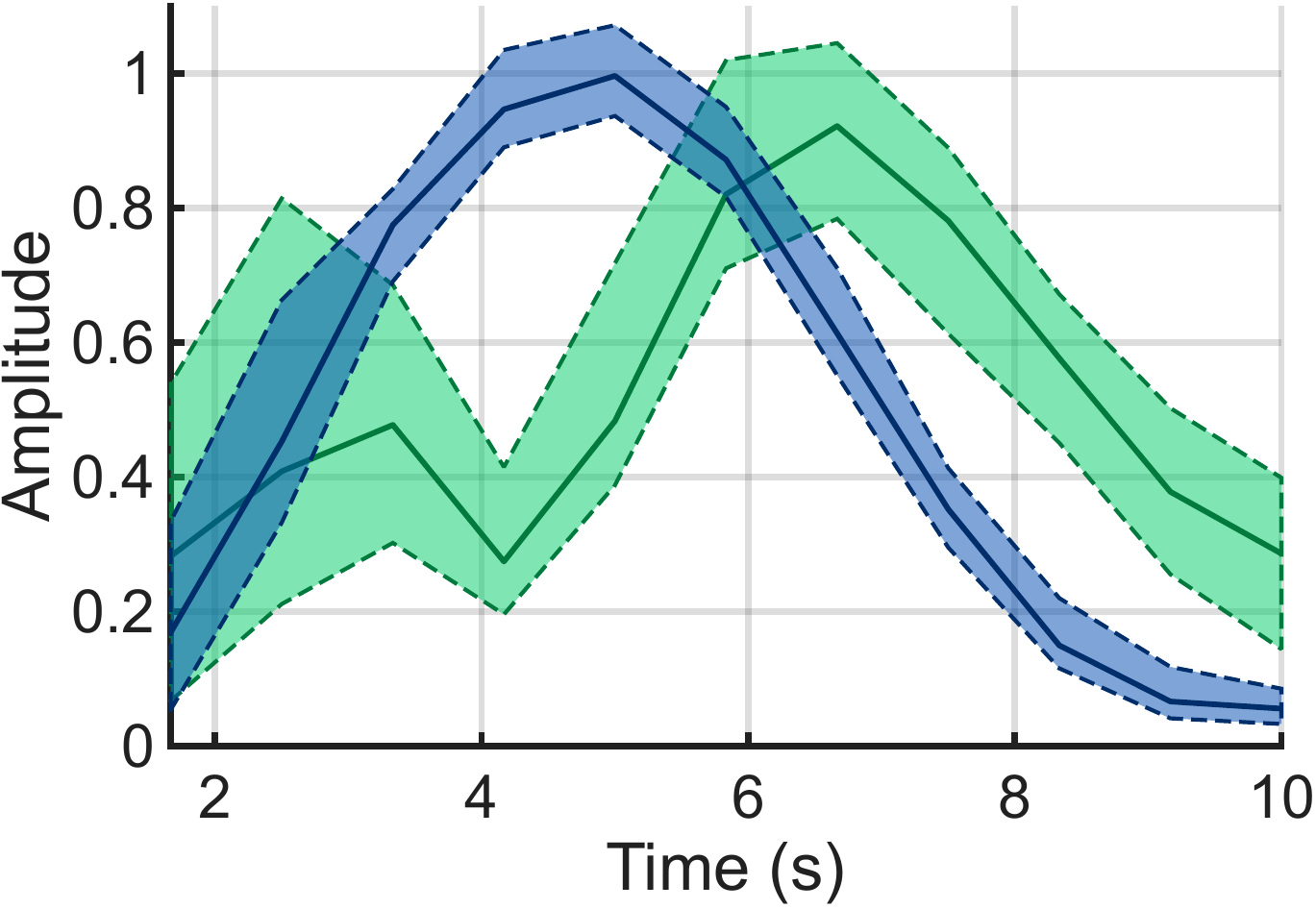}
    \end{minipage}\vspace{0.015\textwidth}
    \begin{minipage}{0.04\textwidth}
            \rotatebox{90}{\textsl{\scriptsize{\bf RW-SKF}}}
        \end{minipage}\begin{minipage}{0.19\textwidth}
        \includegraphics[width=\textwidth]{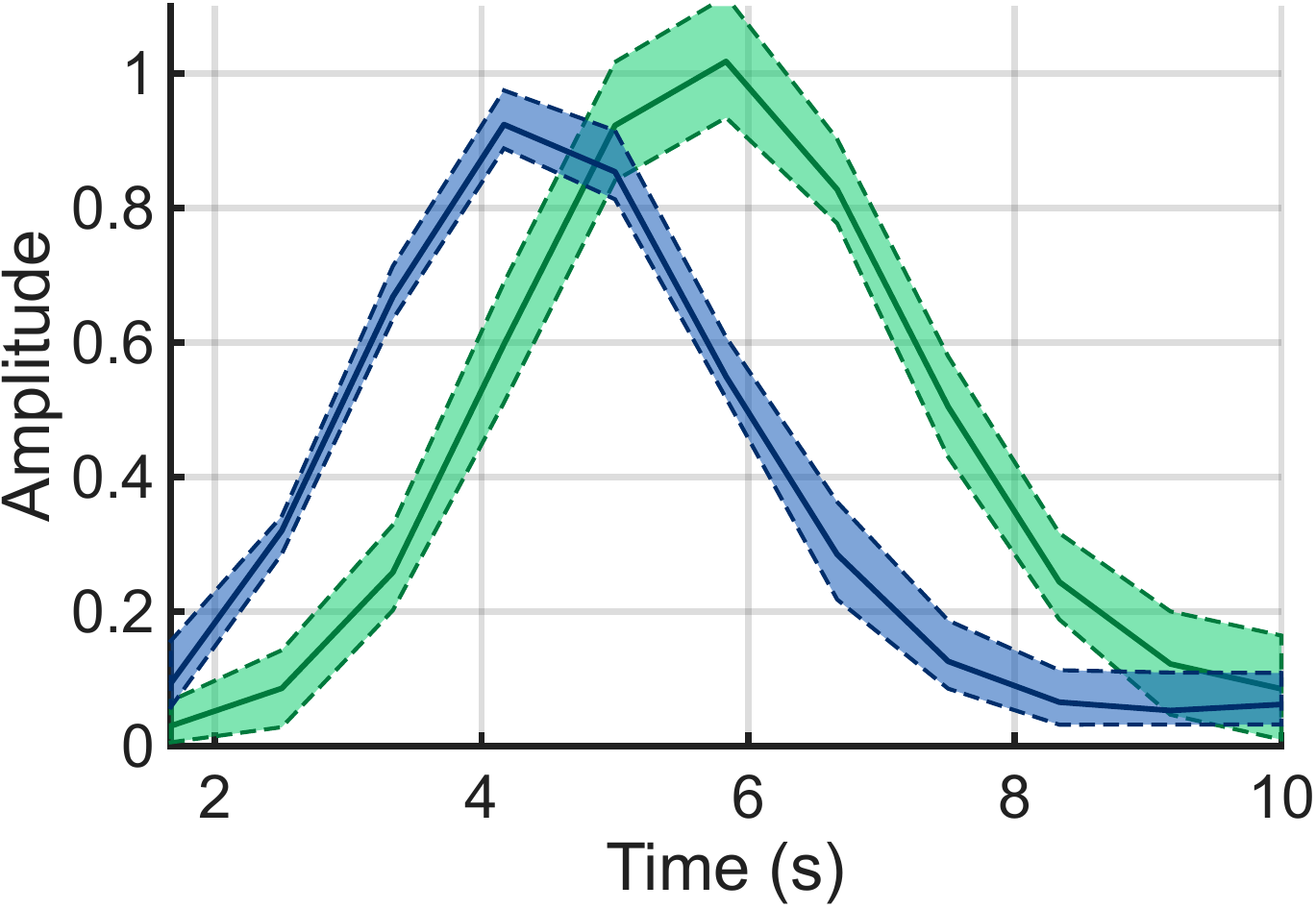}
    \end{minipage}\begin{minipage}{0.19\textwidth}
        \includegraphics[width=\textwidth]{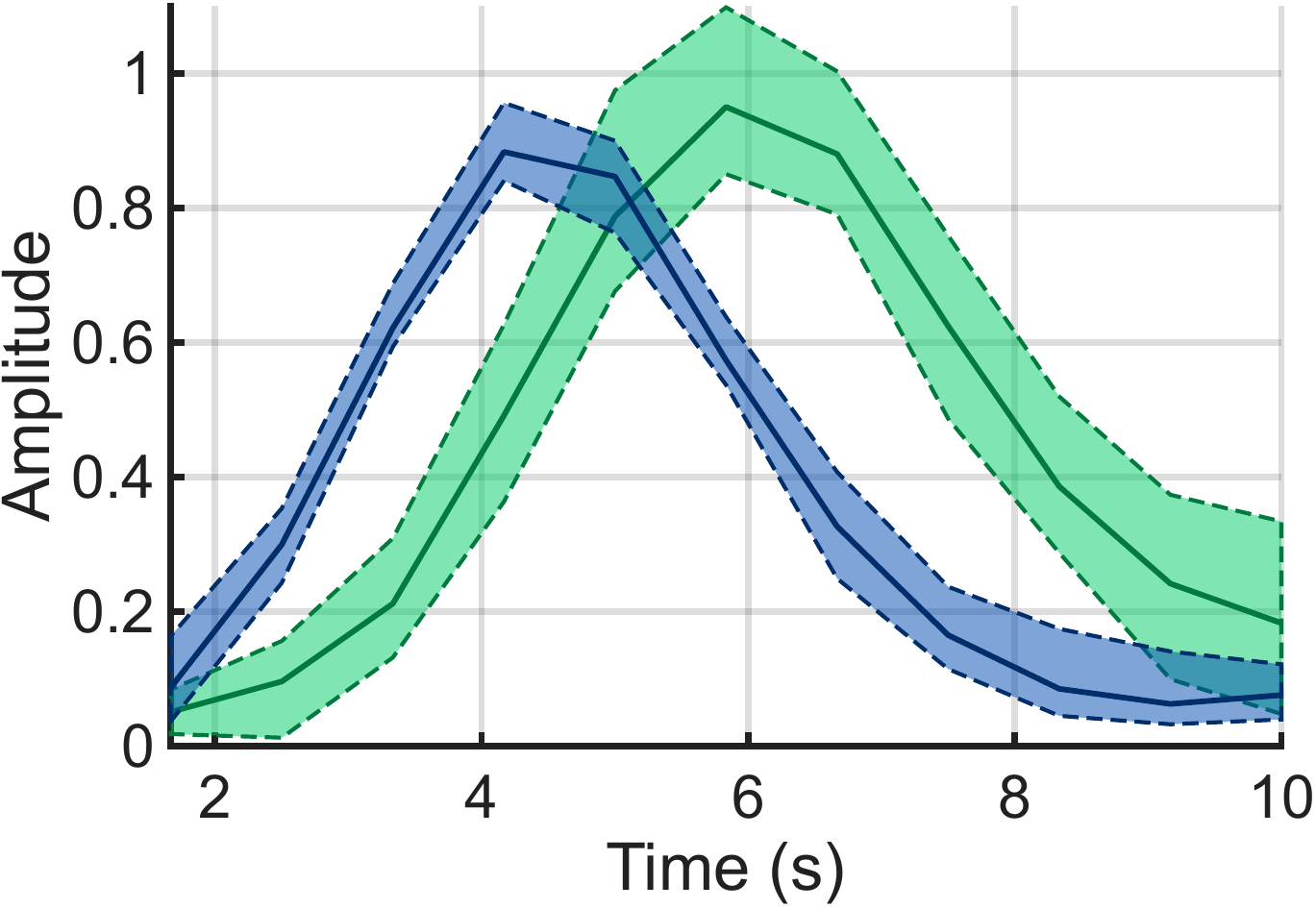}
    \end{minipage}\begin{minipage}{0.19\textwidth}
        \includegraphics[width=\textwidth]{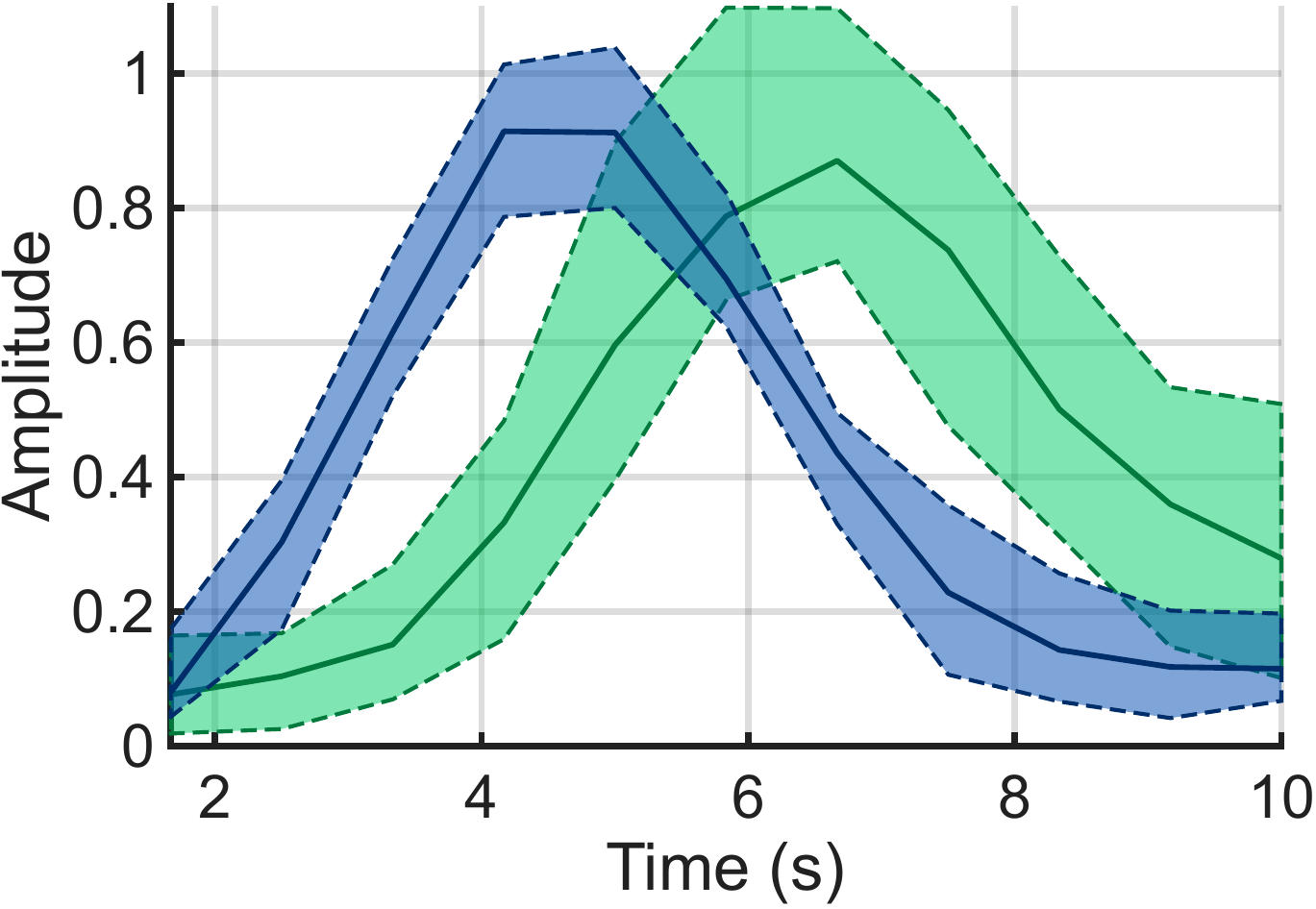}
    \end{minipage}\vspace{0.015\textwidth}
    \begin{minipage}{0.04\textwidth}
            \rotatebox{90}{\textsl{\scriptsize{\bf sLORETA}}}
        \end{minipage}\begin{minipage}{0.19\textwidth}
        \includegraphics[width=\textwidth]{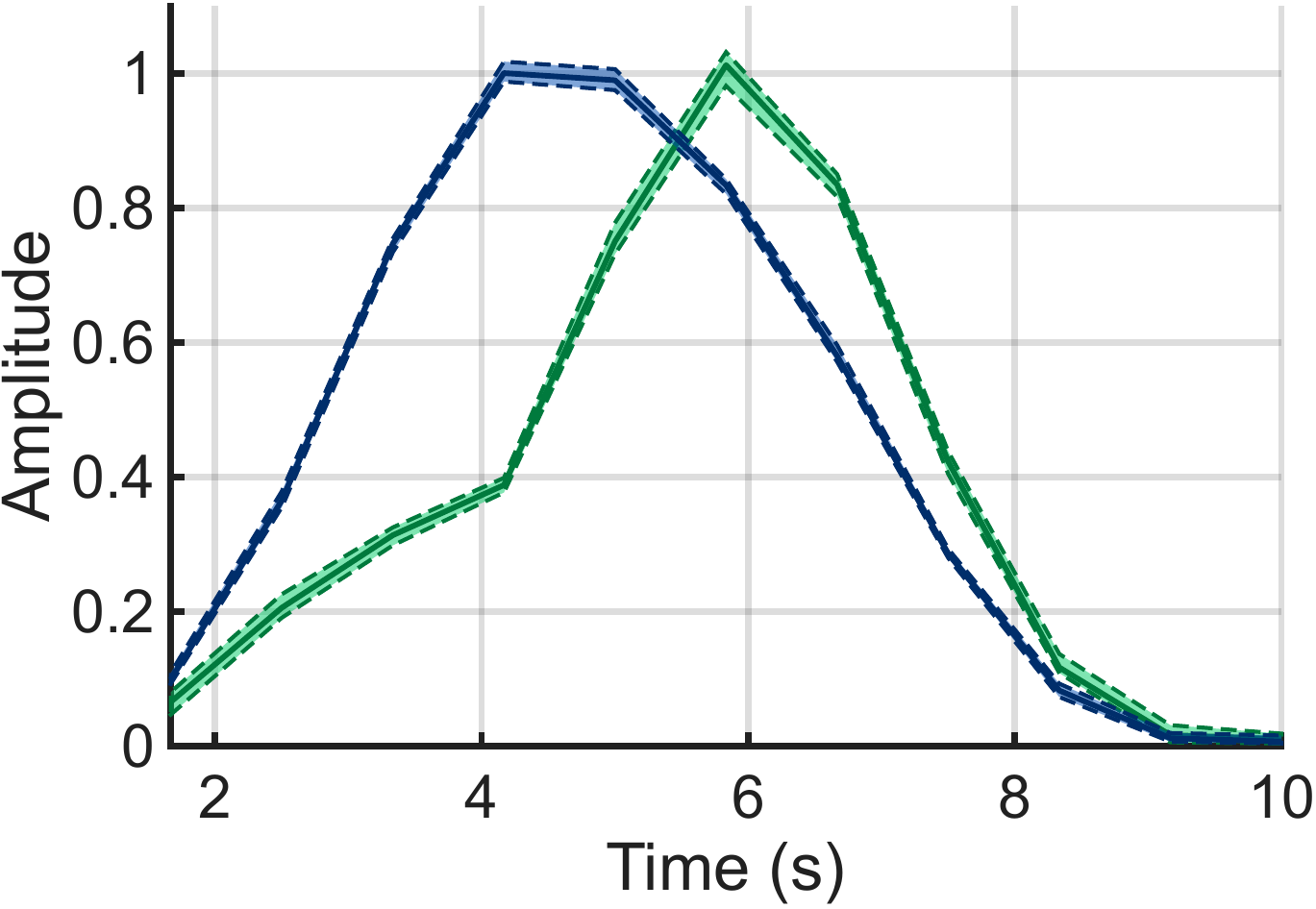}
    \end{minipage}\begin{minipage}{0.19\textwidth}
        \includegraphics[width=\textwidth]{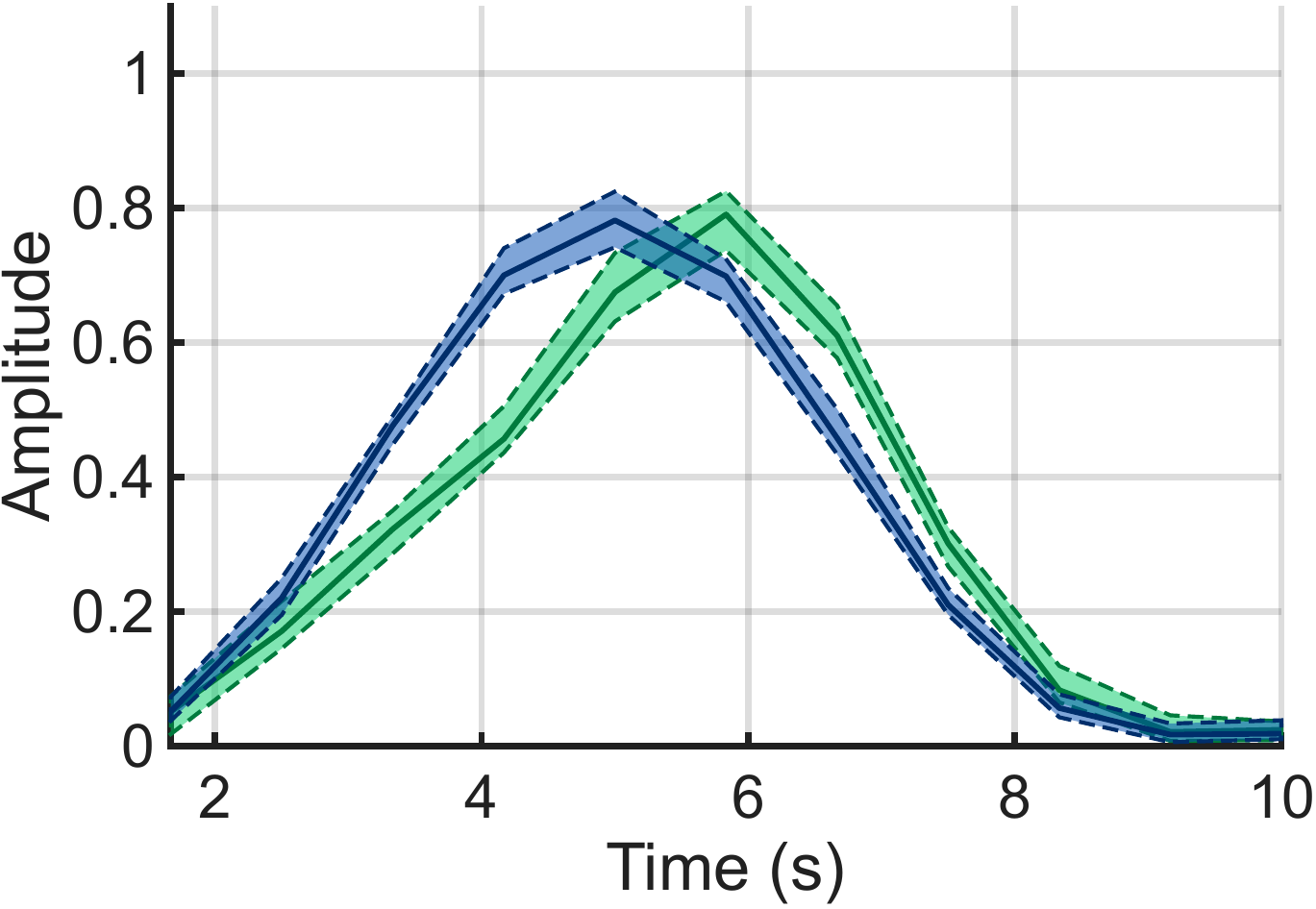}
    \end{minipage}\begin{minipage}{0.19\textwidth}
        \includegraphics[width=\textwidth]{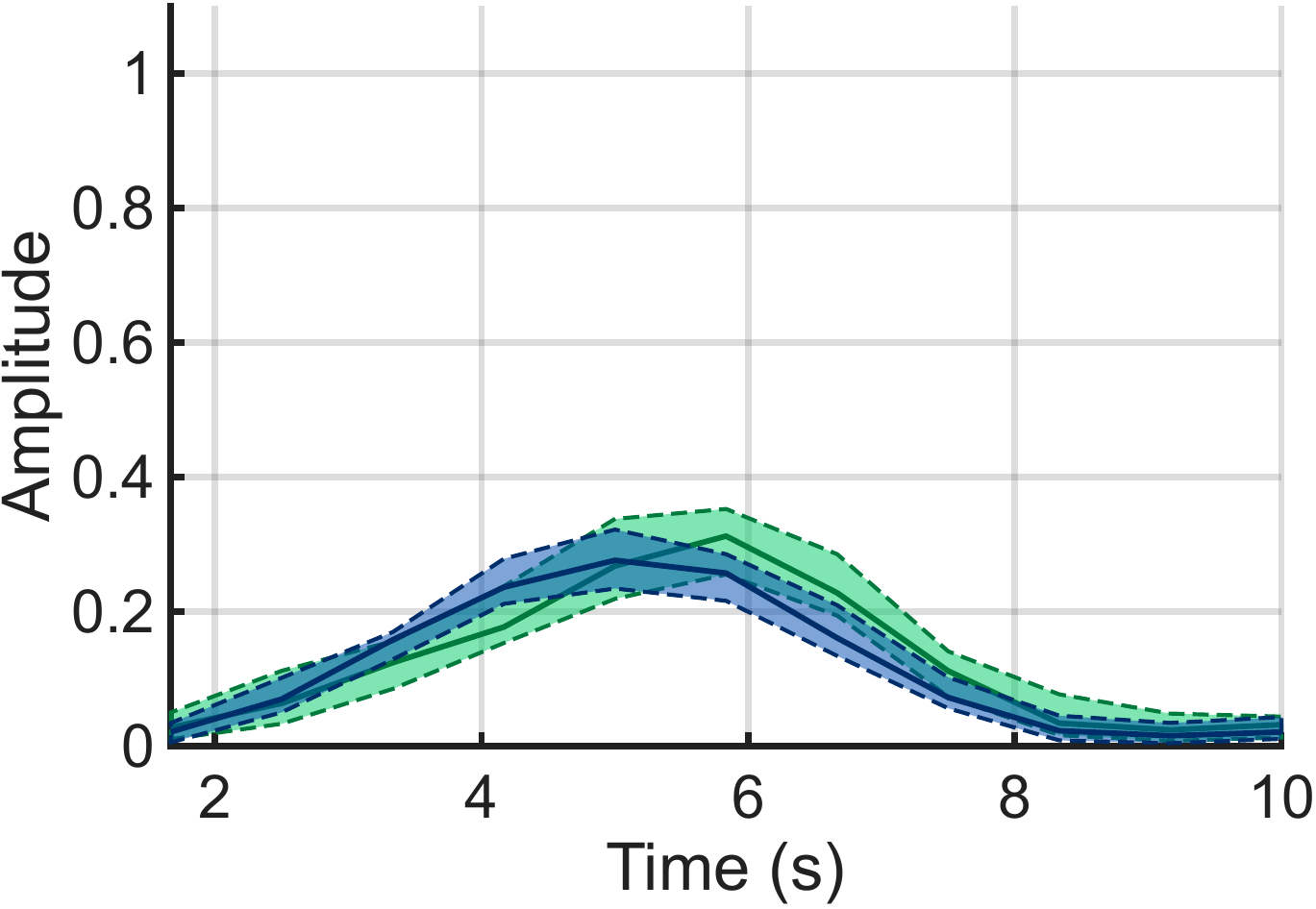}
    \end{minipage}\vspace{0.015\textwidth}
    \hrule\vspace{0.015\textwidth}
    \underline{5000 \unit{\hertz}}
    
    \vspace{0.02\textwidth}
    \begin{minipage}{0.04\textwidth}
            \rotatebox{90}{\textsl{\scriptsize{\bf CR-SKF}}}
        \end{minipage}\begin{minipage}{0.19\textwidth}
        \includegraphics[width=\textwidth]{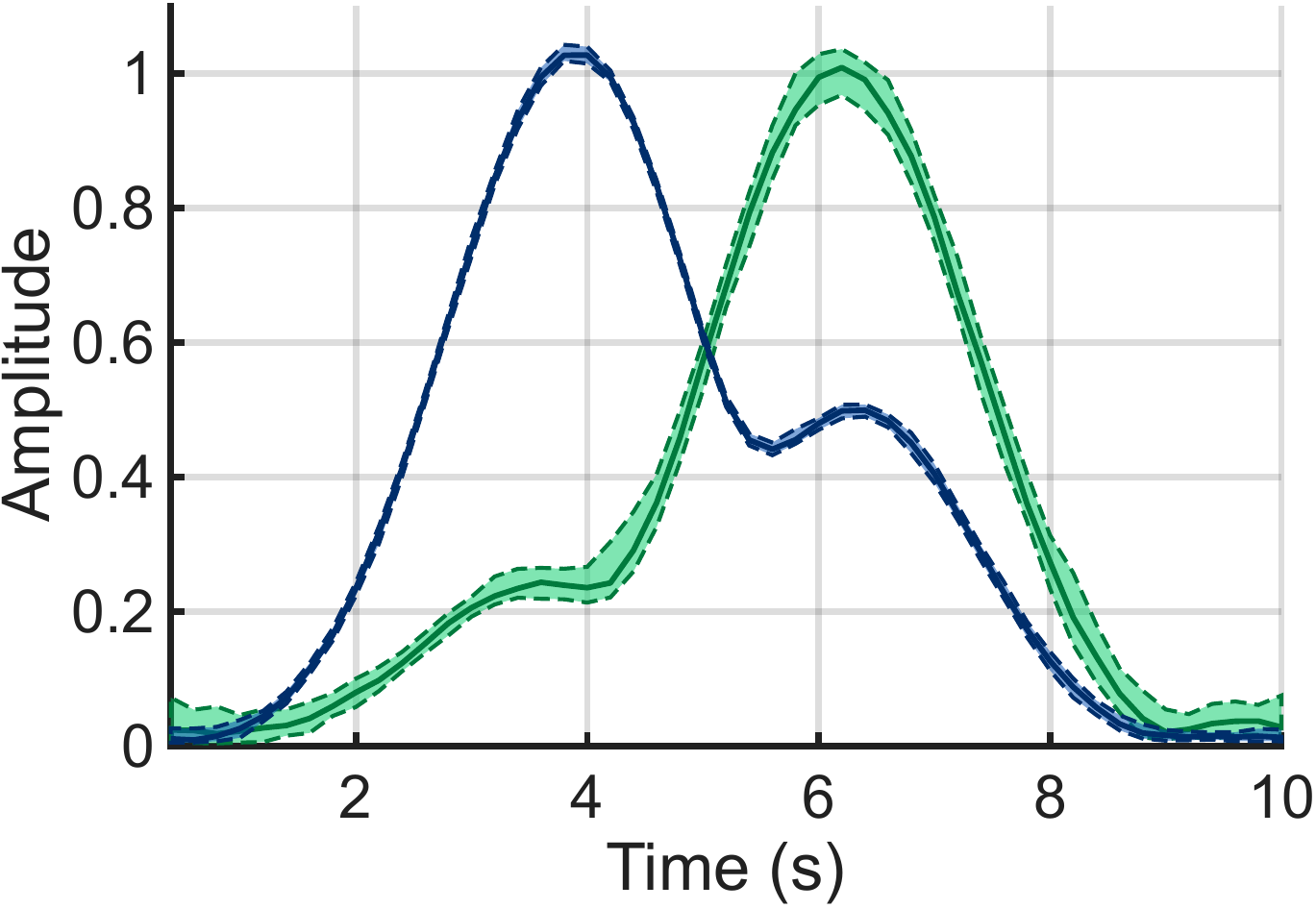}
    \end{minipage}\begin{minipage}{0.19\textwidth}
        \includegraphics[width=\textwidth]{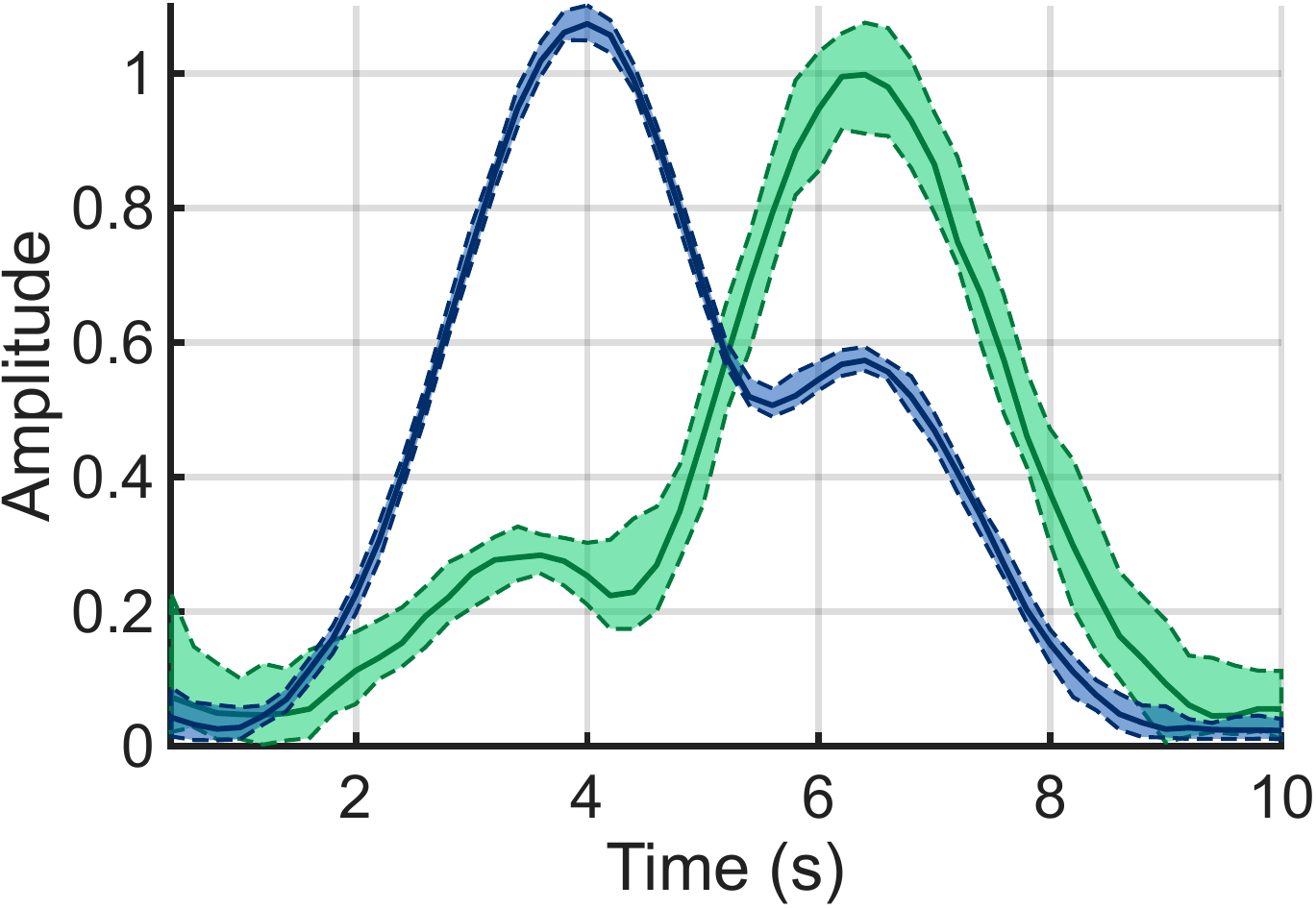}
    \end{minipage}\begin{minipage}{0.19\textwidth}
        \includegraphics[width=\textwidth]{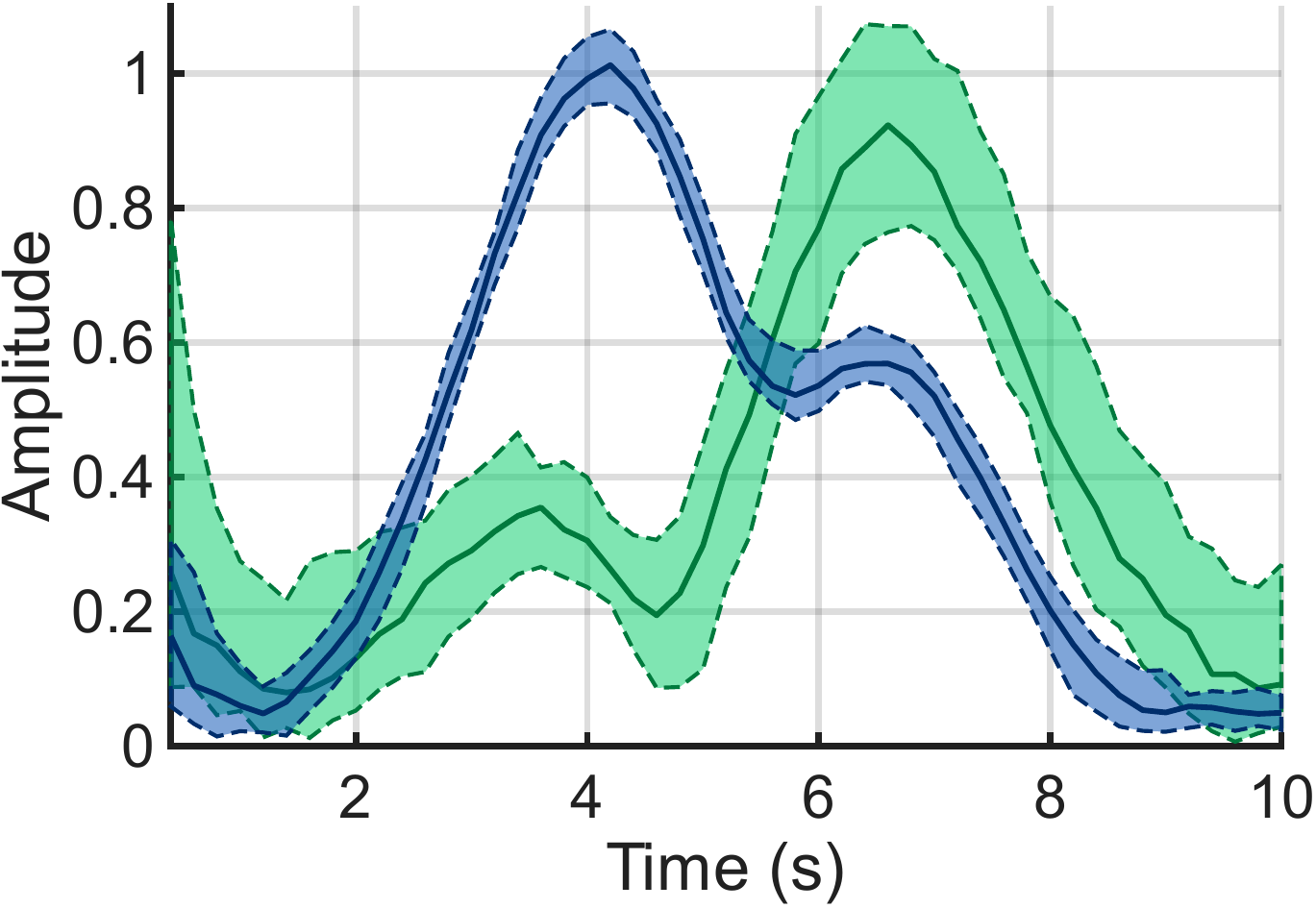}
    \end{minipage}\vspace{0.015\textwidth}
    \begin{minipage}{0.04\textwidth}
            \rotatebox{90}{\textsl{\scriptsize{\bf RW-SKF}}}
        \end{minipage}\begin{minipage}{0.19\textwidth}
        \includegraphics[width=\textwidth]{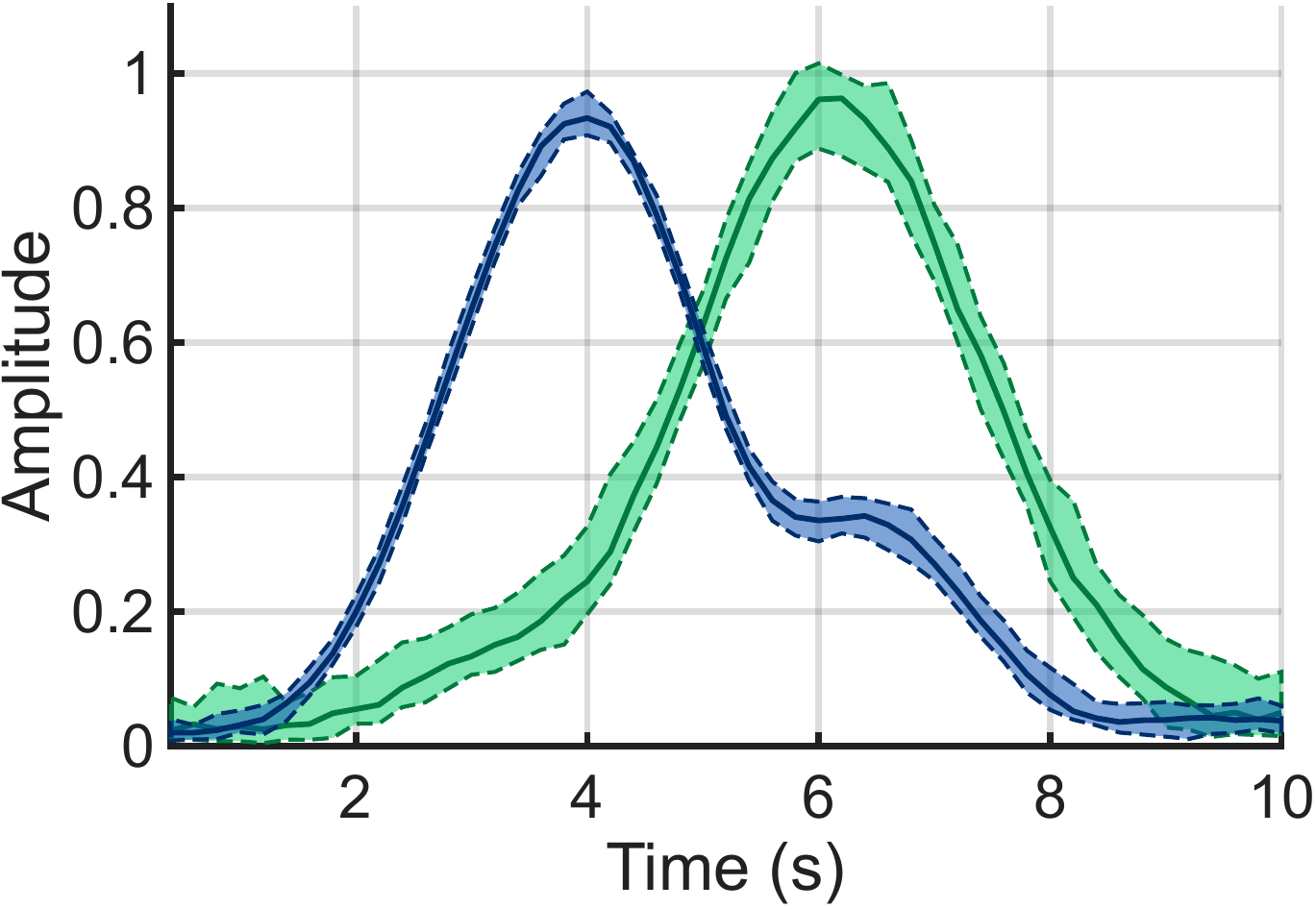}
    \end{minipage}\begin{minipage}{0.19\textwidth}
        \includegraphics[width=\textwidth]{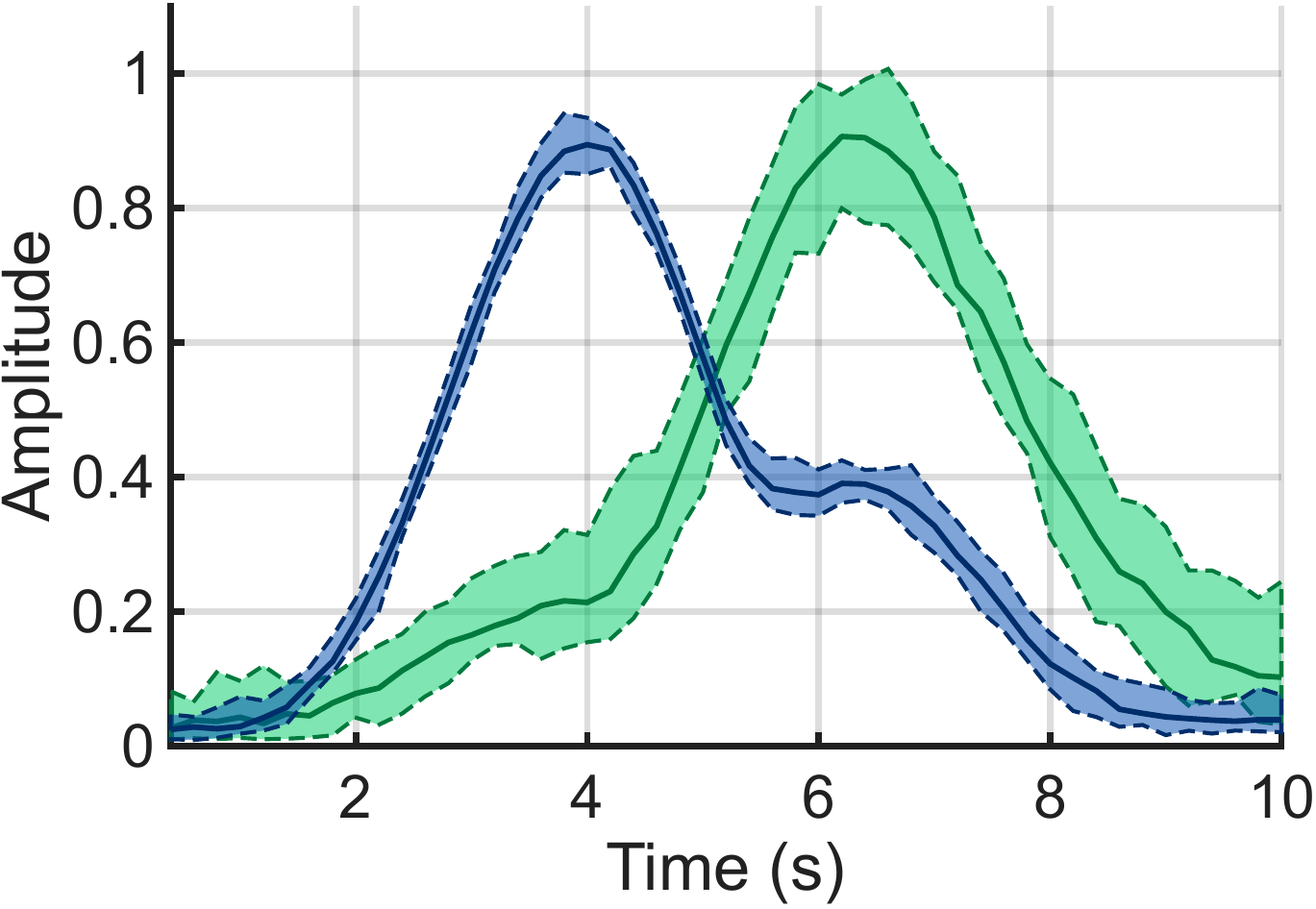}
    \end{minipage}\begin{minipage}{0.19\textwidth}
        \includegraphics[width=\textwidth]{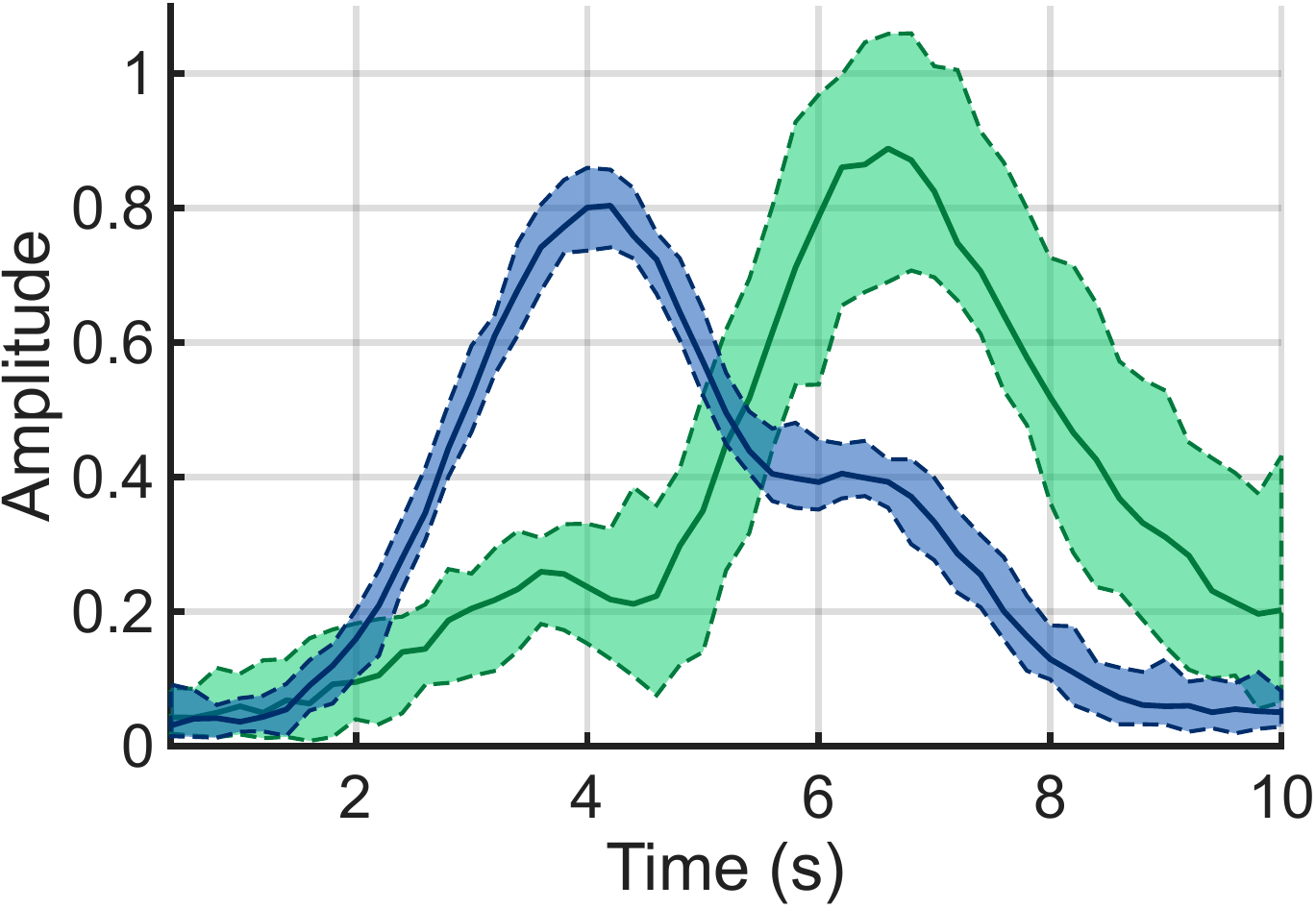}
    \end{minipage}\vspace{0.015\textwidth}
    \begin{minipage}{0.04\textwidth}
            \rotatebox{90}{\textsl{\scriptsize{\bf sLORETA}}}
        \end{minipage}\begin{minipage}{0.19\textwidth}
        \includegraphics[width=\textwidth]{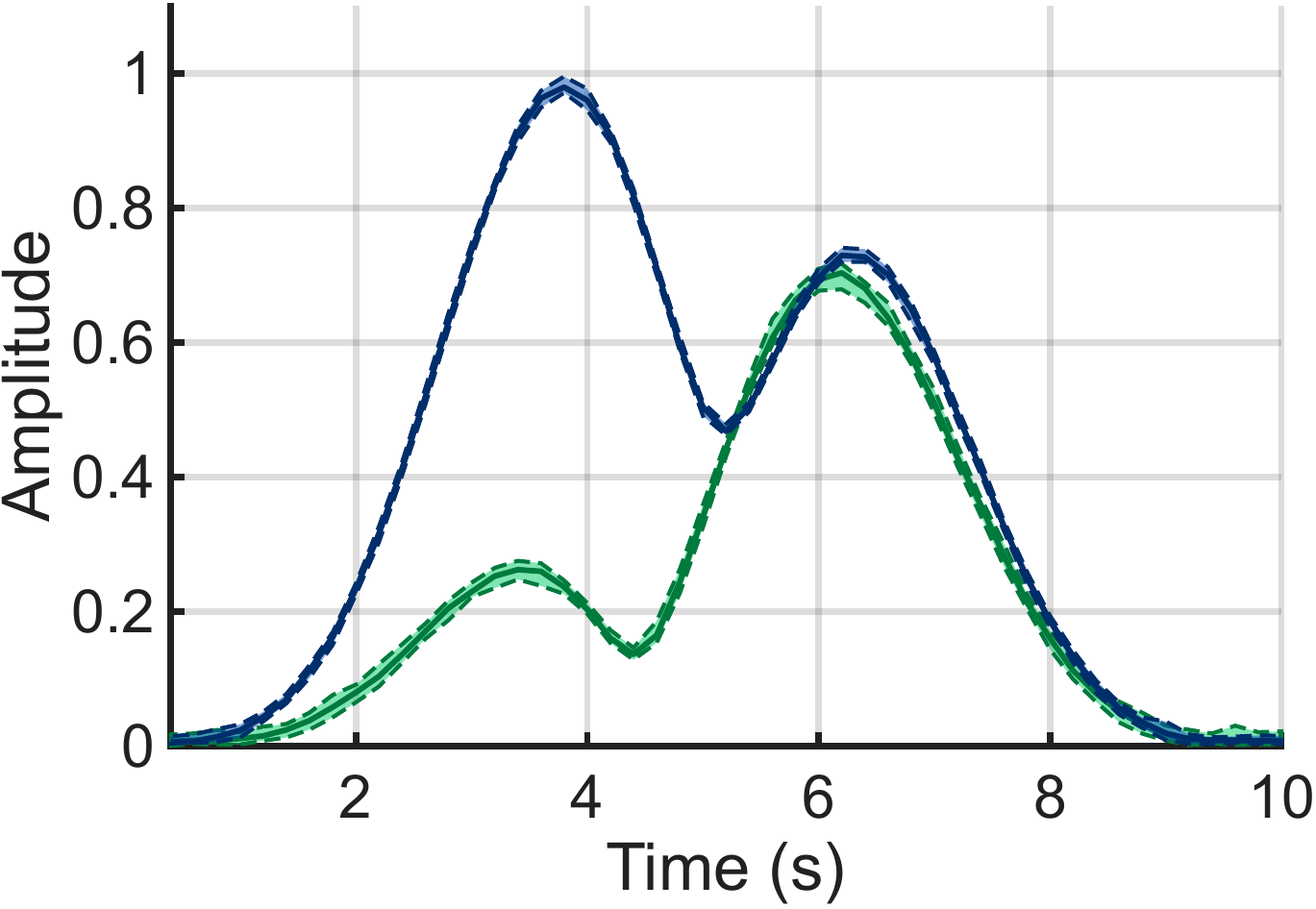}
    \end{minipage}\begin{minipage}{0.19\textwidth}
        \includegraphics[width=\textwidth]{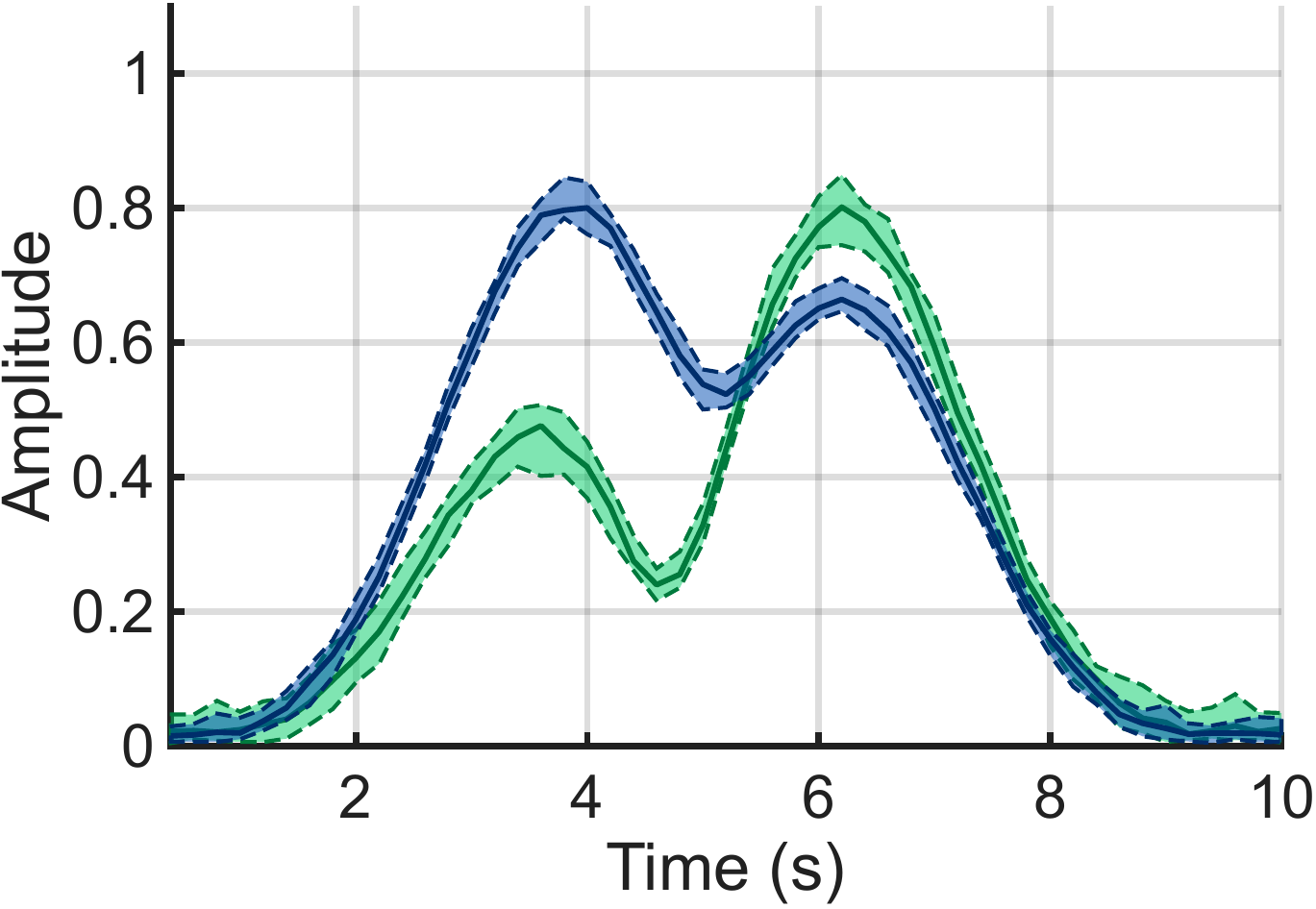}
    \end{minipage}\begin{minipage}{0.19\textwidth}
        \includegraphics[width=\textwidth]{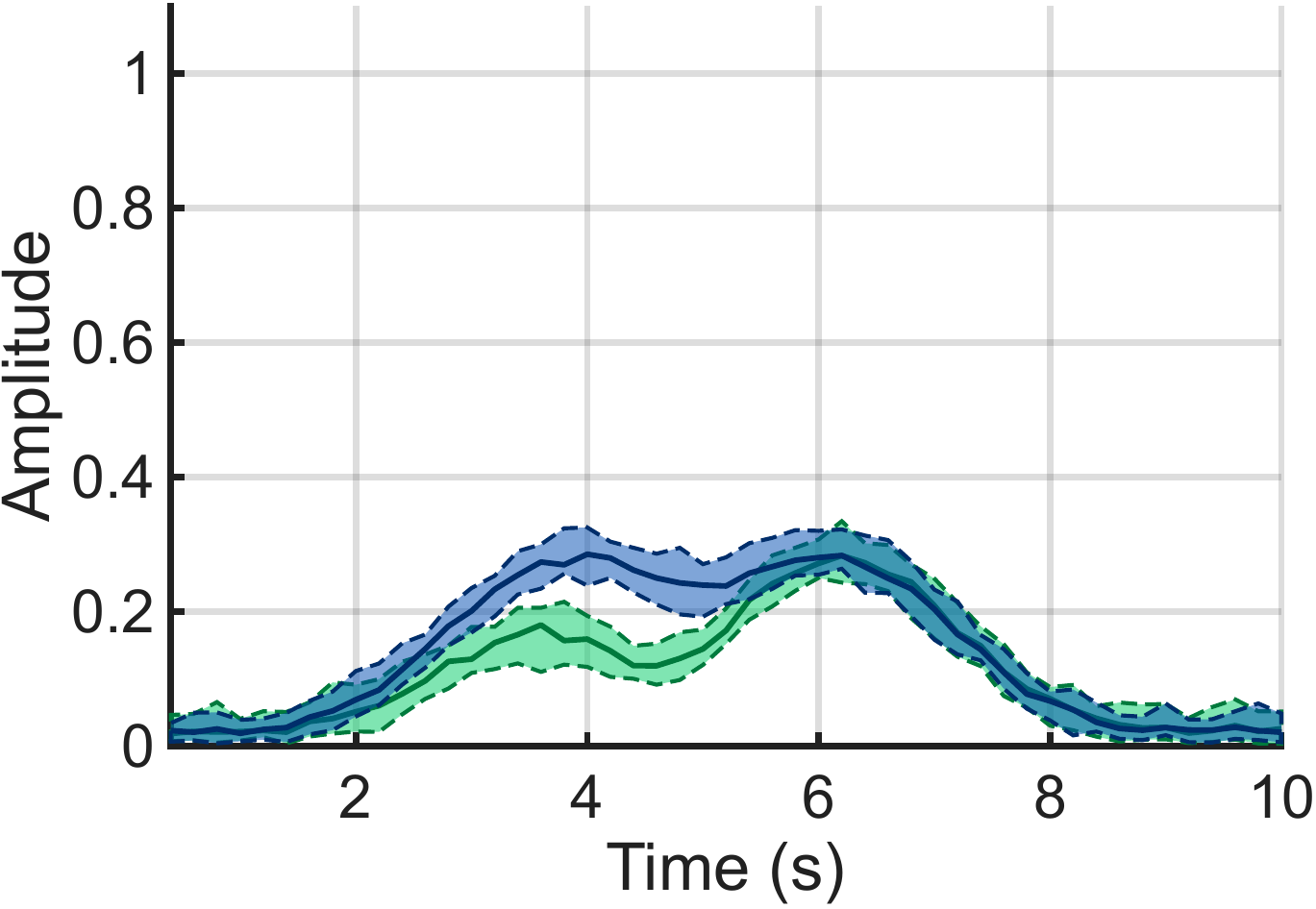}
    \end{minipage}
    \caption{Tracking results with a lower sampling rate of 1250 \unit{\hertz} (upper group) and a higher sampling rate of 5000 \unit{\hertz}. Columns show the results for different measurement noise levels, and the rows indicate whether the track is estimated with the change rate model (CR-SKF), the random walk model (RW-SKF), or sLORETA. The blue curve indicates the deep activity, and the green curve shows the surface activity. The same 25 noise realizations are used midst a group.}
    \label{fig:samplingtracks}
\end{figure}

The results with two extremely different evolution prior parameter $\rho$ values (Figure \ref{fig:mistracks}) show really stable tracks with the value 0 \unit{\decibel}. However, the peak strength is weakening with increasing noise, and both tracks keep shifting forward in time and are delayed even at 25 \unit{\decibel}. It is also noteworthy that the enmeshment of the tracks has increased compared to the tracking with suitable parameters discussed in the previous paragraph.

Based on the results, the CR-SKF track is less affected by noise due to its lower delay and attenuation compared to the RW-SKF. The two bottom rows display the tracks with a $\rho$ value of 70 \unit{\decibel}, in which we see a case where the tracks are overly sensitive to changes in the data. In this case, the RW-SKF tracks start to vary more than the tracks of the CR-SKF. This is especially true of the deep activity represented in blue.

Result obtained with 44 \unit{\decibel} parameter value and two different sampling rates, 1250 and 5000 \unit{\hertz}, shown in Figure \ref{fig:samplingtracks}, display a congruent result with 5000 \unit{\hertz} compared to the results in Figure \ref{fig:tracks}. The only obtainable major difference is the increased enmeshment of deep and cortical activity for CR-SKF. With a low sampling rate, we can clearly obtain a decrease in tracking quality. However, both peaks are clearly separable with the SKF variant. Interestingly, sLORETA, as a method without a dynamical model, exhibits increasing enmeshment as the noise level increases, and activities cannot be separated when the SNR is 5 \unit{\decibel}. We can also see false cortical activity rise at the first steps of CR-SKF at 15 and 5 \unit{\decibel} of SNR.

Results indicate that CR-SKF is the most robust against measurement noise out of the three methods. It is also stable against even large deviations in its parametrization. The sampling rates 1250, 2500, and 5000 \unit{\hertz} are observed to cause minor changes in tracking, even with time-independent sLORETA. Both activities are still distinguishable with both variants of the Standardized Kalman filter.

\subsection{Spatial accuracy in numerical simulation and with real subjects' data}
\begin{figure*}[b!]
    \centering
    \includegraphics[width=0.98\linewidth]{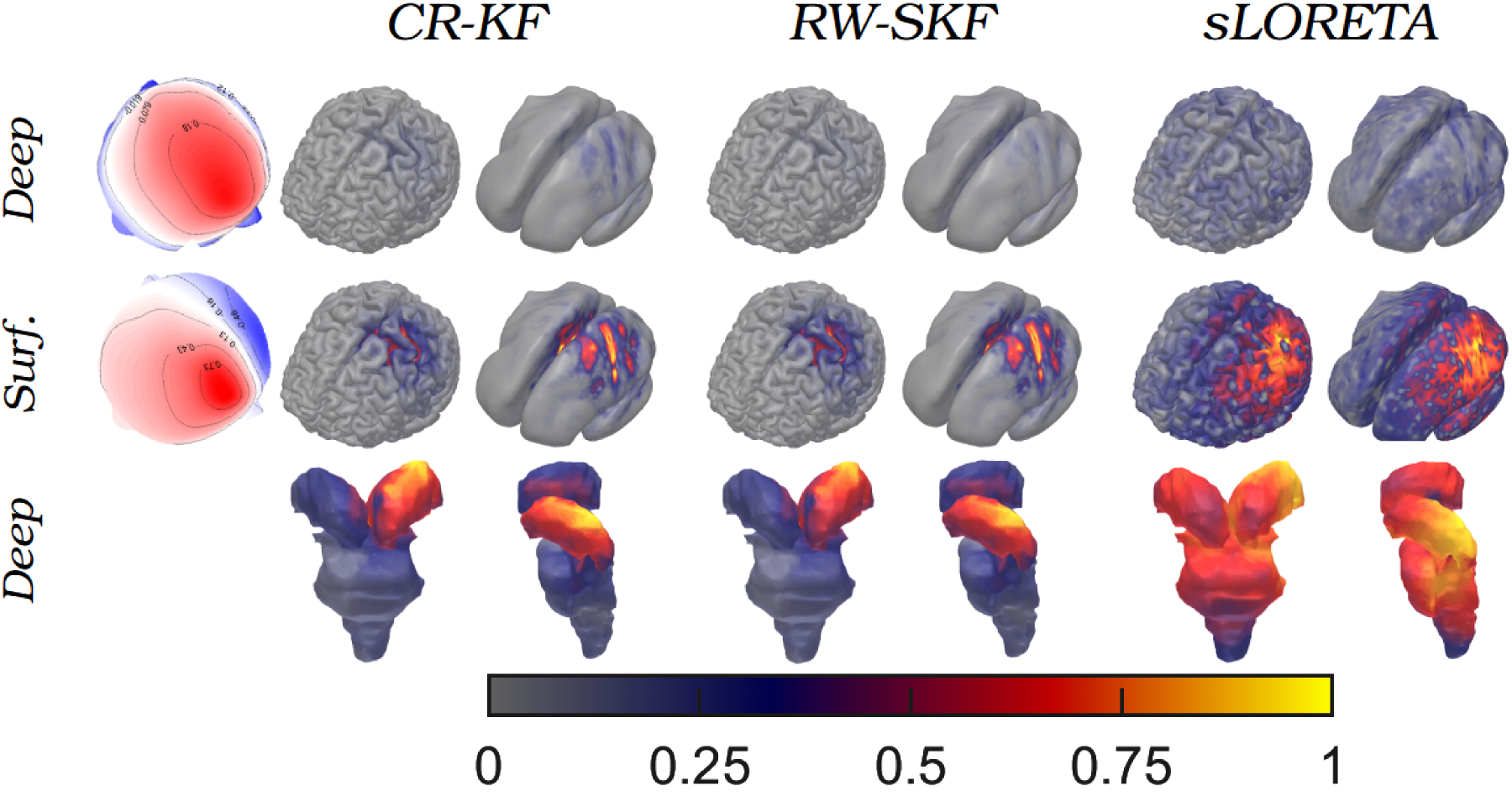}
    \caption{Visualization of spatial distribution of the Change Rate Standardized Kalman filter (CR-SKF), Random Walk-SKF (RW-SKF), and sLORETA estimations of the simulated track peaks presented for each method in adjacent columns. The words "Surf." and "Deep" indicate whether the row displays estimations of surface or deep activity. Next to these, topographies of the normalized peak activity are presented.}
    \label{fig:SynthBrains}
\end{figure*}

The spatial estimation of simulated activity presented in Figure \ref{fig:SynthBrains} indicates that all of the compared methods are capable of localizing the potentials correctly to the somatosensory cortex and left thalamus, as the yellow regions are interpreted as the most prominent location of a point source. CR-SKF and RW-SKF have highly similar estimations, except that the strength of the surface activity at the location of the true source is higher for CR-SKF. sLORETA estimation is significantly wider than the filtering ones, and the strength of cortical projection of the deep activity is stronger, but it does not give false activity on the neocortex, as the coloring stays blue, which indicates about 10 \% of the maximum. Localization errors and earth mover's distances for each method are presented in Table \ref{tab:synthDLEEMD}.

\begin{table}[h!]
    \centering
    \begin{tabular}{|c|c|c|}
         \hline
       \diagbox{Method}{Component}  &  deep & cortical\\ \hline
         \multicolumn{3}{|c|}{Localization error}\\ \hline
        RW-SKF & 0.0 & 4.7\\
        CR-SKF & 0.0 & 4.5\\
        sLORETA & 0.8 & 4.2\\ \hline
        \multicolumn{3}{|c|}{Earth mover's distance}\\ \hline
        RW-SKF & 5.4 & 12\\
        CR-SKF & 5.3 & 12\\
        sLORETA & 6.7 & 13\\ \hline
    \end{tabular}
    \caption{Localization errors and earth mover's distances (in millimeters) obtained with synthetic data for Random Walk-Standardized Kalman filter (RW-SKF), change rate-modeled SKF (CR-SKF), and sLORETA. }
    \label{tab:synthDLEEMD}
\end{table}

\begin{figure*}[t!]
    \centering
    \includegraphics[width=0.98\linewidth]{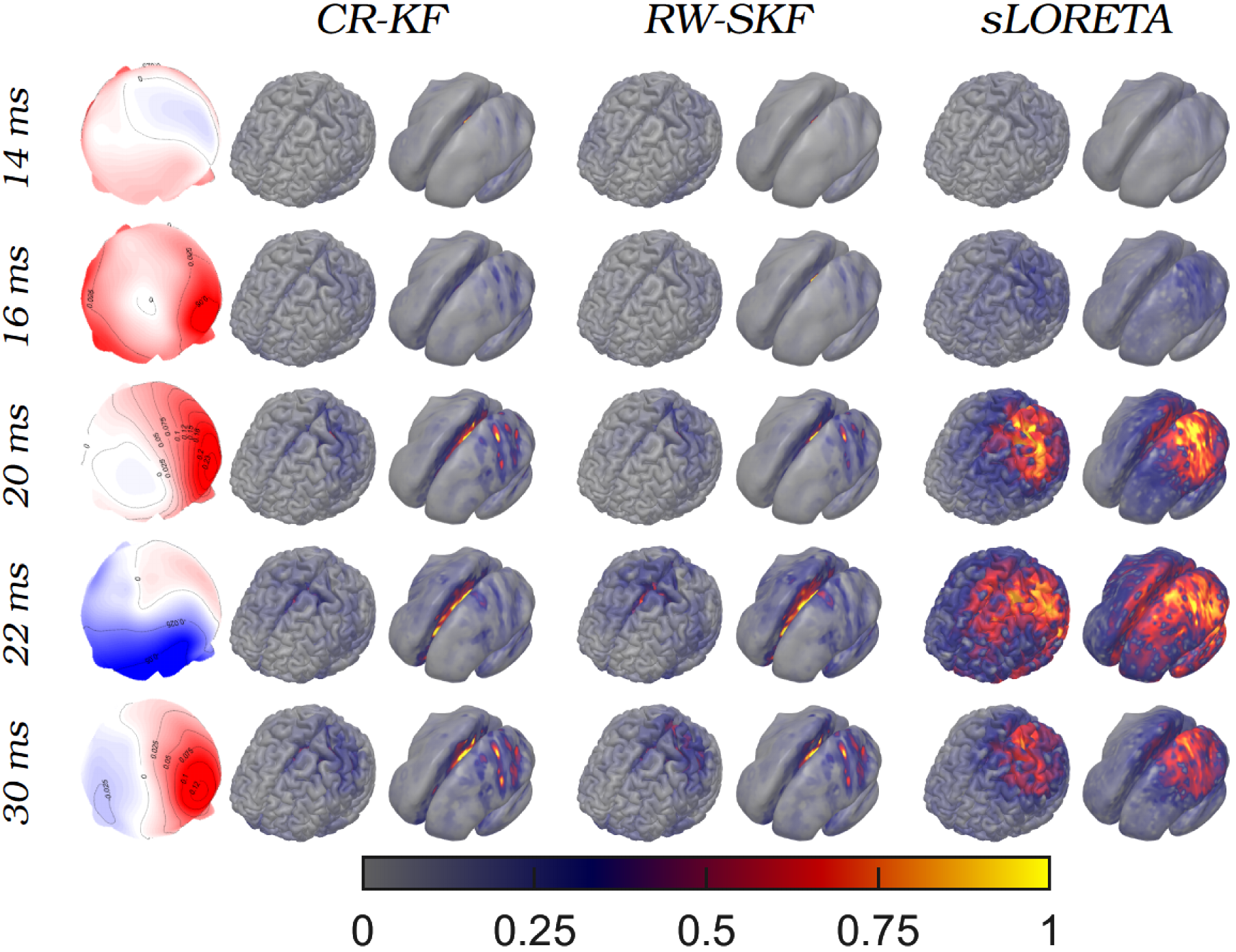}
    \caption{Visualization of spatial distribution of the Change Rate Standardized Kalman filter (CR-SKF), Random Walk-SKF (RW-SKF), and sLORETA estimations of the 1st subject's somatosensory evoked potential peaks presented for each method in adjacent columns. The rows display the average post-stimulus time points. Next to these, topographies of the normalized peak activity are presented.}
    \label{fig:realCortex}
\end{figure*}

\begin{figure*}[t!]
    \centering
    \includegraphics[width=0.55\linewidth]{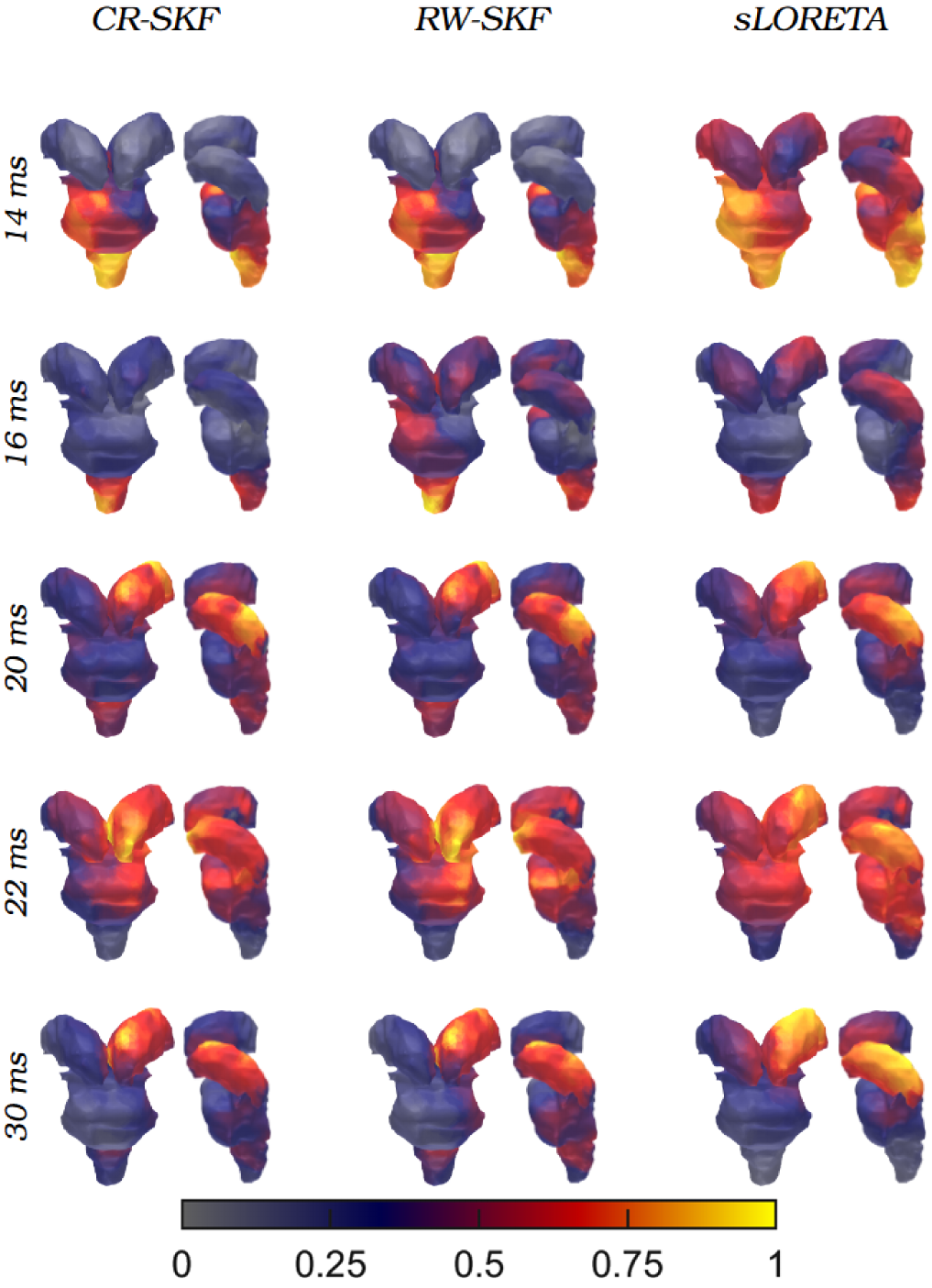}
    \caption{Visualization of spatial distribution of the Change Rate Standardized Kalman filter (CR-SKF), Random Walk-SKF (RW-SKF), and sLORETA estimations in the deep brain structures using the 1st subject's data. Different rows present the activity at different somatosensory evoked potential peaks.}
    \label{fig:realDeep}
\end{figure*}

\begin{figure*}[t!]
    \centering
    \includegraphics[width=0.98\linewidth]{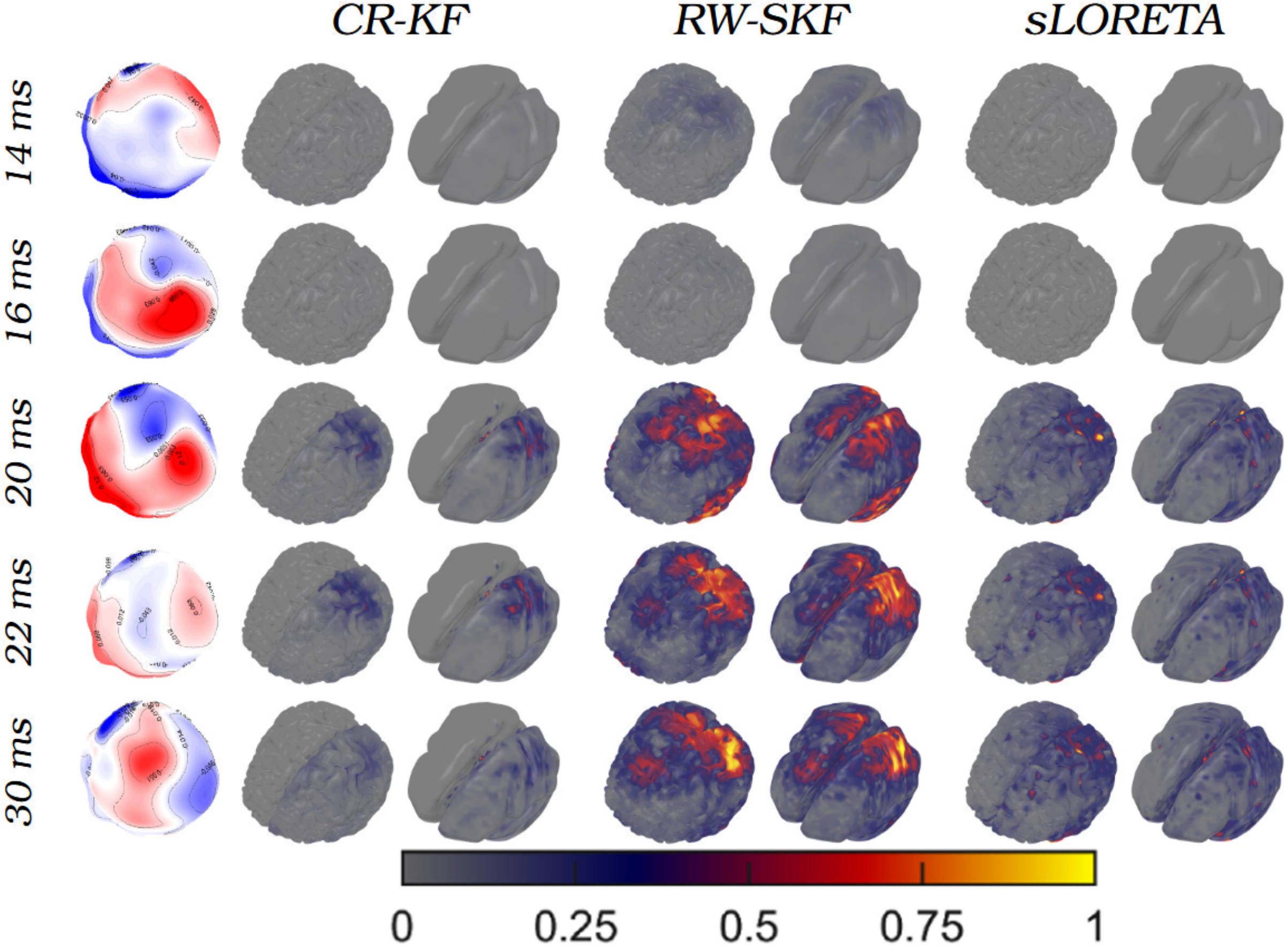}
    \caption{Visualization of spatial distribution of the Change Rate Standardized Kalman filter (CR-SKF), Random Walk-SKF (RW-SKF), and sLORETA estimations of the 2nd subject's somatosensory evoked potential peaks presented for each method in adjacent columns. The rows display the average post-stimulus time points. Next to these, topographies of the normalized peak activity are presented.}
    \label{fig:S2Cortex}
\end{figure*}

\begin{figure*}[t!]
    \centering
    \includegraphics[width=0.55\linewidth]{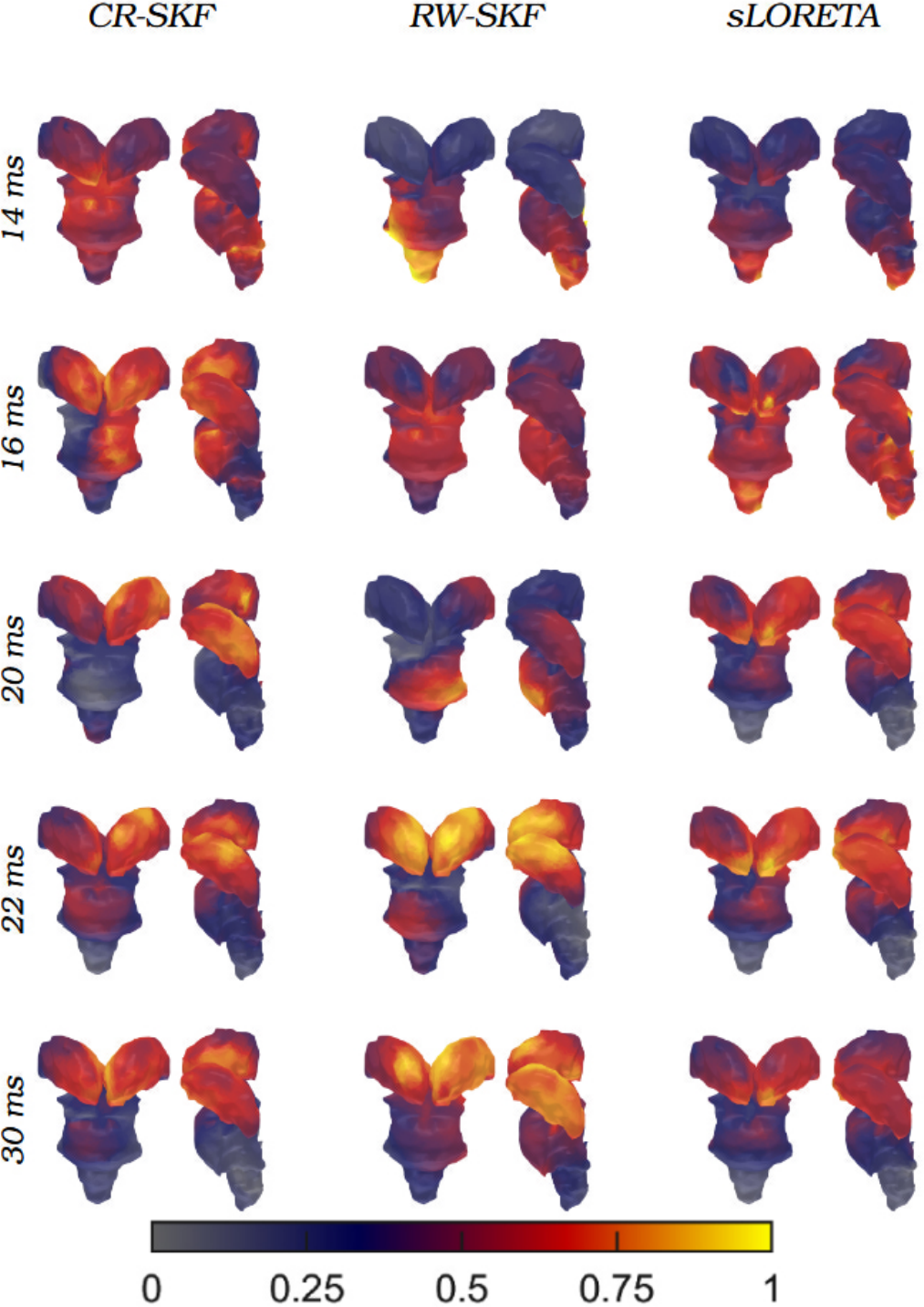}
    \caption{Visualization of spatial distribution of the Change Rate Standardized Kalman filter (CR-SKF), Random Walk-SKF (RW-SKF), and sLORETA estimations in the deep brain structures using the 2nd subject's data. Different rows present the activity at different somatosensory evoked potential peaks.}
    \label{fig:S2Deep}
\end{figure*}

The estimates of the spatial locations of somatosensory evoked potentials from the 1st real subject are presented for the surface layer in Figure \ref{fig:realCortex} and for deep structures, i.e., the brainstem and thalamus, in Figure \ref{fig:realDeep}. Based on the results and the literature, sLORETA is localizing each potential nearly perfectly. At 14 \unit{\milli\second}, the estimated activity at pons is wrongly extended to the bottom of the brainstem. The 16 \unit{\milli\second} bi-activity at the bottom of the brainstem and left thalamus is expected due to the montage-averaging attributed to the data since the signal is traveling from the bottom of the brainstem to the left thalamus during that time period. In a single event, the 16 \unit{\milli\second} component can be seen flashing at the bottom of the brainstem or at the left thalamus. At 20, 22, and 30 \unit{\milli\second}, we observe activity in the left thalamus, as expected. 20 and 30 \unit{\milli\second} components can be expected to be similar based on the literature description of the component. In the cortex, we see the concentrated spot at the sulcus wall at 20 and 30 \unit{\milli\second} and an extended region in the central sulcus at 22 \unit{\milli\second}.

Similarly to the simulated cases, both SKFs yield highly similar estimates, with CR-SKF showing greater strength for the 20 and 30 \unit{\milli\second} components in the somatosensory cortex. In both cases, the 14 \unit{\milli\second} mislocalized to the bottom of the brainstem, and the surface component at 22 \unit{\milli\second} went undetected. The accuracy of localization of the 16 \unit{\milli\second} component cannot be confidently compared among the methods for the previously mentioned reason. However, we can state that the RW-SKF and sLORETA results are more plausible than the CR-SKF estimate. Interestingly, both SKFs exhibit a small bright spot in the thalamus at 20, 22, and 30 \unit{\milli\second}. From the better-known deep components, at 20 \unit{\milli\second}, the spot is at the ventral posterolateral portion of the thalamus, which is the correct sub-region for the activity. Also, the spot on the anterior part at 30 \unit{\milli\second} is according to the literature (Figure \ref{fig:LiteratureComponents}).

The results for the 2nd subject, presented in Figures \ref{fig:S2Cortex} and \ref{fig:S2Deep}, show correct localization for cortical sources; however, the source extends over the primary somatosensory cortex in the case of RW-SKF and sLORETA. CR-SKF restricts the activity in the somatosensory cortex and extends to the primary motor cortex when 22 \unit{\milli\second} activity is peaking, but the estimation strength for cortical sources is significantly weaker than with the 1st subject. sLORETA exhibits secondary localization at the parietal lobe, posterior to the somatosensory cortex. 

The deep 14 \unit{\milli\second} component is estimated to be at the bottom of the brainstem by RW-SKF and sLORETA, while extending to the pons for RW-SKF. CR-SKF localizes the deep activity to the pons. The 16 \unit{\milli\second} component is localized on the pathway between the bottom of the brainstem and the left thalamus. The methods compared show activity at the left thalamus at the 20 \unit{\milli\second} peak time point, although the maximum of RW-SKF is localizable to the medulla. For the 22 \unit{\milli\second} component, RW-SKF and sLORETA show activity in both thalami, whereas CR-SKF is on the left. The activity is shared among the thalami also at the 30 \unit{\milli\second}. Localization error and earth mover's distances are presented for both subjects in Table \ref{tab:realDLEEMD}.

In conclusion, the numerical case shows that the localization of the activities with SKF variants is nearly identical, whereas sLORETA is overly spread across the cortex. With the first subject, sLORETA localizes the sources perfectly, and both SKFs exhibit estimation error. With the second subject, sLORETA localizes cortical activity in line with the literature, RW-SKF provides wide estimates, and CR-SKF is cortically focal and concordant with the literature. Deep sources are estimated according to the literature only by CR-SKF. Overall, CR-SKF localizes activities best because, when errors occur, they are significant in sLORETA.

\begin{table}[h!]
    \centering
    \begin{tabular}{|c|c|c|c|c|c|}
    \hline
       \diagbox{Method}{Component} & 14 \unit{\milli\second} & 16 \unit{\milli\second} & 20 \unit{\milli\second} & 22 \unit{\milli\second} & 30 \unit{\milli\second} \\ \hline
         \multicolumn{6}{|c|}{Subject 1: Localization error}\\ \hline
        RW-SKF & 21 & 0.0 & 0.0, 5.7, & 0.0, 32 & 0.0, 14\\
        CR-SKF & 21 & 0.0 & 0.0, 5.7 & 0.0, 32 & 0.0, 12\\
        sLORETA & 0.0 & 13 & 0.0, 5.7 & 0.8, 17 & 0.0, 0.0\\ \hline
        \multicolumn{6}{|c|}{Subject 1: Earth mover's distance}\\ \hline
        RW-SKF & 12 & 26 & 24 & 24 & 23\\
        CR-SKF & 11 & 24 & 24 & 25 & 23\\
        sLORETA & 12 & 25 & 24 & 24 & 24\\ \hline
        \multicolumn{6}{|c|}{Subject 2: Localization error}\\ \hline
        RW-SKF & 22 & 19 & 26, 24 & 0.0, 35 & 0.0, 2.8\\
        CR-SKF & 0.6 & 0.0 & 0.0, 0.0 & 0.0, 0.0 & 0.0, 22\\
        sLORETA & 6.9 & 11 & 8.6, 33 & 8.6, 33 & 5.8, 33\\ \hline
        \multicolumn{6}{|c|}{Subject 2: Earth mover's distance}\\ \hline
        RW-SKF & 14 & 30 & 26 & 25 & 24\\
        CR-SKF & 15 & 35 & 26 & 26 & 25\\
        sLORETA & 13 & 32 & 26 & 26 & 26\\ \hline
    \end{tabular}
    \caption{Localization errors and earth mover's distances obtained in the cases of the first and second subject. In rows dedicated to components from 22 \unit{\milli\second} to 30 \unit{\milli\second}, there are two sources; hence, the cells contain two localization errors: first, for the localization of the source at the thalamus, and second, for the cortical source.}
    \label{tab:realDLEEMD}
\end{table}

\clearpage

\begin{figure}
    \centering
    \includegraphics[width=0.45\linewidth]{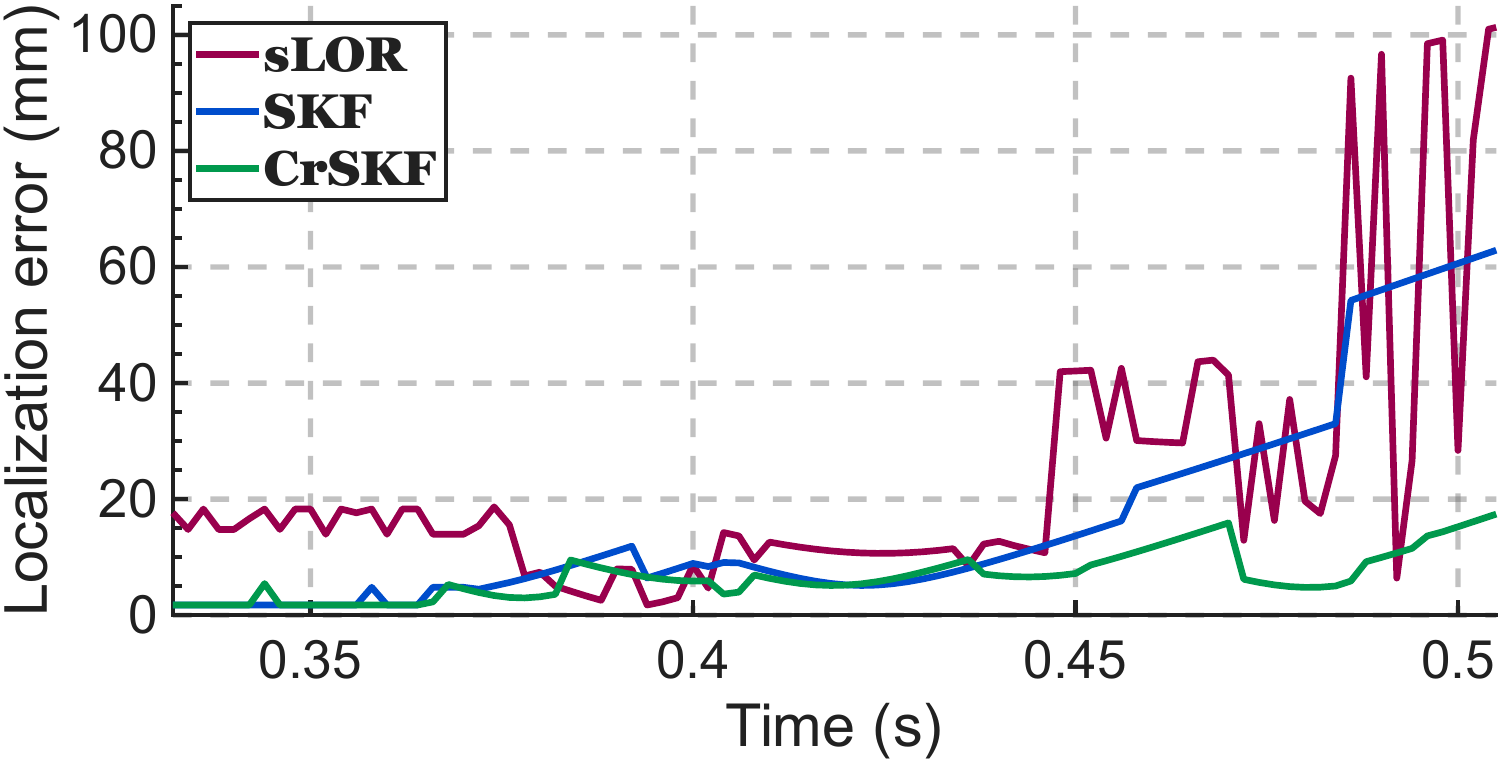}

    \includegraphics[width=0.45\linewidth]{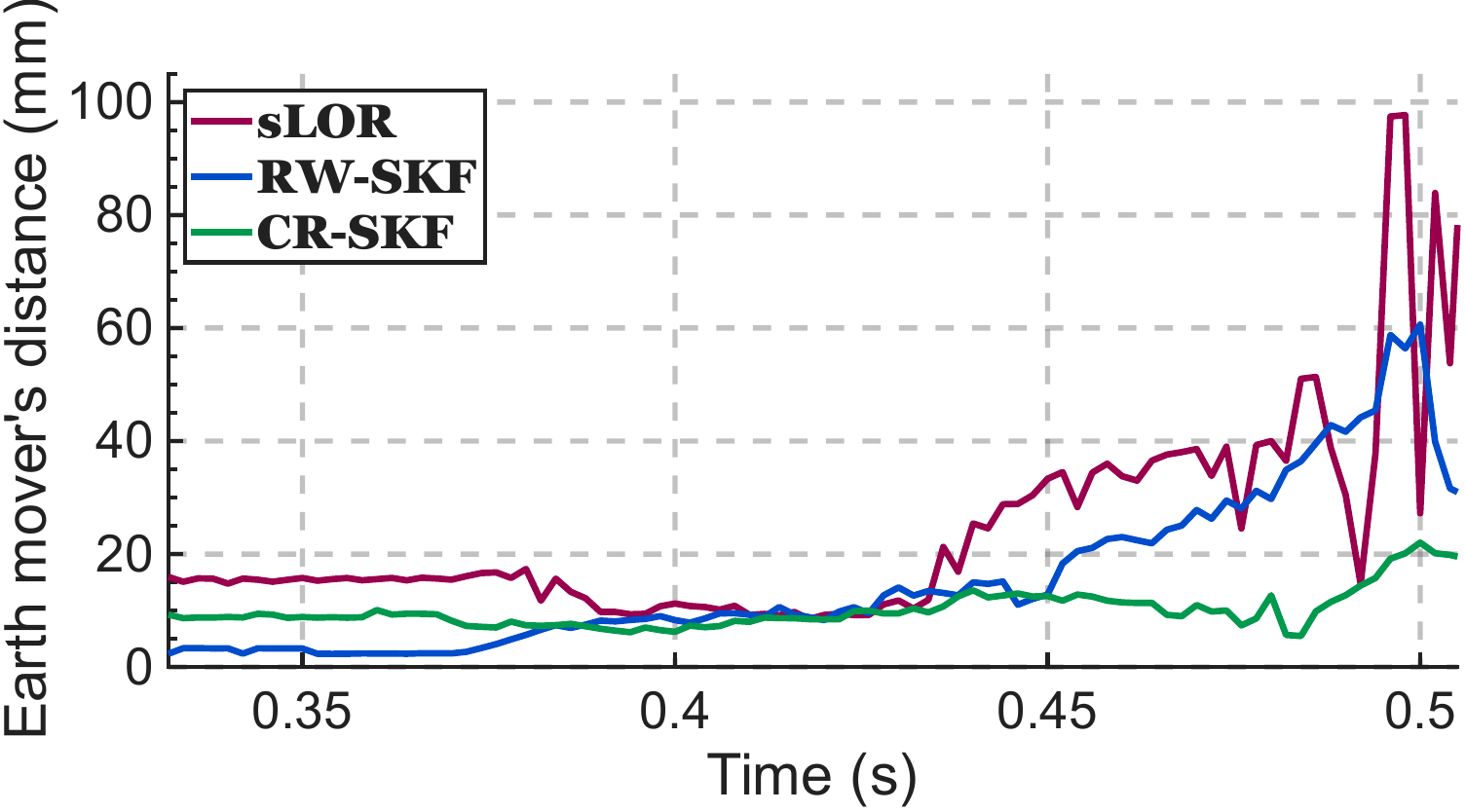}
    \caption{Localization error (top) and earth mover's distance (bottom) for sLORETA (red), Standardized Kalman filter (blue), and Change-rate Standardized Kalman filter (green) when estimating the simulated epileptic activity modeled using the nonlinear Jansen-Rit neural mass model.}
    \label{fig:EpilepsyDLEMD}
\end{figure}

\subsection{Numerical simulation of nonlinearly modeled epilepsy activity}

The results with Jansen-Rit modeled epileptic activity, presented in Figure \ref{fig:EpilepsyDLEMD}, show the time course of localization error and earth mover's distance for all three compared methods. At the beginning of the activity, when the signal-to-noise ratio is poor and the simulated activity is most focal, both Standardized Kalman filtering approaches achieve perfect localization, with small errors caused by the separate source spaces used in forward modeling and inversion. Slightly before 0.4 \unit{\second} (0.35 \unit{\second}, to be exact), the localization accuracy of sLORETA improves from 18 \unit{\milli\meter} to near zero when the signal-to-noise ratio is largest. During that time, the localization error of the standardized Kalman filter increases to near 1 \unit{\centi\meter}. After that, the simulated activity begins to spread, and its centre of mass shifts. This results in a spike-like increase of localization error for sLORETA and a modest increase with RW-SKF. Conversely, the localization error of CR-SKF does not exceed 1.8 \unit{\centi\meter}, thus providing the most stable localization. EMD provides a similar picture of the situation, where the CR-SKF estimation does not begin to scatter, unlike the compared methods. Moreover, a change in the estimation spread obtained with sLORETA and RW-SKF due to a mere increase in signal-to-noise ratio at the interval from the beginning to 0.35 \unit{\second} is not desirable behavior.

The experiment demonstrates the estimation power of the linear Kalman filter approach against a nonlinear observation model and the stability of the change-rate model against spatiotemporal changes in activity.

\section{Discussion}
In this paper, we aim to improve the evolution modeling of Standardized Kalman filtering (SKF). In practice, we introduced the change rate as an additional trackable variable and the second-order backward difference formula to improve tracking smoothness and robustness against measurement noise, thereby enabling otherwise challenging brain activity estimation and tracking from electroencephalography SEP data. In addition, the averaged sensitivity weighting \cite{Calvetti2019SensitivityWeight} was used for initializing the algorithms.

The results demonstrate the benefit of time-dependent tracking, which compares favorably with time-independent sLORETA, a method that was successful in finding the epileptic focus \cite{LealAlberto2008_4,deGooijer2013_14,Coito2019_18}. The Kalman filter can, indeed, separate deep and surface tracks and follow them accurately, unlike sLORETA, thus reducing the chance that separate activities will become enmeshed and more correlated than they really are. The results also show that tracking the change rate of the standardized current densities with posed dynamical data conditioning improves robustness against measurement noise and against poor parametrization of the evolution model, namely the selected process noise variance. This can be seen throughout, with the narrower quantile width of the CR-SKF tracks compared to that of the RW-SKF tracks.

Considering the 1st subject, the only difference was in the estimation of the 16 \unit{\milli\second} component, which could yield different outcomes at the bottom of the brainstem and in the left thalamus depending on the sampling rate and the montage-averaging approach. Both filtering methods failed to localize the 14 \unit{\milli\second} component to the brainstem and to detect surface activity at 22 \unit{\milli\second}. The reason could lie in the dynamic transition from the very focal surface-level 20 \unit{\milli\second} component to the wider 22 \unit{\milli\second} component. If the 22 \unit{\milli\second} component is not well-localizable as suggested in \cite{buchner1995somatotopy}, the montage averaging could make the peak lower as the measurements are more sensitive to the variations of surface-level activity, and thus, get greatly overshadowed by the standardized thalamic component.

The introduced prior parametrization enabled sLORETA to localize somatosensory evoked potentials nearly in textbook fashion in the first subject. The only observed flaw was in the 14 \unit{\milli\second} component, which extended from the pons to the bottom of the brainstem. The unpredictability of sLORETA becomes apparent in the second subject, where some cortical and thalamic activities are mislocalized. Another disadvantage of sLORETA is its wide estimations, which are hard to interpret visually. This problem is addressed in a paper proposing high-resolution sLORETA \cite{Sadat-NejadYounes2021HighRessLOR}. Standardized Kalman filtering, as a more focal method, does not suffer from this. By examining the thalamic estimates from Standardized Kalman filters, we see concentrated subthalamic spots indicating the source location. This suggests that if the data is clear enough, Standardized Kalman filtering could localize activities within the thalamus. This could then be used as a guidance method for deep brain stimulation of the anterior nucleus of the thalamus since the anterior nucleus of the thalamus is suspected to be involved in refractory epilepsy \cite{BouwensvanderVlisTimA.M.2019Dbso}. 

For the 2nd subject, the change-rate-modeled version of the Standardized Kalman filter estimates the activity with significantly greater concordance with the literature. The reason lies in the data quality, which is poorer for the 2nd subject, as evident in the comparison of the highly dipolar peak brain topographies with those of the 1st subject. As shown by the 1-dimensional toy experiment and numerical results, the backward difference formulation increases measurement noise robustness, which is the main differentiator here in favor of the Change Rate-SKF. As the subjects are notably different in age (49 and 32 years old), differences in the conductivity of the skull and brain tissue can be expected. The conductivity discrepancies between the model and reality are one of the most significant sources of error in the inverse brain source estimation \cite{Antonakakis2020SkullCondu,Lahtinen2023EIT,KoulouriAlexandra2025Bmpl}. Yet, given congruent results with CR-SKF, we could expect the method to be measurement-noise-robust in this regard as well. Also, the measurement noise robustness of CR-SKF is valuable on its own: Considering clinical patient data, e.g., interictal spike data from an epilepsy patient, we do not usually have thousands of spikes to montage-average or ideal measurement scenarios, but instead highly noisy data. Noisiness is one of the main challenges in epilepsy localization, which is why CR-SKF could be more attractive in clinical settings than RW-SKF.

Comparing the numerical results with the results obtained with two subjects, the major limiting factor in the investigation is the low sampling frequency of 1200 \unit{\hertz} used with the real data. The signal lacks the information provided by ultra-high-frequency signals, and some components are meshed together at a low signal-to-noise ratio due to missing sample points in between. 

When experimenting with the simulation of epilepsy activity, where the center of activity mass is moving, and the observations are modeled using a nonlinear Jansen-Rit neural mass model with spatially spreading activity, the localization accuracy and earth mover's distance indicate that the estimation with CR-SKF is more measurement-noise-robust than with the compared methods. Also, the estimation spread of CR-SKF does not deteriorate, whereas RW-SKF and sLORETA do, and they lose accuracy, leading to estimation failure in later states with those two.

While the models used in Kalman filters are linear, they provide high estimation power against complex and nonlinear systems. Besides the fact that so-called Extended Kalman filtering (EKF) is proposed for nonlinear models, it is still a linearization of otherwise nonlinear observation and evolution models -- similarly, the change-rate model is a second-order approximation of the underlying dynamic behavior. Due to this resemblance, it can be proved for CR-SKF as for EKF that it will converge to the correct estimation, when the second-order changes of the evolutions are bounded, the initial state is guessed relatively close, and noises are following Gaussian distributions \cite{KRENER2003ConvergenceofEKF}. If we were able to model the neural behavior more accurately, we could construct a nonlinear filter that is more reliable or converges faster. However, major computational challenges exist: nonlinear filters provide intractable solutions where no analytic solution exists or when computation is infeasible. In both cases, extensive use of numerical methods requires substantial computational power and time to find the solution, which could be considered impractical. In the style of EKF, one could build an SKF method that assumes a nonlinear Jansen-Rit neural mass model as the evolution model. With this approach, another danger is overspecification, since neural mass models are suitable for certain types of activity; for example, the Jansen-Rit model can model single-frequency oscillations, which limits its ability to model disorders with multifrequency alterations, such as Alzheimer’s \cite{Sanchez-TodoRoser2023JRmodel}. The aim of the proposed method is to have wide applicability, as no patient is the same, while the computation is light enough for potential practical use.

In contrast to the results of the original article proposing SKF \cite{Lahtinen2024SKF}, where the deep and surface tracks are highly enmeshed, we observe that the new parametrization significantly reduces track enmeshment and makes the tracks more robust against measurement noise. Moreover, comparing the sLORETA "tracks", we see that in the original paper, the tracks are essentially the same, except that the deep activity track is attenuated, unlike in the results of this paper. This, on its own, works as evidence for the balancing effect of sensitivity weighting.

\subsection{Suggested Parameter tuning and estimation setup workflow}
Results show the superiority of 44 \unit{\decibel} parameter value for $\rho$ from Equation (\ref{eq:qparameter}) throughout the experiments conducted. If the signal-to-noise ratio is high due to carefully conducted, reproducible measurements that can be montage-averaged, a higher decibel value can be used. In these experiments, real and numerical, 44 \unit{\decibel} strikes the balance between uncertainty and traceability of brain activity.

Based on the results of the experiment conducted with two significantly different data sampling frequencies, the longer time window from the beginning, with a low sampling rate, is advisable, as some of the initial steps of the Change Rate-SKF may show false activity until it settles into the right track.

\subsection{Method Limitations and Future Work}
Although the method proposed in this study has demonstrated promising results, it is essential to acknowledge its limitations. The Kalman filter is a linear filtering strategy that works ideally when the observation model and the underlying dynamics of the states (here, brain activity) can be described by linear models, and the perturbing Gaussian noises are mutually independent across time steps \cite{KaipioSomersalo}. We should note that the evoked potentials are neurophysiologically complex, and accurate simulations of these are nonlinear \cite{JansenRit1993,StephanieJones2007}. Accurate connections between brain regions and neural masses could be obtained by diffusion tensor imaging and fiber tractography to build an anatomical connectivity map \cite{Skudlarski2008DTI}, for example. However, incorporating empirical data into mathematical neuron models remains a challenge, particularly in balancing realism and computational efficiency \cite{BassettD2018NeuralModels}. Another possible front to improve the methodology is to use more advanced Kalman filtering approaches, such as data driven approach with input-output feedback or adaptive invariant Kalman filter used in robotics \cite{NettoMarcos2018DataDrivenKF,LiXin2024AdaptiveInvKF,KimKyung-Hwan2025AdaptiveInvKF}; Koopman-Kalman filter \cite{ZhaoDongdong2024KoopmanKF} that uses a finite approximation of the infinite-dimensional linear representation of the Koopman-von Neumann operator \cite{KordaMilan2018Koopman}, which is the dynamic operator appearing in classical mechanics. If general and accurate enough, nonlinear measurement and evolution models will be developed, and the Unscented Kalman filter (UKF) \cite{JulierUhlmann1997UKF,JulierUhlmann2004UKF} would be an attractive alternative for approximating the solution to the corresponding nonlinear filtering problem. For example, simulated epilepsy activity modeled using the Jansen-Rit neural mass model could be tracked using the UKF to estimate the corresponding model parameters for the observation model \cite{Liang2023UKFNMM}.

The Kalman filter, as a linear and simplistic model, does not suffer from computational inefficiency. However, the bottleneck of the Standardized Kalman filtering algorithm lies in the "standardization step," which requires computing the matrix square root of the full predictive source covariance matrix, i.e., performing the Schur decomposition \cite{BjorckA1983ASqrtm}. As a frame of reference, on a 4-core system with an Intel Core i5-7300HQ CPU at 2.50 \unit{\giga\hertz}, Kalman filtering takes 7 seconds, while standardization takes 50 seconds, about 7 times longer.

Approximating the square root is possible but challenging for large matrices: the correlations between sources are captured in the predictive covariance, so that any approximation would reduce the information about brain states. One option is to use the Denman-Beavers iteration \cite{DenmanBeavers1976}, which provides the inverse of a square root matrix directly, as used in the standardization step. If we have confidence in the subregions where the activity should occur, we could reduce the source space and, consequently, the dimension of the problem. Adaptive source space density having a close resemblance to the shrinkage operation used in \cite{SSLOFO2005Liu}, or Randomized Multiresolution Scanning \cite{rezaei2020randomized}, where the source space is denser around the regions in which the activity is predicted to be and coarser in regions with less importance. The Parallel Kalman filter directly addresses the computational effort problem of the Kalman filter \cite{Lantz2020ParallelKF,KoulouriAlexandra2022ParallelKF,LiuChongwei2024ParallelKF}, which could improve the computation time of the standardization step simultaneously.

Overall, more subjects and a higher sampling rate are needed to draw stronger conclusions, and the applicability of the proposed parametrization and change-rate evolution model used with the Standardized Kalman filter.

As future work, the model and parametrization can be tested using interictal data from multiple epilepsy patients or other evoked potentials, such as visually or auditorily evoked potentials, at higher sampling rates to allow better distinguishability and tracking of the individual activity components. The use of epilepsy data and the aforementioned other evoked potentials can provide further insight into the breadth of applicability of the proposed methodology. A stochastic model of a human head conductivity atlas, including the skull and conductivity changes around blood vessels, can be used to generate large synthetic subject populations with realistic head conductivity variations between subjects \cite{Moura_2021,Lahtinen2023EIT}. These synthetic subjects can be used further to investigate the capabilities of the Change rate-Standardized Kalman filter. Additionally, the potential of a multimodal approach combining a Kalman filter and diffusion tensor imaging, as well as of a nonlinear filtering approach using a neural mass model-based evolution model, could be tested as more accurate alternatives. 

\section{Conclusions}
While electroencephalography has a high temporal rate, only a few methods utilize this property \cite{Knosche2022}. Partly, the reason lies in the extensive variable space that accurate forward modeling requires \cite{rullmann2009eeg}, and partly in the dynamics, where the activity becomes "invisible" as it travels along axons from one neural mass to another. 

Kalman filtering for dipole field distribution estimation provides a time-recursive algorithm that is light enough to estimate spatiotemporal neural signals in a highly realistic manner. Standardization weighting allows recovery of deeper activity by reducing the Kalman filter's depth bias. However, good predictions require a relatively accurate estimate of the brain's activity dynamics and its statistical properties at the initial state. These models and parameters can be tedious to select, as they require multimodal consideration, e.g., the use of diffusion tensor imaging and fiber tractography to build an anatomical connectivity map \cite{Skudlarski2008DTI} for more accurate state transition modeling. 

The proposed change rate modeling provides an alternative approach. Using backward-differentiation formulas to construct an ad hoc observation model enhances stability against invalid parametrization and measurement noise. Together with sensitivity-weighted prior variances, these advances could make standardized methodologies, spatiotemporal SKF, or state-of-the-art sLORETA, attractive guiding tools for the placement of stereo encephalography sensors or implanted electrodes for deep brain stimulation.

\section{Acknowledgment}
\subsection*{Informed consent and ethics related to the used data}
The institution's ethical review board (Ethik Kommission der Ärztekammer Westfalen–Lippe und der WWU) approved all experimental procedures on 02.02.2018 (Ref. No. 2014-156-f-S). The subjects gave written informed consents before the experiments. Permission to use the data is granted for this project.

\subsection*{Declaration of competing interest}
The author declares no known competing financial interests or personal relationships that could have appeared to influence the work reported in this paper.

\subsection*{Funding}
The author is funded by the Jenny and Antti Wihuri Foundation. The article project is supported by PerEpi: 'Personalised diagnosis and treatment for refractory focal paediatric and adult epilepsy (PerEpi)', Research Council of Finland, number 344712; DAAD: 'Non-invasively reconstructing and inhibiting activity in focal epilepsy', Research Council of Finland, number 354976; and 'Flagship of Advanced Mathematics for Sensing, Imaging and Modelling', Research Council of Finland, number 359185.

\appendix

\section{Evolution prior derivation}
\label{app:evolutionVar}
Let us assume that the continuous activity $\tilde{\bf x}(t)$ follows a Wiener process. In such a case, the subsequent samples at time steps $t$ and $t+\Delta t$ have the following property:
\begin{equation}\label{eq:fact1}
    \mathrm{Var}\left[\tilde{x}_k(t+\Delta t)-\tilde{x}_k(t)\right]\propto \Delta t=f^{-1},
\end{equation}
as for a constant sampling rate, the time difference between samples is the reciprocal of the sampling frequency $f$. 

Now, by using the parametrization method for Bayesian methods \cite{Calvetti2019SensitivityWeight}, we can consider hypothetical{\em differential signal-to-noise ratio} $\mathrm{dSNR}$ that we define as
\begin{equation}
    \mathrm{dSNR}=\frac{\mathbb{E}\left[\left\|\Delta \tilde{\bf y}_t\right\|^2\right]}{\mathbb{E}\left[\left\|\Delta \tilde{\bf r}_t\right\|^2\right]}=\frac{\mathbb{E}\left[\left\|L\Delta \tilde{\bf x}_t+\Delta \tilde{\bf r}_t\right\|^2\right]}{\mathbb{E}\left[\left\|\Delta \tilde{\bf r}_t\right\|^2\right]}=\frac{q\left\|L\right\|_F^2}{\mathrm{Tr}\left(\mathrm{cov}\left[\Delta \tilde{\bf r}_t\right]\right)}+1.
\end{equation}
We cannot get access to the $\mathrm{dSNR}$ or $\mathrm{cov}\left[\Delta {\bf r}_t\right]$ in the continuous time interval; therefore, we consider those as a constant and deduce
\begin{equation}\label{eq:fact2}
    q\propto \left\|L\right\|_F^{-2}.
\end{equation}
By combining (\ref{eq:fact1}) and (\ref{eq:fact2}) and introducing decibel value $\rho$ to be the tuning parameter reflecting the stochastic behavior of constant measurements, we get
\begin{equation}
    q=\frac{10^{\rho/20}}{\left\|L\right\|_F^{2}f}.
\end{equation}

\section{Change Rate Standardized Kalman filter algorithm}\label{app:CRSKFalgo}
\begin{algorithmic}[1]
\State Compute initial parameters measurement noise covariance $R$, parameters $\bm{\theta}$ and $q$.
\State Set $P_0=\mathrm{Diag}(\bm{\theta},\: \bm{\theta}/\Delta t^2)$, $Q=\mathrm{Diag}(2q/3\,I\; \; 2q/(3\Delta t^2)\, I)$, $R=\mathrm{Diag}(R,\: 6.5R/\Delta t^2)$.
\State Compute modified measurements for BDF: ${\bf f}_t=(3{\bf y}_t/3-2{\bf y}_{t-1}+{\bf y}_{t-2}/2)/\Delta t$ for $t=3,\cdots, T$.
\State Extend the data vector ${\bf Y}_t=[{\bf y}_t^\mathrm{T}\; {\bf f}_t^\mathrm{T}]^\mathrm{T}$
\State Set modified lead field: $H=\mathrm{Diag}(L,\: L)$
\State Denote the state vector ${\bf X}_t=[{\bf x}_t^\mathrm{T}\; {\bf v}_t^\mathrm{T}]^\mathrm{T}$
\For{$t=3,\cdots,T$} 
    \State $\hat{\bf X}_{t\mid t-1}=\hat{\bf X}_{t-1\mid t-1}$
    \State $P_{t\mid t-1}=P_{t-1\mid t-1}+Q$
    \State $S_t=HP_{t\mid t-1}H^\mathrm{T}+R$ 
    \State $K_t=P_{t\mid t-1}H^\mathrm{T}S_t^{-1}$
    \State $\hat{\bf X}_{t\mid t}=\hat{\bf X}_{t\mid t-1}+K_t({\bf Y}_t-H\hat{\bf X}_{t\mid t-1})$
    \State $P_{t\mid t}=P_{t\mid t-1}-K_tS_tK_t^\mathrm{T}$
    \State $E=[I,\: O]$, $\Sigma_{t\mid t-1}=EP_{t\mid t-1}E^\mathrm{T}$,
    \State $W_t =\mathrm{Diag}\left(EP_{t\mid t-1}^{-1/2}K_tS_tK_t^\mathrm{T}P_{t\mid t-1}^{-1/2}E^\mathrm{T}\right)^{-1/2}\Sigma_{t\mid t-1}^{-1/2}$
    \State $\hat{\bf z}_{t\mid t}=W_t\hat{\bf x}_{t\mid t}$
\EndFor 
\end{algorithmic}

With steps 9 -- 13, we can use the symmetry of covariance matrices whose structure can be divided into the following $2\times 2$ block formulation
\begin{equation}
    \begin{bmatrix}
        A & C^\mathrm{T}\\ C&B
    \end{bmatrix}
\end{equation}
yielding 3 smaller linear matrix equations, where sub-matrices $A$, $B$, and $C$ need to be only considered.

\section{Jansen-Rit neural mass model}
\label{app:JRNMM}
The neural mass model derived by Jansen and Rit \cite{JansenRit1993} describes the electrical firing behavior of a neural population at the mesoscopic scale. The model uses three interacting population types to model the cortical column, which are pyramidal neurons, excita-
tory interneurons, and inhibitory interneurons. The main pyramidal population activates both interneuron groups, which then project back onto the pyramidal cells.

The behavior of each population depends on two distinct transformations. The first maps the mean rate of incoming action potentials to a corresponding average postsynaptic membrane potential, either excitatory or inhibitory. For excitatory inputs $f^{(e)}(t)$, this relationship is described by a second-order differential equation,
\begin{equation}
    \ddot{u}^{(e)}(t)+2a\dot{u}^{(e)}(t)+a^2u^{(e)}(t)=aAf^{(e)}(t),
\end{equation}
and for inhibitory inputs,
\begin{equation}
    \ddot{u}^{(i)}(t)+2b\dot{u}^{(i)}(t)+b^2u^{(i)}(t)=bBf^{(i)}(t),
\end{equation}
where $a$ and $b$ are the aggregated form of the sums of the inverses of the passive membrane time constant, and $A$ and $B$ are the amplitudes of the excitatory and inhibitory post-synaptic
potentials.

Following \cite{JansenRit1993,Escuain-PooleLara2018JRmodelCoupld}, a multiple-column system can be expressed for $k$th column as
\begin{equation}
    \begin{split}
        \ddot{u}^{(0)}_k(t)+2a\dot{u}_k^{(0)}(t)+a^2u_k^{(0)}&=Aa\mathrm{Sigm}\left(u_k^{(e)}(t)-u_k^{(i)}(t)\right),\\
        \ddot{u}^{(e)}_k(t)+2a\dot{u}_k^{(e)}(t)+a^2u_k^{(e)}&=Aa\Bigg(\sum_{j=1}^NK_{kj}\mathrm{Sigm}\left(u_k^{(e)}(t-\tau_{kj})-u_k^{(i)}(t-\tau_{kj})\right)\\
        &+C_2\mathrm{Sigm}\left(C_1u_k^{(0)}(t)\right)+p_k(t)\Bigg),\\
        \ddot{u}^{(i)}(t)+2b\dot{u}^{(i)}(t)+b^2u^{(i)}(t)&=bBC_4\mathrm{Sigm}\left(C_3u_k^{(0)}(t)\right),
    \end{split}
\end{equation}
where $K$ is the adjacent coupling matrix, $C_1$ to $C_4$ are connectivity constants, $p_k$ is the possible external input assigned here as non-zero for one cortical column as the "igniting" impulse. The function $\mathrm{Sigm}(\cdot)$ is the sigmoid function
\begin{equation}
    \mathrm{Sigm}(\nu)=\frac{2e_0}{1+e^{r(\nu_0-\nu)}},
\end{equation}
where $e_0$ is the maximum firing rate, $r$ is the slope parameter, and $\nu_0$ is the post-synaptic potential for which half of the firing rate is obtained. Table \ref{tab:JRvalues} provides the used values for each of these parameters.

\begin{table}[h!]
    \centering
    \begin{tabular}{|c|c|}\hline
        $a$ & 100 \unit{\per\second} \\
        $b$ & 50 \unit{\per\second} \\
        $A$ &  3.25 \unit{\milli\volt}\\
        $B$ &  22.0 \unit{\milli\volt}\\
        $C_1$ &  135\\
        $C_2$ &  108\\
        $C_3$ &  33.75\\
        $C_4$ &  33.75\\
        $e_0$ &  2.5 \unit{\per\second}\\
        $\nu_0$ & 6 \unit{\milli\volt} \\
        $r$ & 0.56 \unit{\per\milli\volt} \\
        $K_{ij}$ &  10 or 0\\
        $p_k$ & 200 \unit{\per\second} \\\hline
    \end{tabular}
    \caption{Used values of coupled Jansen-Rit multicolumn model.}
    \label{tab:JRvalues}
\end{table}

\clearpage

\bibliographystyle{elsarticle-num} 
\bibliography{bibliography}



%\begin{thebibliography}{00}

%% For numbered reference style

%\bibitem{lamport94}
%  \textit{\LaTeX: a document preparation system},
%  Addison Wesley, Massachusetts,

%\end{thebibliography}
\end{document}